\documentclass[10pt]{article}
\usepackage{amssymb,amsmath,natbib,amsthm,graphicx,anysize,epstopdf,hyperref,dsfont,geometry,comment,float,color}
\hypersetup{colorlinks=false}
\newcommand\kh{K_1\left( \frac{\mathbf{X}_t-\mathbf{x}}{h_{1n}} \right)}
\newcommand\kv{K_2\left( \frac{\mathbf{X}_t-\mathbf{x}}{h_{2n}} \right)}
\newcommand\kl{K_3\left( \frac{\hat{\varepsilon}_t-y}{h_{3n}} \right)}

\newcommand\logg{log\,g}
\newcommand\convd{\stackrel{d}{\rightarrow}}
\newcommand\convp{\stackrel{p}{\rightarrow}}
\newcommand\sumin{\overset{n}{\underset{i=1}{\sum}}}
\newcommand\sumtn{\overset{n}{\underset{t=1}{\sum}}}

\newcommand\ntoi{n \rightarrow \infty}

\begin{document}
\newtheorem{df}{Definition}
\newtheorem{lem}{Lemma}
\newtheorem{thm}{Theorem}
\newtheorem{cor}{Corollary}

\setlength{\baselineskip}{12pt}
\begin{center} \large \sc Nonparametric estimation of conditional value-at-risk and expected shortfall based on extreme value theory\normalsize\footnote{We thank Peter C. B. Phillips, Eric Renault, an Associate Editor and an anonymous referee for comments that improved the paper substantially.  Any remaining errors are the authors' responsibility.} \rm  \\[0.25in]

\begin{tabular}{lcl}
\multicolumn{3}{c}{ \sc Carlos Martins-Filho}\\[.1in] 
Department of Economics &  &IFPRI \\  
University of Colorado &  & 2033 K Street NW  \\
Boulder, CO 80309-0256, USA& \&&Washington, DC 20006-1002, USA\\ 
email: carlos.martins@colorado.edu& & email: c.martins-filho@cgiar.org\\
Voice: + 1 303 492 4599 & &Voice: + 1 202 862 8144\\
\end{tabular}\\[.1in]

\begin{tabular}{lcl}
\multicolumn{3}{c}{ \sc Feng Yao}\\[.1in]
Department of Economics & &School of Economics and Trade  \\  
West Virginia University& &Guangdong University of Foreign Studies\\
Morgantown, WV 26505, USA& \& & Guangzhou, Guangdong 510006, P. R. China\\ 
email: feng.yao@mail.wvu.edu & & email: 201470006@oamail.gdufs.edu.cn\\
Voice: +1 304 293 7867 & &  Voice: + 86 20 3932 8858
\end{tabular}\\[.1in]

\begin{tabular}{l}
\multicolumn{1}{c}{ \sc Maximo Torero}\\[.1in]
IFPRI \\  
2033 K Street NW \\
Washington, DC 20006-1002, USA\\ 
email: m.torero@cgiar.org\\
Voice: + 1 202 862 5662\\[.1in]
\end{tabular}

October, 2016\\[.1in]
\end{center}
\noindent \bf Abstract. \rm We propose nonparametric estimators for conditional value-at-risk (CVaR) and conditional expected shortfall (CES) associated with conditional distributions of a series of returns on a financial asset.  The return series and the conditioning covariates, which may include lagged returns and other exogenous variables, are assumed to be strong mixing and follow a nonparametric conditional location-scale model.  First stage nonparametric estimators for location and scale are combined with a generalized Pareto approximation for distribution tails proposed by \cite{Pickands1975} to give final estimators for CVaR and CES.  We provide consistency and asymptotic normality of the proposed estimators under suitable normalization.  We also present the results of a Monte Carlo study that sheds light on their finite sample performance.  Empirical viability of the model and estimators is investigated through a backtesting exercise using returns on future contracts for five agricultural commodities.\\[.1in]

\noindent \bf Keywords and phrases. \rm conditional value-at-risk, conditional expected shortfall, extreme value theory, nonparametric location-scale models, strong mixing.\\[.1in]

\noindent \bf JEL classifications. \rm C14, C15, C22, G10.\rm \\[.1in]
\noindent \bf AMS-MS classifications. \rm Primary: 62G32, 62G07, 62G08, 62G20.\rm
\clearpage
\setlength{\baselineskip}{24pt}
\pagestyle{plain}
\setcounter{page}{1}
\setcounter{footnote}{0}
\section{Introduction}
Conditional value-at-risk (CVaR) and conditional expected shortfall (CES) are two of the most used synthetic measures of market risk  in empirical finance (see \cite{Duffie2003}, \cite{McNeil2005} and \cite{Danielsson2011}).  From a statistical perspective they have straightforward definitions.  Let $\{Y_t\}$ denote a stochastic process representing the returns\footnote{Let $P_t$ denote the price of a financial asset at time $t$.  In this paper a ``return" is defined as $Y_t =- \text{log} \, \frac{P_t}{P_{t-1}}$.  We adopt this definition because in practice, regulators, portfolio and risk managers are mostly concerned with the distribution of losses, i.e., negative values of $\text{log} \, \frac{P_t}{P_{t-1}}$.} on a given stock, portfolio or market index, where $t \in \mathds{Z}$ indexes a discrete measure of time, and $F_{Y_t|\mathbf{X}_t=\mathbf{x}}$ denote the conditional distribution of $Y_t$ given $\mathbf{X}_t=\mathbf{x}$.  The vector $\mathbf{X}_t \in \mathds{R}^d$ normally includes lag returns $\{Y_{t-\ell}\}_{ 1 \leq \ell \leq p}$, for some $p \in \mathds{N}$, as well as other relevant conditioning variables that reflect economic or market conditions.  Then, for $ a \in (0,1)$,  $a$-CVaR($\mathbf{x}$) is defined to be the $a$-quantile associated with $F_{Y_t|\mathbf{X}_t=\mathbf{x}}$ and $a$-CES($\mathbf{x}$) is defined to be the conditional expectation of $Y_t$ given that $Y_t$ exceeds $a$-CVaR($\mathbf{x}$), i.e., $a$-CES($\mathbf{x}$)=$E(Y_t|Y_t>\mbox{$a$-CVaR($\mathbf{x}$),\,$\mathbf{X}_t=\mathbf{x}$})$.\footnote{For $a\in (0,1)$ and an arbitrary distribution function $F$, we define the $a$-quantile associated with $F$ as $\text{inf}\,\{s:F(s)\geq a\}$.}

In this paper we consider the estimation of $a$-CVaR($\mathbf{x}$) and $a$-CES($\mathbf{x}$) for processes $\{Y_t\}$ that admit a location-scale representation 
\begin{equation}\label{ls}
Y_t=m(\mathbf{X}_t)+ h^{1/2}(\mathbf{X}_t)\varepsilon_t, 
\end{equation}
where $m$ and $h>0$ are nonparametric functions defined on the range of $\mathbf{X}_t$, $\varepsilon_t$ is independent of $\mathbf{X}_t$, and $\{\varepsilon_t\}$ is an independent and identically distributed (IID) innovation process with $E(\varepsilon_t)=0$, $V(\varepsilon_t)=1$ and distribution function $F$ belonging to a suitably restricted class (see section 2).  This representation can be viewed as a nonparametric generalization of certain autoregressive conditionally heteroscedastic (ARCH) structures and has been studied by \cite{Masry1995}, \cite{Hardle1997}, \cite{Masry1997} and \cite{Fan1998}, among others.\footnote{We rule out more general innovation processes, such as those appearing in semi-strong GARCH processes, that are simply stationary, ergodic with $\{\varepsilon_t^2\}$ satisfying a martingale difference condition.  See, e.g., \cite{Drost1993}, \cite{Escanciano2009a} and \cite{Linton2010}.}  Under \eqref{ls} we have
\begin{equation}\label{cvar}
a\mbox{-CVaR}(\mathbf{x}) \equiv q_{Y_t|\mathbf{X}_t=\mathbf{x}}(a)=m(\mathbf{x})+ h^{1/2}(\mathbf{x})q(a)
\end{equation}
and
\begin{equation}\label{ces}
a\mbox{-CES}(\mathbf{x}) \equiv E\left( Y_t| Y_t >q_{Y_t|\mathbf{X}_t=\mathbf{x}}(a),\,\mathbf{X}_t=\mathbf{x}\right)=m(\mathbf{x})+ h^{1/2}(\mathbf{x})E(\varepsilon_t| \varepsilon_t>q(a)),
\end{equation}
where $q_{Y_t|\mathbf{X}_t=\mathbf{x}}(a)$ denotes the conditional $a$-quantile associated with $F_{Y_t|\mathbf{X}_t=\mathbf{x}}$ and $q(a)$ is the $a$-quantile associated with $F$.  

In insurance and empirical finance, where regulators and portfolio managers are interested in high levels of risk, with $a$ in the vicinity of 1 (\cite{Chernozhukov2001,Tsay2010}), an important concern for inference is that conventional asymptotic theory does not apply sufficiently far in the tails of $F_{Y_t|\mathbf{X}_t=\mathbf{x}}$ (see Chernozhukov (2005, p. 807)).  Hence, we focus on proposing and characterizing the asymptotic behavior of nonparametric estimators for $a$-CVaR($\mathbf{x}$) and $a$-CES($\mathbf{x}$) when $a \rightarrow 1$.  In the conditional quantile (regression) literature this is commonly referred to as \it extreme \rm quantile regression (\cite{Chernozhukov2005}) and in empirical finance the corresponding notion is that of \it extreme \rm CVaR, see e.g., Embrechts et al. (1997, p. 349) and Chernozhukov and Umantsev (2001, p. 273).  

Nonparametric estimators of $a$-CVaR and $a$-CES for the non-extreme case, where $a$ is fixed in $(0,1)$, have been proposed and studied in several papers.  Since $a$-CVaR($\mathbf{x}$) is a conditional quantile, estimation can naturally proceed using nonparametric regression quantiles as in \cite{Yu1998}, \cite{Cai2002}, \cite{Cosma2007} or \cite{Cai2008}.  These estimators for $a$-CVaR($\mathbf{x}$) can then be used to produce nonparametric estimators for $a$-CES($\mathbf{x}$) as in \cite{Scaillet2005}, \cite{Cai2008}, \cite{Kato2012} and \cite{Linton2013}.  Also, for the non-extreme case, there exists a large literature on nonparametric estimation of \it unconditional \rm value-at-risk and expected shortfall, see e.g., \cite{Scaillet2004}, \cite{Chen2005a}, \cite{Chen2008}, \cite{Linton2013} and \cite{Hill2014}, that is indirectly related to our work.

We propose a two-stage estimation procedure for $a$-CVaR and $a$-CES.  First, motivated by \eqref{ls}, nonparametric estimators for $m$ and $h$ are obtained and used to produce standardized residuals.  Second, these residuals are used to obtain estimators for $q(a)$ and $E(\varepsilon_t|\varepsilon_t>q(a))$ using a likelihood procedure to estimate the parameters of a Generalized Pareto Distribution (GPD) that approximates the upper right tail of $F$.  This stage is motivated by Theorem 7 in \cite{Pickands1975} and extends the work of  \cite{Smith1987}.  These estimators for $q(a)$ and $E(\varepsilon_t|\varepsilon_t>q(a))$ are then combined with the first stage estimators for $m$ and $h$ to produce estimators for $a$-CVaR($\mathbf{x}$) and $a$-CES($\mathbf{x}$).  To our knowledge, this two stage estimation approach was first proposed by \cite{McNeill2000} in the case where $m$ and $h$ are indexed by a finite dimensional parameter.  They provided no asymptotic characterization or finite sample properties for the resulting estimators of conditional value-at-risk or expected shortfall.  However, their backtesting exercise on several time series of selected market indexes provided encouraging evidence of the estimators' performance.  \cite{Martinsfilho2006} generalized the estimation framework of McNeil and Frey to the case where $m$ and $h$ are nonparametric functions.  They demonstrated via an extensive Monte Carlo simulation, and through backtesting, that accounting for nonlinearities in $m$ and $h$ can be important in improving the estimators' finite sample performance.  \cite{Martinsfilho2012} provides an asymptotic characterization of the two stage estimation procedure for $a$-CVaR when $a \rightarrow 1$  in a model with constant and unknown variance ($h(\mathbf{X}_t)=\theta$) and a process $\{(Y_t,\mathbf{X}_t^T)\}_{t\in \mathds{Z}}$ ($\mathbf{x}^T$ indicates the transposition of the vector $\mathbf{x}$) that is IID.  Their results, however, are of limited use in empirical finance where the IID assumption is untenable and $h$ cannot be adequately modeled as a constant function of $\mathbf{X}_t$.  Furthermore, by restricting attention to the case where $\mathbf{X}_t \in \mathds{R}$, i.e., a scalar, they failed to elucidate the restrictions that the dimension $d$ may impose on nonparametric estimation of conditional value-at-risk and expected shortfall.    

Here, we extend \cite{Martinsfilho2012} in three important directions:  a) we relax the assumption that $\{(Y_t,\mathbf{X}_t^T)\}_{t\in \mathds{Z}}$ is an IID process and instead consider the case where the process is strictly stationary and strong mixing of a suitable order.  This allows for the presence of lagged values of $Y_t$ in the conditioning vector $\mathbf{X}_t$, a possibility not covered in our earlier paper and of significant practical interest; b) we allow the conditional variance $h$ to be a non-constant function of $\mathbf{X}_t$; and c) we consider the estimation of $a$-CVaR($\mathbf{x}$) \it and \rm $a$-CES($\mathbf{x}$).  We first establish consistency and asymptotic normality of the estimators for $q(a)$ and $E(\varepsilon_t|\varepsilon_t>q(a))$ based on the maximum likelihood estimators for the parameters of the (approximating) GPD using residuals from the first stage estimation.  From a technical perspective, this extends the results in \cite{Smith1987} to the case where only \it residuals\rm, rather than the actual sequence $\{\varepsilon_t\}$, are observed.  These results are used to obtain consistency and asymptotic normality of our proposed estimators for $a$-CVaR($\mathbf{x}$) and $a$-CES($\mathbf{x}$).  Estimators, like ours, that rely on tail approximations based on the GPD are normally asymptotically biased and require bias correction for valid inference.  Hence, we provide bias-corrected versions of our estimators that can be easily used for inference as their asymptotic distributions are correctly centered.  

Besides this introduction, this paper has five more sections and two appendices.  Section 2 provides a discussion of the main restrictions we impose on $F$, as well as a motivation, description and discussion of the estimation procedure.  Section 3 contains a list of assumptions needed to study the estimators for $m$ and $h$ and the main theorems that describe the asymptotic behavior of our estimators.  Section 4 contains a Monte Carlo study that sheds light on the finite sample behavior of our estimators and contrasts their performance with the estimators proposed by \cite{Cai2008}.  Section 5 provides an empirical application in which $a$-CVaR and $a$-CES are estimated using time series of returns on future contracts for five widely traded agricultural commodities.  A backtesting exercise is also conducted for each of the time series.  Section 6 provides concluding remarks and gives some directions for future research.  Tables and figures associated with the Monte Carlo study and the empirical exercise are provided in Appendix 1.  The proofs for Theorems 2 through 5 are provided in Appendix 2.  Supporting lemmas and their proofs, as well as the proof of Theorem 1, can be found in the online supplement to this article available at Cambridge Journals Online (\texttt{journals.cambridge.org/ect}).
  

\section{Estimation of $a$-CVaR and $a$-CES}
As stated in the introduction, our estimation procedure has two stages.  In the first stage we produce a sequence of standardized nonparametric residuals based on the estimation of $m$ and $h$.  Given a sample $\{(Y_t,\mathbf{X}_t^T)\}_{t=1}^n$, we consider a local linear (LL) estimator for $m$, denoted by $\hat{m}(\mathbf{x}) \equiv \hat{\beta}_0$, where 
$
(\hat{\beta}_0,\hat{\beta}) \equiv \underset{\beta_0,\beta}{\mbox{argmin}} \sum_{t=1}^n \left(Y_t-\beta_0 -(\mathbf{X}_t^T-\mathbf{x}^T) \beta\right)^2\kh
$, 
$K_1(\cdot)$ is a multivariate kernel function and $h_{1n}>0$ is a bandwidth.  For the estimation of $h$, we follow the procedure proposed in \cite{Fan1998}, where we obtain a sequence $\{\hat{U}_t \equiv Y_t - \hat{m}(\mathbf{X}_t)\}_{t=1}^n$ and define $\hat{h}(\mathbf{x}) \equiv \hat{\eta}$, where $(\hat{\eta},\hat{\eta}_1) \equiv \underset{\eta,\eta_1}{\mbox{argmin}} \sum_{t=1}^n \left(\hat{U}_t^2-\eta -(\mathbf{X}_t^T-\mathbf{x}^T) \eta_1 \right)^2\\ \kv$, $K_2(\cdot)$ is a multivariate kernel function and $h_{2n}>0$ is a bandwidth, both potentially different from those used in the definition of $\hat{m}$.\footnote{Since $\mathbf{X}_t$ may contain up to $p$ lagged values of $Y_t$, the effective sample size used in the estimation of $\hat{m}(\mathbf{x})$ and $\hat{h}(\mathbf{x})$ is $n-p$.  However, for notational ease, we assume that $Y_0,Y_{-1},\cdots $ are observed as needed to define the relevant sums of length $n$.}  The estimators $\hat{m}(\mathbf{x})$ and $\hat{h}(\mathbf{x})$ are used to produce a sequence of standardized nonparametric residuals $\{\hat{\varepsilon}_t\}_{t=1}^n$, where 
\begin{equation}\label{sr}
\hat{\varepsilon}_t= \left\{ 
\begin{array}{ccc}
  \frac{Y_t-\hat{m}(\mathbf{X}_t)}{\hat{h}^{1/2}(\mathbf{X}_t)},& \mbox{ if $\hat{h}(\mathbf{X}_t)>0$}  \\
 0, &   \mbox{ if $\hat{h}(\mathbf{X}_t)\leq 0$} 
\end{array} \right., \mbox{ for $t=1,\cdots , n$}.
\end{equation}
In the second stage, we use these residuals to construct estimators for $q(a)$ and $E(\varepsilon_t| \varepsilon_t>q(a))$, appearing in \eqref{cvar} and \eqref{ces}, which are combined with $\hat{m}(\mathbf{x})$ and $\hat{h}(\mathbf{x})$ to produce our estimators for $a\mbox{-CVaR}(\mathbf{x})$ and $a\mbox{-CES}(\mathbf{x})$.  Whereas to motivate the first stage of estimation, the only restrictions imposed on $F$ were that it has mean zero and variance one, the motivation for the second stage requires additional restrictions.  First, we assume 

\noindent \bf Assumption FR1\rm: a) $F$ is strictly monotonic and absolutely continuous with positive density $f$ such that for some $k_0<0$, $\underset{x \rightarrow \infty}{\mbox{lim}}\frac{xf(x)}{1-F(x)}=-\frac{1}{k_0}$; b) $f$ is such that $\int |\varepsilon|^{4+\epsilon}f(\varepsilon) d \varepsilon< \infty$ for some $\epsilon>0$; c) $f$ is $m_1 \geq 2$-times continuously differentiable with $\left| \frac{d^j}{d u^j}f(u)\right|<C$ for some constant $C$ and $j=1,\cdots , m_1$.

\noindent \bf Remarks on FR1: \rm1. By Proposition 1.15 in \cite{Resnick1987}, if $F$ satisfies FR1 a), then it belongs to the maximum domain of attraction of a Fr\'echet distribution with parameter $-1/k_0$, denoted here by $F \in D(\Phi_{-1/k_0})$.\footnote{See \cite{Leadbetter1983}, \cite{Resnick1987} or \cite{Embrechts1997a} for the definition of maximum domains of attraction.}  This, in turn, is equivalent to $L(x)=(1-F(x))x^{-1/k_0}$ being slowly varying as $x \rightarrow \infty$ (see  \cite{Gnedenko1943}).  Thus, any $F$ satisfying FR1 a) is such that $(1-F(x))x^{-1/k_0}$ is slowly varying at infinity.

\noindent 2. The restriction that $F$ belongs to the domain of attraction of a Fr\'echet distribution is not entirely arbitrary.  There are only two other possibilities: a) $F$ belongs to the domain of attraction of a (reverse) Weibull distribution, in which case $F$ has a finite right endpoint, a restriction that is not commonly placed on the innovation associated with location-scale models; b) $F$ belongs to the domain of attraction of a Gumbel distribution, in which case, when $F$ has an infinite right endpoint $1-F$ is rapidly varying at infinity (\cite{Resnick1987}), a case we must avoid to derive the asymptotic properties of our estimators.  Thus, distribution functions where $1-F(x)$ decays exponentially fast as $x \rightarrow \infty$ are ruled out by FR1 a).

\noindent 3. We note that assumptions FR1 a) and b) imply that $-1/(4 + \epsilon)<k_0<0$.  This ensures that the (right) tail $1-F(x)$ decays sufficiently fast as $x \rightarrow \infty$.  It rules out distribution functions with right tails that are ``too thick" in the sense of $k_0 \leq -1/4$.  Hence, $F$ is in a class of distributions that can have thick tails, but not so thick as to prevent the existence of moments slightly larger than four.  The restriction is critical for the asymptotic results we derive and can be empirically binding in that many financial time series appear to have heavy tails with $k_0 \leq -1/2$ (\cite{Embrechts1997a}).  

\noindent 4. FR1 c) is needed to provide asymptotic characterizations of our proposed estimators (see, e.g., the proof of Lemma 5), but is not required to provide a motivation for their definition.

Theorem 7 in \cite{Pickands1975} shows that if $F \in D(\Phi_{-1/k_0})$, then its \it extreme \rm upper tail is uniformly close to a generalized Pareto distribution (GPD).  Formally, for $k_0<0$ and some function $\sigma_0(\xi)>0$ with $\xi \in \mathds{R}$,
\begin{equation}\label{4}
F \in D(\Phi_{-1/k_0}) \iff \underset{ \xi \rightarrow \infty}{\mbox{lim}}\,\, \underset{\xi<\xi+u<\infty}{\mbox{sup}}\left|F_\xi(u)-G(u;\sigma_0(\xi),k_0)\right|=0,
\end{equation}
where $F_\xi(u)=\frac{F(u+\xi)-F(\xi)}{1-F(\xi)}$ and $G(u;\sigma,k)=1-(1-ku/\sigma)^{1/k}$ with $0<u<\infty$.  As in Davis and Resnick (1984, p. 1471) and Smith (1987, p. 1176), we use the equivalence in \eqref{4} to motivate our estimator for $q(a)$.  To this end, let $N_n$ be a nonstochastic subsequence on $n$ such that $N_n \rightarrow \infty$ and $\frac{N_n}{n} \rightarrow 0$ as $n\rightarrow \infty$.  For notational simplicity put $N \equiv N_n$ and define $a_N=1-\frac{N}{n}$.  Then, $a_N \rightarrow 1$ and $q(a_N) \rightarrow \infty$ as $n \rightarrow \infty$.  Since \eqref{4} is valid for any sequence $\xi \rightarrow \infty$, we put $\xi \equiv q(a_N)$ and note that by strict monotonicity of $F$, $1-F(q(a_N))=N/n$.  Then, putting $a \equiv  F(u+q(a_N))$ and noting that $\frac{1-F(u+q(a_N))}{1-F(q(a_N))}=1-F_{q(a_N)}(u)$ we have
$
\frac{n}{N}(1-a)=1-F_{q(a_N)}(u),
$ 
where $u=q(a)-q(a_N)>0$ provided $a_N<a$.  Note that since $a_N \rightarrow 1$, we have $a \rightarrow 1$ as $n \rightarrow \infty$.  By \eqref{4} we have that for $n$ sufficiently large
$
\frac{n}{N}(1-a) \approx \left( 1- \frac{k_0}{\sigma_0(q(a_N))} ( q(a) -q(a_N) ) \right)^{1/k_0}.
$
Rearranging the terms, we have
\begin{equation}\label{motiv}
q(a) \approx q(a_N)+\frac{\sigma_0(q(a_N))}{k_0}\left( 1- \left( \frac{n}{N}(1-a) \right)^{k_0}\right),
\end{equation}
which motivates our proposed estimator for $q(a)$.  We first use the residuals $\{\hat{\varepsilon}_t\}_{t=1}^n$ to estimate $F$ by integrating a Rosenblatt kernel estimator for the density $f$, i.e.,
\begin{equation}\label{6} 
\tilde{F}(u)=\frac{1}{nh_{3n}}\sumtn \int_{-\infty}^u\kl dy
\end{equation}
where $K_3(\cdot)$ is a univariate kernel and $h_{3n}>0$ is a bandwidth.  Then, we define the estimator $\tilde{q}(a_N)$ for $q(a_N)$ as the solution for $\tilde{F}(\tilde{q}(a_N))=a_N$.  We note that a simpler estimator is $\hat{q}_n(a_N)$, the $a_N$-quantile associated with the empirical distribution of the nonparametric residuals $\{\hat{\varepsilon}_t\}_{t=1}^n$.  We prefer $\tilde{q}(a_N)$ because it is well known from the unconditional distribution and quantile estimation literature (\cite{Azzalini1981,Falk1985,Yang1985,Bowman1998,Martinsfilho2008}) that smoothing beyond that attained by the empirical distribution can produce significant gains in finite samples with no impact on asymptotic rates of convergence.
    
To estimate $k_0$ and $\sigma_0(q(a_N))$ we follow the maximum likelihood procedure suggested by \cite{Smith1987} using the approximation provided by $G(u;\sigma_0(q(a_N)),k_0)$.  To this end we define the ascending order statistics $\{\hat{\varepsilon}_{(t)}\}_{t=1}^n$ and construct a sequence of exceedances  $\{\tilde{Z}_i\}_{i=1}^{N_s} \equiv \left\{ \hat{\varepsilon}_{(n-N_s+i)}-\tilde{q}(a_N)\right\}_{i=1}^{N_s}$, which are used to obtain maximum likelihood estimators for $\sigma_0(q(a_N))$ and $k_0$ based on the density $g(z;\sigma,k)=\frac{1}{\sigma}\left(1-\frac{kz}{\sigma}\right)^{1/k-1}$ associated with the GPD.  Here, it is important to note that the number of residuals that exceed $\tilde{q}(a_N)$, i.e., $N_s$ is stochastic and generally different from $N$.  Our estimators are a solution $(\tilde{\sigma}_{\tilde{q}(a_N)},\tilde{k})$ for the likelihood equations
\begin{equation}\label{foc}
\frac{\partial}{\partial \sigma}\tilde{L}_N(\sigma,k)=0 \,\,\, \mbox{ and } \,\,\,
\frac{\partial}{\partial k}\tilde{L}_N(\sigma,k)=0,
\end{equation}
where $\tilde{L}_N(\sigma,k)=\frac{1}{N} \sum_{i=1}^{N_s}  \logg (\tilde{Z}_i; \sigma,k)$.  Then, based on \eqref{motiv} our estimator $\hat{q}(a)$ for ${q}(a)$ is given by
\begin{equation}\label{10}
\hat{q}(a)=\tilde{q}(a_N)+\frac{\tilde{\sigma}_{\tilde{q}(a_N)}}{\tilde{k}}\left(1-\left(\frac{n}{N}(1-a)\right)^{\tilde{k}}\right).
\end{equation}
To motivate our estimator for $E(\varepsilon_t| \varepsilon_t>q(a))$ we place the following additional restriction on $F$,

\noindent \bf Assumption FR2\rm : For $L(x)=(1-F(x))x^{-1/k_0}$ we have $\frac{L(tx)}{L(x)}=1+k(t)\phi(x)+o(\phi(x))$ for each $t>1$, where $0<\phi(x) \rightarrow 0$ as $x\rightarrow \infty$ is regularly varying with index $\rho < 0$ and $k(t)=\frac{t^\rho-1}{\rho}$.

If the exceedances over the quantile $q(a)$ were distributed \it exactly \rm as $g(z;\sigma_0,k_0)$, then integration by parts gives $E(\varepsilon_t| \varepsilon_t>q(a))=\frac{q(a)}{1+k_0}\left(1+\frac{\sigma_0}{q(a)}\right)$ (Embrechts et al. (1997, p. 165)).  In the general case where the exceedances are not distributed as $g(z;\sigma,k)$, but $F$ satisfies assumptions FR1 a), b) and FR2 it can be easily shown (see Lemma 8 in the online supplement) that $E(\varepsilon_t| \varepsilon_t>q(a))= \frac{q(a)}{1+k_0}(1+o(1))$ .  This motivates our proposed estimator for $E(\varepsilon_t| \varepsilon_t>q(a))$ which is given by
\begin{equation}\label{10aa}
\widehat{E}(\varepsilon_t| \varepsilon_t>q(a))=\frac{\hat{q}(a)}{1+\tilde{k}}.
\end{equation}

\noindent \bf Remarks on FR2\rm: 1. Assumption FR2 is equivalent to requiring that the error in approximating the tail $1-F(x)$ by a Pareto distribution be given by $\phi(x)\left( \frac{1}{\rho} +o(1)\right)$ as $x \rightarrow \infty$  (see Theorem 2.2.2, in Goldie and Smith (1987, p. 48)).  Assumptions on how the approximating error decays are necessary for an asymptotic characterization of estimators for the parameters of the GPD.  Our FR2 is similar to the condition SR2 in \cite{Smith1987}, whereas a stronger version of it is assumed by \cite{Hall1982}.  \cite{Goldie1987} provides a comprehensive discussion of these second order assumptions on the characterization of estimators that, like ours, depend on Extreme Value Theory.

\noindent 2. FR2 is necessary in providing a first order approximation for the expected value of the score associated with the likelihood procedure, which is used in characterizing and correcting the asymptotic bias.  For example, in the case of the parameter $k_0$ (see Appendix 2) 
\begin{equation}\label{escorek}
E\left(\frac{\partial }{\partial k} \logg (Z;\sigma_0(q(a_N)),k_0) \right)=\frac{k_0^{-1} \phi(q(a_N))}{(-k_0^{-1}-\rho)(1-k_0^{-1}-\rho)} + o(\phi(q(a_N)),
\end{equation}
providing a handle on how the asymptotic bias of the estimators we propose depends on $k_0$ and $\rho$.

\noindent 3.  A zero-mean and suitably scaled (to have variance 1) Student-t distribution with degrees of freedom $v>4$ satisfies assumptions FR1, FR2 and the restrictions on the innovation in our location-scale model.  In this case, FR1 is satisfied with $k_0=-v^{-1}$.

Combining the estimators $\hat{m}$, $\hat{h}$, (\ref{10}) and (\ref{10aa}) into equations (\ref{cvar}), (\ref{ces}), we define the estimators $\hat{q}_{Y_t|\mathbf{X}_t=\mathbf{x}}(a)=\hat{m}(\mathbf{x})+\hat{h}^{1/2}(\mathbf{x})\hat{q}(a)$ and $\widehat{E}\left( Y_t| Y_t >q_{Y_t|\mathbf{X}_t=\mathbf{x}}(a),\mathbf{X}_t=\mathbf{x}\right)=\hat{m}(\mathbf{x})+ \hat{h}^{1/2}(\mathbf{x})\widehat{E}(\varepsilon_t| \varepsilon_t>q(a))$ for $a$-CVaR($\mathbf{x}$) and $a$-CES($\mathbf{x}$) associated with the series $\{Y_t\}$ and the conditioning set $\{\mathbf{X}_t=\mathbf{x}\}$.  In the next section we study the asymptotic behavior of these estimators.   
\section{Asymptotic characterization of the proposed estimators }
\subsection{Assumptions and existence of $\tilde{\sigma}_{\tilde{q}(a_N)}$ and $\tilde{k}$}

We begin the study of the asymptotic behavior of our estimators by establishing that a solution for equation (\ref{foc}) exists and corresponds to a local maximum of the likelihood function.  Our strategy is to show that score functions associated with $\tilde{L}_N(\sigma,k)=\frac{1}{N} \sum_{i=1}^{N_s}  \logg (\tilde{Z}_i; \sigma,k)$ are uniformly asymptotically equivalent in probability to those associated with $L'_N(\sigma,k)=\frac{1}{N}\sum_{i=1}^{N_1} log \, g(Z_i'; \sigma,k)$, where $Z_i'=\varepsilon_{(n-N_1+i)}-q(a_N)$,  $\{\varepsilon_{(t)}\}_{t=1}^n$ are ascending order statistics associated with $\{\varepsilon_t\}_{t=1}^n$ and $N_1$ is the stochastic number of exceedances over the nonstochastic threshold $q(a_N)$.  

This is accomplished in two steps.  First, we show in Lemma 1 (see online supplement) that the score functions associated with $L'_N(\sigma,k)$ are uniformly asymptotically equivalent in probability to those associated with $L_N(\sigma,k)=\frac{1}{N}\sum_{i=1}^{N} log \, g(Z_i; \sigma,k)$ where 
$\{Z_i\}_{i=1}^N = \left\{ \varepsilon_{(n-N+i)}-q_n\left(a_N\right)\right\}_{i=1}^N$,
and $q_n(a)=\left\{ 
\begin{array}{cc}
\varepsilon_{(na)} & \mbox{ if $na \in \mathds{N}$ }     \\
\varepsilon_{([na]+1)} &   \mbox{ if $na \notin \mathds{N}$ }  
\end{array}
\right.
$.  That is, $q_n(a_N)$ is the quantile of order $a_N$ associated with the empirical distribution $F_{n}(u)=\frac{1}{n}\sum_{t=1}^n\chi_u(\varepsilon_t)$, where $\chi_u(\varepsilon)=\left\{ 
\begin{array}{cc}
1 &   \mbox{ if $\varepsilon \leq u $ }     \\
0 &   \mbox{ if $\varepsilon >u $  }  
\end{array}
\right.$.  This, together with Lemma 5 in \cite{Smith1985}, establishes the important result that Theorem 3.2 in \cite{Smith1987} is valid for the case where a stochastic threshold, in this case $q_n(a_N)$, is used in conjunction with our nonstochastic $N$.\footnote{Smith (1987, pp. 1180-1181) observes that the use of Theorem 3.2 normally involves taking either $N$ or $q(a_N)$ as being stochastic and the other as being nonstochastic.  Throughout this paper we take $N$ as nonstochastic and let thresholds be sample dependent (stochastic).  The validity of Theorem 3.2 for stochastic thresholds was discussed in Smith (1987, pp. 1180-1181), but no formal proof was given.}  

Second, we show in the proof of Theorem \ref{thm1} that the score functions associated with $\tilde{L}_N(\sigma,k)$ and $L_N(\sigma,k)$ are uniformly asymptotically equivalent in probability.  This, in combination with Lemma 5 in \cite{Smith1985}, establishes the existence of $\tilde{\sigma}_{\tilde{q}(a_N)}$ and $\tilde{k}$, and characterizes them as a local maximum for $\tilde{L}_N(\sigma,k)$.  Since establishing this equivalence involves the nonparametric residuals $\{\hat{\varepsilon}_t\}_{t=1}^n$ that appear in $\tilde{L}_N(\sigma,k)$, additional assumptions are needed to ensure that the nonparametric estimators $\hat{m}(\mathbf{x})$ and $\hat{h}(\mathbf{x})$ converge uniformly in probability to $m(\mathbf{x})$ and $h(\mathbf{x})$ at suitable rates.  

We adopt the following notation in our assumptions and proofs: a) $0<C<\infty$ will represent an inconsequential and arbitrary constant taking different values; b) $\mathcal{G}$ denotes a compact subset of $\mathds{R}^d$; c) $[x]$ denotes the integer part of $x\in \mathds{R}$; d) $P(A)$ denotes the probability of event $A$ associated with a probability space $(\Omega,\mathcal{F},P)$ or a probability measure, depending on the context; e) for any function $m: \mathds{R}^d \rightarrow \mathds{R}$ whose $s$ order partial derivatives exist, we denote by $D_im(\mathbf{x}):\mathds{R}^d \rightarrow \mathds{R}$ the first order partial derivatives of $m$ with respect to its $i^{th}$ argument for $i=1,\cdots,d$ and the $s$-order partial derivatives are denoted by $D_{i_1\cdots i_s}m(\mathbf{x}):\mathds{R}^d \rightarrow \mathds{R}$ for $i_1,\cdots, i_s=1,\cdots,d$.  The gradient of the function $m$ is denoted by $m^{(1)}(\mathbf{x})$ and its Hessian by $m^{(2)}(\mathbf{x})$; f) the joint density of the vector of conditioning variables $\mathbf{X}_t$ is denoted by $f_\mathbf{X}$.  For a vector $\mathbf{j} \in \mathds{R}^d$ with components $j_i$ that are non-negative integers, we write $\mathbf{x}^\mathbf{j}=\overset{d}{\underset{i=1}{\prod}}x_i^{j_i}$ and $|\mathbf{j}|=\overset{d}{\underset{i=1}{\sum}}j_i$.

\noindent \bf Assumption A1\rm: $K(\mathbf{x}): \mathds{R}^d \rightarrow \mathds{R}$ is a product kernel $K(\mathbf{x})=\prod_{j=1}^d\mathcal{K}(x_j)$ with $\mathcal{K}(x): \mathds{R}\rightarrow \mathds{R}$ such that: 1) $\left|\mathcal{K}(x)x^j \right| \leq C$ for all $x \in \mathds{R}$ and $j=0,1,2,3$; 2) $\int |x^j\mathcal{K}(x)|dx \leq C$ for $j=0,1,2,3$; 3) $\int \mathcal{K}(x)dx=1$, $\int x^j\mathcal{K}(x)dx=0$ for $j=1,\cdots,s-1$, $\int x^s\mathcal{K}(x)dx=\mu_{\mathcal{K},s}<\infty$; 4) $\mathcal{K}(x)$ is continuously differentiable on $  \mathds{R}$ with $|x^j  \frac{d}{dx} \mathcal{K}(x)| \leq C$ for all $x \in \mathds{R}$ and $j=0,1,2,3$; 5) The kernel $K_3$ is symmetric and twice continuously differentiable in $\mathds{R}$ with $|K_3(x)|\leq C$, $\int |K_3(x)|dx \leq C$, $\int K_3(x)dx=1$, $\int x^j K_3(x)dx=0$ for $j=1,\cdots,m_1$, $\int |x^{m_1+1}K_3(x)|dx\leq C$, $ \int \left| \frac{d}{dx}K_3(x)\right| dx \leq C$, $\left| \frac{d^2}{dx^2}K_3(x)\right| \leq C$ and $m_1 \geq 2$.

The kernel $\mathcal{K}(x)$ is used to construct $K_i(\mathbf{x}):\mathds{R}^d \rightarrow \mathds{R}$ where $K_i(\mathbf{x})=\prod_{j=1}^d\mathcal{K}(x_j)$ for $i=1,2$.  Furthermore, for $j=1,\cdots, d$ we have  $\int_{\mathds{R}^d} K_i(\mathbf{x})d\mathbf{x}=1$, $\int x_j^l K_i(\mathbf{x})d\mathbf{x}=0$ for $l=1,\cdots , s-1$, $\int x_j^sK_i(\mathbf{x})d\mathbf{x}=\mu_{\mathcal{K},s}$ and $\int x_{i_1} \cdots x_{i_r} K_i(\mathbf{x})d\mathbf{x}=0$ whenever $r<s$ or $i_j \neq i_k$ for some $j,k \leq s$.  The order $s$ for $K_1$ and $K_2$ is needed to establish that the biases for $\hat{m}$ and $\hat{h}$ are, respectively, of order $O(h_{in}^s)$ for $i=1,2$ in Lemmas 3 and 4.  The order $m_1$ for $K_3$ is necessary in the proof of Lemma 5.  All other assumptions are common in the nonparametric estimation literature and are easily satisfied by a variety of commonly used kernels.

\noindent \bf Assumption A2\rm: 1) $\{\mathbf{X}_t \}_{t=1,2,\cdots}$ is a strictly stationary $\alpha$-mixing process with $\alpha(l) \leq C \, l^{-B}$ for some $B>2$; 2) $f_\mathbf{X}(\mathbf{x})$ and all of its partial derivatives of order $<$ s are differentiable and uniformly bounded on $\mathds{R}^d$; 3) $ 0<\underset{\mathbf{x} \in \mathcal{G}}{\mbox{inf}}f_\mathbf{X}(\mathbf{x})$.  

A2 1) implies that for some $\delta >2$ and $a>1-\frac{2}{\delta}$, $\sum_{j=1}^\infty j^a\alpha(j)^{1 -\frac{2}{\delta}}<\infty$, a fact that is needed in our proofs.  We note that $\alpha$-mixing is the weakest of the mixing concepts (\cite{Doukhan1994}) and its use here is only possible due to Lemma A.2 in \cite{Gao2007}, which plays a critical role in the proof of Lemma 5.

\noindent \bf Assumption A3\rm: 1) $m(\mathbf{x})$ and all of its partial derivatives of order $<s$  are differentiable on $\mathds{R}^d$.  The partial derivatives are uniformly bounded on $\mathds{R}^d$; 2) $0<h(\mathbf{x})$ and all of its partial derivatives of order $<s$  are differentiable and uniformly bounded on $\mathds{R}^d$; 3) $E(h(\mathbf{X})^\zeta)\leq C$ for some $\zeta>2$.

The degree of smoothness $s$ of $m$, $h$ and $f_\mathbf{X}$ (in A2 and A3), the dimension $d$ and the mixing size $B$ are, as expected, tightly connected with the speed at which $\hat{m}$ and $\hat{h}$ converge (uniformly) to $m$ and $h$.  These parameters also interact in specific ways to determine the asymptotic behavior of $\hat{q}(a)$ and $\hat{E}(\varepsilon_t|\varepsilon_t>q(a))$.

A3 3) and FR1 b) imply, by the $c_r$-Inequality, that $E\left(|Y_t|^{4+\epsilon}\right)<\infty$.  \cite{Linton2013} and \cite{Hill2014} have proposed nonparametric estimators for \it unconditional \rm expected shortfall (ES) and studied their asymptotic behavior for fixed $a \in (0,1)$ for cases where $E(|Y_t|)<\infty$ and $V(Y_t)=\infty$.  Our model is more restrictive regarding tail behavior than theirs, but in contrast we are able to study \it conditional \rm VaR and ES when $a \rightarrow 1$.  

\noindent \bf Assumption A4\rm: 1) The joint density of $\mathbf{X}_i,\mathbf{X}_t,\varepsilon_i$, denoted by $f_{\mathbf{X}_i,\mathbf{X}_t,\varepsilon_i}(\mathbf{X}_i,\mathbf{X}_t,\varepsilon_i)$ is continuous; 2) The joint density of $\mathbf{X}_i, \mathbf{X}_j,\mathbf{X}_t,\varepsilon_i,\varepsilon_j,\varepsilon_t$, denoted by $f_{\mathbf{X}_i,\mathbf{X}_j,\mathbf{X}_t,\varepsilon_i,\varepsilon_j,\varepsilon_t}(\mathbf{X}_i,\mathbf{X}_j,\mathbf{X}_t,\varepsilon_i,\varepsilon_j,\varepsilon_t)$ is continuous.

Assumption A4 is necessary in Lemma 5 and is directly related to the verification of the existence of bounds required to use Lemma A.2 in \cite{Gao2007}.

\noindent \bf Assumption A5\rm: $h_{1n}\propto n^{-\frac{1}{2s+d}}$, $h_{2n}\propto n^{-\frac{1}{2s+d}}$, $h_{3n}\propto n^{-\frac{s}{2(2s+d)}+\delta}$, $N\propto n^{\frac{2s}{2s+d}-\delta}$ for some $\delta>0$ and $s\geq 2d$.

The following Theorem \ref{thm1} establishes the existence of $\tilde{\sigma}_{\tilde{q}(a_N)}$ and $\tilde{k}$ and characterizes them as a local maximum.  It will be convenient to re-parametrize the likelihood functions and represent arbitrary values $\sigma$ as $\sigma_{N}(1+\tau_1\delta_{N})$, where $\sigma_{N} \equiv \sigma_0(q(a_N))$, and $k$ as $k_0+\tau_2 \delta_{N}$ for $\tau_1, \tau_2 \in \mathds{R}$ with $\delta_{N} \rightarrow 0$ as $N \rightarrow \infty$.  Note that these arbitrary values belong to a shrinking neighborhood ($\delta_N \rightarrow 0$ as $N \rightarrow \infty$) of the true values $k_0$ and $\sigma_0(q(a_N))$.  Hence, we write $\tilde{L}_{TN}(\tau_1,\tau_2) = \frac{1}{N}\sum_{i=1}^{N_s} \logg(\tilde{Z}_i;\sigma_N(1+\tau_1\delta_N),k_0+\tau_2 \delta_N)$.

\begin{thm} \label{thm1}
Assume FR1, FR2 and A1-A5.  Let $\tau_1,\tau_2 \in \mathds{R}$, $0<\delta_N \rightarrow 0$, $\delta_NN^{1/2} \rightarrow \infty$, $N^{1/2}\phi(q(a_N))=O(1)$ as $N \rightarrow \infty$ and denote arbitrary $\sigma$ and $k$ by $\sigma_N(1+\tau_1\delta_N)$ and $k_0+\tau_2 \delta_N$, respectively.  We define the log-likelihood function
$
\tilde{L}_{TN}(\tau_1,\tau_2) = \frac{1}{N}\sum_{i=1}^{N_s} \logg(\tilde{Z}_i;\sigma_N(1+\tau_1\delta_N),k_0+\tau_2 \delta_N), 
$ 
where $\tilde{Z}_i=\hat{\varepsilon}_{(n-N_s+i)}-\tilde{q}(a_N)$, $a_N$, $\tilde{q}(\cdot)$ and $\hat{\varepsilon}_{(n-N_s+i)}$ are as defined in section 2.  Then, as $n \rightarrow \infty$ (and consequently $N \rightarrow \infty$), $\frac{1}{\delta_N^2}\tilde{L}_{TN}(\tau_1,\tau_2)$ has, with probability approaching $1$, a local maximum $(\tau_1^*,\tau_2^*)$ on $S_T=\{(\tau_1,\tau_2): \tau_1^2+\tau_2^2<1\}$ at which $\frac{1}{\delta_N^2}\frac{\partial}{\partial \tau_1}\tilde{L}_{TN}(\tau_1^*,\tau_2^*)=0$ and $\frac{1}{\delta_N^2}\frac{\partial}{\partial \tau_2}\tilde{L}_{TN}(\tau_1^*,\tau_2^*)=0$.  
\end{thm}
The vector $(\tau_1^*,\tau_2^*)$ implies values $\tilde{\sigma}_{\tilde{q}(a_N)}$ and $\tilde{k}$ which are solutions for the likelihood equations
\begin{equation*}
\frac{\partial}{\partial \sigma}\frac{1}{N}\sum_{j=1}^{N_s} \logg (\tilde{Z}_j; \tilde{\sigma}_{\tilde{q}(a_N)},\tilde{k})=0 \mbox{ and }
\frac{\partial}{\partial k}\frac{1}{N}\sum_{j=1}^{N_s} \logg (\tilde{Z}_j; \tilde{\sigma}_{\tilde{q}(a_N)},\tilde{k})=0.
\end{equation*}
Hence,  there exists, with probability approaching 1, a local maximum $(\tilde{\sigma}_{\tilde{q}(a_N)}=\sigma_N(1+\tau_1^*\delta_N),\, \tilde{k}=k_0+\tau_2^*\delta_N)$ on $S_R=\{(\sigma, \,k): \| (\frac{\sigma}{\sigma_N}-1, \,k-k_0)\|_E<\delta_N\}$ that satisfies the first order conditions in equation (\ref{foc}).

The proof of Theorem \ref{thm1} (see online supplement) depends critically on two sets of results.  First, since $\varepsilon_t$ is unobserved and is estimated by $\hat{\varepsilon}_t$, we must obtain convergence of both $\hat{m}(\mathbf{x})$ and $\hat{h}(\mathbf{x})$ to the true $m(\mathbf{x})$ and $h(\mathbf{x})$ uniformly in $\mathcal{G}$ at suitable rates.  This is addressed in Lemmas 3 and 4.  Second, Lemma 5 shows that $\tilde{q}(a_N)$ is asymptotically close to $q_n(a_N)$ by satisfying $\frac{\tilde{q}(a_N)-q_n(a_N)}{q(a_N)} =O_p(N^{-1/2})$.  It is in this lemma that the stochasticity of the estimated threshold $\tilde{q}$ is explicitly handled and where the full set of restrictions in FR1 and FR2 on the class of functions to which $F$ belongs are needed.  It is also in Lemma 5 that the stochasticity of $N_s$, and the fact that it may differ from $N$ in finite samples, is handled by showing that $\frac{N_s-N}{N^{1/2}}=O_p(1)$.  

We note that, as in \cite{Smith1987}, we require $N^{1/2}\delta_N \rightarrow \infty$ and $N^{1/2}\phi(q(a_N))=O(1)$.  The restriction that  $N^{1/2}\delta_N \rightarrow \infty$ is for convenience and places no stochastic constraint on our model, whereas $N^{1/2}\phi(q(a_N))=O(1)$ is necessary to provide first order approximations for the expected value of the scores associated with the likelihood function (see, e.g., equation \eqref{limu} below).

The influence of the dimension $d$ of the conditioning space manifests itself on the asymptotic results in a strong manner via the requirement that the degree of smoothness of the functions $m$ and $h$ be such that $s\geq 2d$.  We believe that alleviation of this strong requirement can only result from further constraints on the class of functions containing $m$ and $h$. 

\subsection{Asymptotic normality of $ \tilde{\gamma} = ( \tilde{\sigma}_{\tilde{q}(a_N)}, \tilde{k})^T$}

The following theorem shows that, under suitable normalization, $\tilde{\gamma}$ is asymptotically distributed as a bivariate normal random vector.  The theorem has two parts.  Part a) shows that $ \tilde{\gamma}$ carries a bias that does not decay to zero at the rate $\sqrt{N}$, a problem that is similar to that encountered in Theorem 3.2 in \cite{Smith1987}.  Inspired by \cite{Peng1998}, who proposed a moments based bias corrected version of the traditional \cite{Hill1975} estimator, we provide in part b) a bias corrected version of $ \tilde{\gamma}$, which we denote by $(\tilde{\sigma}^{(b)}_{\tilde{q}(a_N)}, \tilde{k}^{(b)})^T$. 

It will be useful to make two clarifying comments before stating Theorem \ref{thm2}.  First, by Theorem 3.4.5 (b) in \cite{Embrechts1997a} and Theorem 7 in \cite{Pickands1975}, $\sigma_0(q(a_N))\equiv \sigma_N$ is a function satisfying $\frac{\sigma_N}{q(a_N)}\rightarrow -k_0$ as $N \rightarrow \infty$.  Hence, as in \cite{Smith1987} we set, without loss of generality, $\frac{\sigma_N}{q(a_N)}=-k_0$.  Second, since the only requirements on $\phi(x)$ in FR2 are that it decays to zero and be regularly varying at infinity with $\rho<0$, a full characterization of the limiting normal distribution of $(\tilde{\sigma}_{\tilde{q}(a_N)}, \tilde{k})$ requires, as in \cite{Smith1987}, that $\sqrt{N}\phi(q(a_N))$ have a limit.  Given FR2 and the expectation of the scores associated with $\tilde{L}_N$ (see, e.g., equation \eqref{escorek}) we set
\begin{equation}\label{limu}
\frac{\sqrt{N}\phi(q(a_N))}{-k_0^{-1}-\rho} \rightarrow \mu \in \mathds{R}.
\end{equation}
for some $\mu \in \mathds{R}$.  As a limit, $\mu$ is unique for given $\phi$, but since there exist many $\phi$ that satisfy FR2 for a given pair $(\rho,k_0)$, there correspondingly exist many $\mu$ associated with such a pair.  Thus, the bias we encounter due to approximating the upper right tail of $F$ by a GPD is, as a result of assumption FR2, a function of $\mu$, $\rho$ and $k_0$.  

\begin{thm}\label{thm2}  
Assume FR1, FR2, A1-A5 and $\mu$ as defined in \eqref{limu}.  Then, for $\sigma_N=-k_0q(a_N)$ we have
 
\noindent a) $\sqrt{N}
\left( \begin{array}{c}
	\frac{\tilde{\sigma}_{\tilde{q}(a_N)}}{\sigma_{N}}-1\\
	\tilde{k}-k_0
\end{array}\right) \stackrel{d}{\rightarrow}\mathcal{N}\left(\left( \begin{array}{c}
	\frac{\mu(1-k_0)(1+2k_0\rho)}{1-k_0+k_0\rho}\\
	\frac{\mu(1-k_0)k_0(1+\rho)}{1-k_0+k_0\rho}
\end{array}\right) ,H^{-1}(k_0)V_2(k_0)H^{-1}(k_0)\right)
$ 
where 
 $$H(k_0) =\frac{1}{(1-2k_0)(1-k_0)}\left(\begin{array}{cc} 1-k_0 & -1\\-1 & 2 \end{array}\right),\mbox{ and }
V_2(k_0)=\left(\begin{array}{cc}
\frac{k_0^2-4k_0+2}{(2k_0-1)^2}&\frac{-1}{k_0(k_0-1)}\\
\frac{-1}{k_0(k_0-1)}&\frac{2k_0^3-2k_0^2+2k_0-1}{k_0^2(k_0-1)^2(2k_0-1)}
\end{array}\right).$$

\noindent b) Let $\hat{k}(N_s)=-\frac{1}{N_s}\sum_{t=1}^{N_s} \text{log} \left( \frac{\hat{\varepsilon}_{(n-N_s+t)}}{\tilde{q}(a_N)}\right)$, $M_n(N_s) = \frac{1}{N_s}\sum_{t=1}^{N_s} \left( \text{log} \left( \frac{\hat{\varepsilon}_{(n-N_s+t)}}{\tilde{q}(a_N)}\right)\right)^2$, $\hat{\rho}$ be a consistent estimator for $\rho$ and define
\begin{align*}
\tilde{k}^{(b)}&=\tilde{k}-\frac{M_n(N_s)-2( \hat{k}(N_s))^2}{(1-\tilde{k}^{-1}-\hat{\rho})\hat{d}}\left( \begin{array}{cc} 0& 1\end{array}\right)H^{-1}(\tilde{k})\left( \begin{array}{c}1\\\tilde{k}^{-1}(-\tilde{k}^{-1}-\hat{\rho})^{-1}\end{array}\right),\\
\tilde{\sigma}^{(b)}_{\tilde{q}(a_N)}&=\tilde{\sigma}_{\tilde{q}(a_N)} \left(1-\frac{M_n(N_s)-2 (\hat{k}(N_s))^2}{(1-\tilde{k}^{-1}-\hat{\rho})\hat{d}}\left( \begin{array}{cc} 1& 0\end{array}\right)H^{-1}(\tilde{k})\left( \begin{array}{c}1\\\tilde{k}^{-1}(-\tilde{k}^{-1}-\hat{\rho})^{-1}\end{array}\right) \right),
\end{align*}
where $\hat{d}= \frac{2\tilde{k}^4\hat{\rho}}{(1+\hat{\rho}\tilde{k})^2}$, an estimator for $d= \frac{2k_0^4\rho}{(1+\rho k_0)^2}$.  Then,
$$\sqrt{N}
\left( \begin{array}{c}
	\frac{\tilde{\sigma}^{(b)}_{\tilde{q}(a_N)}}{\sigma_{N}}-1\\
	\tilde{k}^{(b)}-k_0
\end{array}\right) \stackrel{d}{\rightarrow}\mathcal{N}\left(\left( \begin{array}{c}
	0\\
	0
\end{array}\right) ,H^{-1}(k_0)V_2^{(b)}(k_0,\rho)H^{-1}(k_0)\right),
$$ 
where $V_2^{(b)}(k_0,\rho)=A(k_0,\rho)V^{(b)}(k_0)A(k_0,\rho)^T$ with 
$$
V^{(b)}(k_0)=\left(\begin{array}{ccccc}
\frac{1}{1-2k_0}&-\frac{1}{(1-k_0)(1-2k_0)}& 0&\frac{4k_0^2-2k_0^3}{(1-k_0)^2}&-\frac{k_0}{1-k_0}\\
-\frac{1}{(1-k_0)(1-2k_0)}&\frac{2}{(1-k_0)(1-2k_0)}& 0& \frac{4k_0^3-6k_0^2}{(1-k_0)^2}&\frac{k_0}{1-k_0}\\
0&0&k_0^2&0 &0 \\
\frac{4k_0^2-2k_0^3}{(1-k_0)^2}&\frac{4k_0^3-6k_0^2}{(1-k_0)^2}&0&20k_0^4&-4k_0^3\\
-\frac{k_0}{1-k_0}&\frac{k_0}{1-k_0}&0&-4k_0^3&k_0^2\\
\end{array}\right),
$$
and
$$
A(k_0,\rho)=\left(\begin{array}{ccccc}
1&0&\frac{1-k_0}{k_0(1-2k_0)}+\frac{(1+2k_0)(1+\rho k_0)^2}{k_0^3\rho((1-\rho)k_0-1)}&-\frac{(1+\rho k_0)^2}{2k_0^3\rho((1-\rho)k_0-1)}&-\frac{2(1+\rho k_0)^2}{k_0^2\rho((1-\rho)k_0-1)}\\
0&1&-\frac{1}{(1-k_0)(1-2k_0)}-\frac{(1+2k_0)(1+\rho k_0)}{k_0^3\rho((1-\rho)k_0-1)}&\frac{(1+\rho k_0)}{2k_0^3\rho((1-\rho)k_0-1)}&\frac{2(1+\rho k_0)}{k_0^2\rho((1-\rho)k_0-1)}\\
\end{array}\right).
$$
\end{thm}
\noindent \bf Remark: \rm  Part b) of Theorem 2 calls for a consistent estimator of $\rho$, which we now provide.  For an arbitrary constant $c>0$ we let $N(c)= c N \text{log}(n)$, $a_{N}(c)=1-\frac{N(c)}{n}$ and define 
\begin{equation*}
\hat{\rho}=-\frac{1}{\hat{k}(N(c))\text{log}(2)} \text{log} \left( \frac{M_n(N(c/2))-2 \hat{k}(N(c/2))^2}{M_n(N(c))-2\hat{k}(N(c))^2}\right).
\end{equation*}
Lemma 9 shows that $\hat{\rho} \convp \rho$.

It is instructive to compare Theorem \ref{thm2} to Theorem 3.2 in \cite{Smith1987}.  There, he obtains 
\begin{equation*}\label{10a}
\sqrt{N}
\left( \begin{array}{c}
	\frac{\hat{\sigma}_{q(a_N)}}{\sigma_N}-1\\
	\hat{k}-k_0
\end{array}\right) \stackrel{d}{\rightarrow}\mathcal{N}\left(\left( \begin{array}{c}
	\frac{\mu(1-k_0)(1+2k_0\rho)}{1-k_0+k_0\rho}\\
	\frac{\mu(1-k_0)k_0(1+\rho)}{1-k_0+k_0\rho}
\end{array}\right) ,H^{-1}(k_0)\right),
\end{equation*}
where $\hat{\sigma}_{q(a_N)}$ and $\hat{k}$ maximize $L_N(\sigma,k)$.  Part a) shows that the use of $\tilde{Z}_i$ instead of $Z_i$ to define the exceedances used in the estimation of the parameters of the GPD impacts the variance of the asymptotic distribution.  It is easy to verify that $H^{-1}(k_0)V_2(k_0)H^{-1}(k_0)-H^{-1}(k_0)$ is positive definite, implying a (expected) loss of efficiency that results from estimating $\varepsilon_t$ nonparametrically.  However, any additional bias introduced by the nonparametric estimation is of second order effect as the asymptotic bias derived in \cite{Smith1987} is precisely the same as the one we obtain in part a) of Theorem \ref{thm2}.  The presence of such bias manifests itself in other estimation procedures for $k_0$ that rely on Extreme Value Theory (\cite{Hill1975,Pickands1975}).  Also, as in \cite{Peng1998}, bias correction may increase the variance of the asymptotic distribution.  Hence, although our bias correction produces correctly centered asymptotic distributions, the cost may be an increase in the variance of the asymptotic distribution.  In our Monte Carlo experiment, we compare the root mean squared error of the uncorrected and bias corrected estimators (see Table 1).  There, it is seen that the benefits from bias correction are far larger than any cost associated with variance increase.      


\subsection{Asymptotic normality of $\hat{q}(a)$, $\hat{E}(\varepsilon_t|\varepsilon_t>q(a))$, $a$-CVaR($\mathbf{x}$) and $a$-CES($\mathbf{x}$) }

The asymptotic distributions of the ML estimators given in parts a) and b) of Theorem \ref{thm2} are  the basis for obtaining the asymptotic distributions of $\hat{q}(a)$ and $\hat{E}(\varepsilon_t|\varepsilon_t>q(a))$.  First, in the case of $\hat{q}(a)$, we rely on the asymptotic properties of $\tilde{q}(a_n)$ and Theorem \ref{thm2} parts a) and b).  Second, since $\hat{E}(\varepsilon_t|\varepsilon_t>q(a))=\frac{\hat{q}(a)}{1+\tilde{k}}$, its asymptotic distribution can be derived directly from the results for $\hat{q}(a)$ and $\tilde{k}$.  The estimators $\hat{q}(a)$ and $\hat{E}(\varepsilon_t|\varepsilon_t>q(a))$ inherit an asymptotic bias that derives from the result in part a) of Theorem \ref{thm2}.  Hence, Theorems \ref{thm3} and \ref{thm4} include bias-corrected versions of the estimators we propose, which we denote by $\hat{q}^{(b)}(a)$ and $\hat{E}^{(b)}(\varepsilon_t|\varepsilon_t>q(a))$.  It is important to emphasize, as mentioned in section 2, that in Theorems \ref{thm3}, \ref{thm4} and \ref{thm5} both $a_N$ and $a$ approach 1 as $n\rightarrow \infty$ with $a_N<a$.  We also note that we assume that $n(1-a) \propto N$ with $N \rightarrow \infty$.  In the extreme (regression) quantile literature this is called the ``intermediate order" type asymptotics (Chernozhukov (2005, p. 809)) and in the extreme CVaR literature this is described as asymptotics for ``high quantiles within the sample" (Embrechts et al. (1997, p. 349)).    
 

\begin{thm}\label{thm3} 
Assume FR1, FR2, A1-A5 and $\mu$ as defined in \eqref{limu}.  Then, for $\sigma_N=-k_0q(a_N)$ and for some $\mathcal{Z}>0$, if $n(1-a) \propto N$ we have

\noindent a) $\sqrt{N}\left(\frac{\hat{q}(a)}{q(a)}-1\right)\convd \mathcal{N}\left(\mu_1,\Sigma_1(k_0)\right)$, where 
\begin{align*}
\mu_1&=k_0 \mu\left((-k_0^{-1}-\rho)\frac{(\mathcal{Z}^\rho-1)}{\rho} +\frac{1}{1-k_0^{-1}-\rho}c_b^T H^{-1} (k_0)\left(\begin{array}{c}
-k_0^{-1}-\rho  \\
k_0^{-1}\\ 
\end{array}\right)\right), \\
\Sigma_1(k_0)&=k_0^2\left( c_b^TH^{-1}(k_0)c_b+k_0^2\left(c_b^TH^{-1}(k_0) \left( \begin{array}{c}b_1\\b_2\end{array}\right)  \right)^2+2k_0\mathcal{Z}^{-1}c_b^TH^{-1}(k_0) \left( \begin{array}{c}b_1\\b_2\end{array}\right)+ \mathcal{Z}^{-2}\right),\\
b_1&=\frac{-(1-k_0)}{k_0(2k_0-1)}, b_2=\frac{-1}{(1-k_0)(1-2k_0)}\mbox{ and }
c_b^T=\left(\begin{array}{cc}
-k_0^{-1}(\mathcal{Z}^{-1}-1)&  k_0^{-2}log(\mathcal{Z})+k_0^{-2}(\mathcal{Z}^{-1}-1)\\
\end{array}\right).
\end{align*}

\noindent b) Let $\hat{q}^{(b)}(a)=\tilde{q}(a_N) \left( 1+ \frac{\tilde{\sigma}^{(b)}_{\tilde{q}(a_N)}}{\tilde{k}^{(b)}\tilde{q}(a_N)} \left( 1- \left(\frac{N}{n(1-a)} \right)^{-\tilde{k}^{(b)}}(1+\hat{B}_q)^{-\tilde{k}^{(b)}}\right)\right)$, where $\hat{B}_q=\frac{\hat{\mathcal{Z}}^{\hat{\rho}}-1}{\hat{\rho}\hat{d}}(M_n(N_s)-2(\hat{k}(N_s))^2)$ and $\hat{\mathcal{Z}}=\frac{\hat{q}(a)}{\tilde{q}(a_N)}$.  Then, 
$
\sqrt{N}\left(\frac{\hat{q}^{(b)}(a)}{q(a)}-1\right)\convd \mathcal{N}\left(0, \Sigma_1^{(b)}(k_0,\rho)\right),
$
where $ \Sigma_1^{(b)}(k_0,\rho)=c_q^TV^{(b)}(k_0)c_q$, $c_q^T=k_0c_b^TH^{-1}(k_0)A(k_0,\rho)+v(k_0,\rho)$ and 
$$
v(k_0,\rho)=\left( \begin{array}{ccccc} 0 & 0 & \mathcal{Z}^{-1}+(\mathcal{Z}^{\rho}-1)\frac{(1+2k_0)(1+\rho k_0)^2}{k_0^3 \rho^2}& -(\mathcal{Z}^{\rho}-1)\frac{(1+\rho k_0)^2}{2k_0^3 \rho^2}& -2(\mathcal{Z}^{\rho}-1)\frac{(1+\rho k_0)^2}{k_0^2 \rho^2} \end{array}\right).
$$
\end{thm}

\noindent \bf Remark\rm:  Under the assumption that $n(1-a) \propto N$, the constant $\mathcal{Z}$ is the limit of $q(a)/q(a_N)$ as $N \rightarrow \infty$.  Thus, it captures the variation of the quantile associated with $F$ as we approach its endpoint, which in this case is infinity.  Bias correction in part b) of the theorem requires a consistent estimator $\hat{\mathcal{Z}}$ for $\mathcal{Z}$.  


\begin{thm}\label{thm4} 
Assume FR1, FR2, A1-A5 and $\mu$ as defined in \eqref{limu}.  Then, for $\sigma_N=-k_0q(a_N)$ and for some $\mathcal{Z}>0$, if $n(1-a) \propto N$ we have

\noindent a) $\sqrt{N}\left(\frac{\hat{E}(\varepsilon_t|\varepsilon_t>q(a))}{\frac{q(a)}{1+k_0}}-1\right)\convd \mathcal{N}\left( \mu_2, \Sigma_2 (k_0)\right)$, where 
$$
\mu_2=\mu k_0 \left( \frac{(\mathcal{Z}^\rho-1)(-k_0^{-1}-\rho)}{\rho} + \left( \frac{1}{1-k_0^{-1}-\rho} c_b^T -  \left(\begin{array}{cc}
0 & \frac{1}{k_0(1+k_0)(1-k_0^{-1}-\rho)}  
\end{array}\right) \right)  H^{-1}(k_0) \left(\begin{array}{c}
-k_0^{-1}-\rho  \\
k_0^{-1}\\ 
\end{array}\right) \right),
$$ 
$\Sigma_2(k_0)=\left(\begin{array}{c}
k_0 \eta^T-\frac{1}{1+k_0}\theta^T 
\end{array}\right)V_1(k_0)\left(\begin{array}{c}
k_0 \eta -\frac{1}{1+k_0}\theta 
\end{array}\right)$,
$\eta^T = \left(\begin{array}{cc}
c_b^T H^{-1}(k_0) &  c_b^T H^{-1}(k_0) \left(\begin{array}{c}
b_1  \\
b_2\\ 
\end{array}\right) +(k_0\mathcal{Z})^{-1}
\end{array}\right),\\
\theta^T = \left(\begin{array}{cc}
 \left(\begin{array}{cc}
0 & 1  
\end{array}\right) H^{-1}(k_0) &  \left(\begin{array}{cc}
0 & 1  
\end{array}\right) H^{-1} (k_0)\left(\begin{array}{c}
b_1  \\
b_2\\ 
\end{array}\right) 
\end{array}\right)$,
$
V_1(k_0)=\left(
\begin{array}{ccc}
\frac{1}{1-2k_0}&  -\frac{1}{(k_0-1)(2k_0-1)}&0\\
-\frac{1}{(k_0-1)(2k_0-1)}& \frac{2}{(k_0-1)(2k_0-1)}&0\\ 
0& 0&k_0^2\\
\end{array}\right), 
$ and $c_b$, $b_1$ and $b_2$ are as defined in Theorem \ref{thm3}.\\
\noindent b) Let $\hat{E}^{(b)}(\varepsilon_t|\varepsilon_t>q(a))=\frac{\hat{q}^{(b)}}{1+\tilde{k}^{(b)}}$, then
$
\sqrt{N}\left(\frac{\hat{E}^{(b)}(\varepsilon_t|\varepsilon_t>q(a))}{q(a)/(1+k_0)}-1\right)\convd\mathcal{N}\left(0,\Sigma_2^{(b)}(k_0,\rho)\right), \mbox{ where }
$
$$
\Sigma_2^{(b)}(k_0,\rho)= \left( c_q^T-\frac{1}{1+k_0} \left(\begin{array}{cc}0&1\end{array}\right)H^{-1}(k_0)A(k_0,\rho)   \right)V^{(b)}(k_0)\left(c_q^T-\frac{1}{1+k_0} \left(\begin{array}{cc}0&1\end{array}\right)H^{-1}(k_0)A(k_0,\rho) \right)^T,
$$
$c_q$ is as defined in Theorem \ref{thm3}, and $A(k_0,\rho)$ and $V^{(b)}(k_0)$ are defined in Theorem \ref{thm2}.
\end{thm}

From Theorems \ref{thm3} and \ref{thm4} we obtain our main results, the asymptotic normality and consistency of $\hat{q}_{Y_t|\mathbf{X}_t=\mathbf{x}}(a)$ and $\hat{E}(Y_t|Y_t>q_{Y_t|\mathbf{X}_t=\mathbf{x}}(a),\mathbf{X}_t=\mathbf{x})$.  Since these estimators also inherit an asymptotic bias, we present only results for estimators of $a$-CVaR($\mathbf{x}$) and $a$-CES($\mathbf{x}$) that are constructed using $\hat{q}^{(b)}(a)$ and $\hat{E}^{(b)}(\varepsilon_t|\varepsilon_t>q(a))$, respectively.  So we define,
$
\hat{q}^{(b)}_{Y_t|\mathbf{X}_t=\mathbf{x}}(a))=\hat{m}(\mathbf{x})+\hat{h}^{1/2}(\mathbf{x})\hat{q}^{(b)}(a)$ and
$\hat{E}^{(b)}(Y_t|Y_t>q_{Y_t|\mathbf{X}_t=\mathbf{x}}(a),\mathbf{X}_t=\mathbf{x})=\hat{m}(\mathbf{x})+\hat{h}^{1/2}(\mathbf{x})\left( \hat{E}^{(b)}(\varepsilon_t|\varepsilon_t>q(a))+\hat{B}_E\right)$,
where $\hat{B}_E=\frac{\hat{q}^{(b)}(a)\hat{\mathcal{Z}}^{\hat{\rho}}\left( M(N_s)-2(\hat{k}(N_s))^2 \right)}{\hat{d}(1+\tilde{k}^{-1}+\hat{\rho})(1+\tilde{k}^{-1})}$, $M(N_s)$, $\hat{k}(N_s)$, $\hat{d}$, $\hat{\rho}$ are as defined in Theorem \ref{thm2} and $\hat{\mathcal{Z}}$ and $\hat{q}^{(b)}(a)$ are as defined in Theorem \ref{thm3}.


\begin{thm}\label{thm5} 
Assume FR1, FR2, A1-A5 and $\mu$ as defined in \eqref{limu}.  Then, for $\sigma_N=-k_0q(a_N)$ and for some $\mathcal{Z}>0$, if $n(1-a) \propto N$ we have

\noindent a) $\sqrt{N}\left(\frac{\hat{q}^{(b)}_{Y_t|\mathbf{X}_t=\mathbf{x}}(a)}{q_{Y_t|\mathbf{X}_t=\mathbf{x}}(a)}-1\right) \convd \mathcal{N}\left( 0,\Sigma_1^{(b)}(k_0,\rho)\right)$, where $\Sigma_1^{(b)}(k_0,\rho)$ is defined in Theorem \ref{thm3}.

\noindent b) $\sqrt{N}\left(\frac{\hat{E}^{(b)}(Y_t|Y_t>q_{Y_t|\mathbf{X}_t=\mathbf{x}}(a),\mathbf{X}_t=\mathbf{x})}{E(Y_t|Y_t>q_{Y_t|\mathbf{X}_t=\mathbf{x}}(a),\mathbf{X}_t=\mathbf{x})}-1\right)\convd \mathcal{N}\left( 0, \Sigma_3^{(b)}(k_0,\rho) \right)$, where 
\begin{align*}\Sigma_3^{(b)}(k_0,\rho)&=\left( c_q^T-\frac{1}{1+k_0} \left(\begin{array}{cc}0&1\end{array}\right)H^{-1}(k_0)A(k_0,\rho)+\upsilon_1(k_0,\rho) \right) V^{(b)}(k_0) \\
&\times\left( c_q^T-\frac{1}{1+k_0} \left(\begin{array}{cc}0&1\end{array}\right)H^{-1}(k_0)A(k_0,\rho) +\upsilon_1(k_0,\rho)  \right)^T,
\end{align*}
$\upsilon_1(k_0,\rho)  = \left( \begin{array}{ccccc} 0&0&\frac{\mathcal{Z}^{\rho}k_0(-2-4k_0)}{d(\rho+k_0^{-1}+1)}&\frac{k_0\mathcal{Z}^{\rho}}{d(\rho+k_0^{-1}+1)}&\frac{4k_0^2\mathcal{Z}^{\rho}}{d(\rho+k_0^{-1}+1)}
\end{array}\right)$, $c_q$ is defined in Theorem \ref{thm3}, $A(k_0,\rho)$, $V^{(b)}(k_0)$ and $d$ are defined in Theorem \ref{thm2}.
\end{thm}
As we have observed following Theorem \ref{thm2}, it is also the case that bias correction in Theorems \ref{thm3}, \ref{thm4} and \ref{thm5} has an impact on the variance of the asymptotic distribution of the estimators.  The following corollary to Theorem \ref{thm5} shows that the benefit of bias correction, in terms of asymptotic mean squared error, depends critically on the parameters $k_0$ and $\rho$, and how they relate to the constants $\mu$ and $\mathcal{Z}$, of which $\mu$ is not uniquely determined by $F$.\footnote{Corollary 2.1 in Peng (1998, p.109) gives a similar result for a much simpler model.  But even in his model, as in the case for the estimators in our Theorem 2, the benefits of bias reduction depend on the interplay of model parameters and his constant $\lambda$ (playing the same role as our $\mu$), which varies with his $A$ (playing the same role as our $\phi$).  See the discussion following our equation \eqref{limu}.}  The corollary is most useful for given $\phi$, in which case the value of $\mu$ is fixed. This theoretical indeterminacy manifests itself in our simulations (see section 4).  For example, when $F$ is a Student-t distribution with $v=6$ degrees of freedom, $k_0=-1/6$ and $\rho=-2$ (\cite{Ling2015}), the benefit of bias reduction in the estimation of the parameters of the GPD seems quite clear.  In contrast, bias reduction in the estimation $q_{Y_t|\mathbf{X}_t=\mathbf{x}}(a)$ and $E(Y_t|Y_t>q_{Y_t|\mathbf{X}_t=\mathbf{x}}(a),\mathbf{X}_t=\mathbf{x})$ is not apparent, at least from the point of view of reduced root mean squared error.  These simulation results suggest that $|\mu|$ satisfies threshold levels for reduced MSE under bias correction for the estimators in Theorem 2 but not for those in Theorem 5.  Evidently, alternative data generating processes may produce different results.     

\begin{cor} Assume the conditions of Theorem \ref{thm5} hold.  Let $MSE(\nu_n)$ denote the mean squared error of the estimator $\nu_n$.  Then,
$$
\underset{N \rightarrow \infty}{\text{lim}}\frac{MSE\left( \frac{\hat{q}^{(b)}_{Y_t|\mathbf{X}_t=\mathbf{x}}(a)}{q_{Y_t|\mathbf{X}_t=\mathbf{x}}(a)} \right)}{MSE\left( \frac{\hat{q}_{Y_t|\mathbf{X}_t=\mathbf{x}}(a)}{q_{Y_t|\mathbf{X}_t=\mathbf{x}}(a)}\right)}
\left\{ \begin{array}{cc} 
> 1& \text{ if } |\mu| < \mathcal{C}_1(k_0,\rho,\mathcal{Z})\\
= 1& \text{ if } |\mu| = \mathcal{C}_1(k_0,\rho,\mathcal{Z})\\
< 1& \text{ if } |\mu| > \mathcal{C}_1(k_0,\rho,\mathcal{Z})\\
\end{array}\right.
$$
where 
$\mathcal{C}_1(k_0,\rho,\mathcal{Z})=\frac{(\Sigma^{(b)}_1(k_0,\rho)-\Sigma_1(k_0))^{1/2}}{-k_0\left| \left(-k_0^{-1}-\rho\right)\frac{\mathcal{Z}^{\rho}-1}{\rho}+\frac{1}{1-k_0^{-1}-\rho}c_b^TH^{-1}(k_0)\left(\begin{array}{c}-k_0^{-1}-\rho\\ k_0^{-1}\end{array}\right)\right|},
$ 
and
$$
\underset{N \rightarrow \infty}{\text{lim}}\frac{MSE\left( \frac{\hat{E}^{(b)}(Y_t|Y_t>q_{Y_t|\mathbf{X}_t=\mathbf{x}}(a),\mathbf{X}_t=\mathbf{x})}{E(Y_t|Y_t>q_{Y_t|\mathbf{X}_t=\mathbf{x}}(a),\mathbf{X}_t=\mathbf{x})} \right)}{MSE\left( \frac{\hat{E}(Y_t|Y_t>q_{Y_t|\mathbf{X}_t=\mathbf{x}}(a),\mathbf{X}_t=\mathbf{x})}{E(Y_t|Y_t>q_{Y_t|\mathbf{X}_t=\mathbf{x}}(a),\mathbf{X}_t=\mathbf{x})}\right)}
\left\{ \begin{array}{cc} 
> 1& \text{ if } |\mu| < \mathcal{C}_2(k_0,\rho,\mathcal{Z})\\
= 1& \text{ if } |\mu| = \mathcal{C}_2(k_0,\rho,\mathcal{Z})\\
< 1& \text{ if } |\mu| > \mathcal{C}_2(k_0,\rho,\mathcal{Z})\\
\end{array}\right.
$$
where 
$\mathcal{C}_2(k_0,\rho,\mathcal{Z})=\frac{(\Sigma^{(b)}_3(k_0,\rho)-\Sigma_2(k_0))^{1/2}}{-k_0\left| \left(-k_0^{-1}-\rho\right)\frac{\mathcal{Z}^{\rho}-1}{\rho}+ \left(  c_b^T -    \left(\begin{array}{cc}0 &  \frac{1}{k_0(1+k_0)} \end{array}\right)   \right) H^{-1}(k_0)  \left(\begin{array}{c}  \frac{-k_0^{-1}-\rho}{1-k_0^{-1}-\rho}\\  \frac{k_0^{-1}}{1-k_0^{-1}-\rho}\end{array}\right) + \frac{k_0^{-1}+\rho}{1+k_0^{-1}+\rho}\mathcal{Z}^\rho \right|}.
$
\end{cor}

Theorem \ref{thm5} provides the basis for inference regarding extreme CVaR($\mathbf{x}$) and CES($\mathbf{x}$).  The covariance matrices depend on $k_0$, $\rho$ and $\mathcal{Z}$ which can be consistently estimated as described in Theorem \ref{thm3} and Lemma 9.  In section \ref{montecarlo} we implement these estimators, construct confidence intervals and report on empirical coverage probabilities.  As a direct consequence of Theorem \ref{thm5} we have 
$
\frac{\hat{q}^{(b)}_{Y_t|\mathbf{X}_t=\mathbf{x}}(a)}{q_{Y_t|\mathbf{X}_t=\mathbf{x}}(a)}=1+o_p(1) \mbox{ \,\,\,and \,\,\,} \frac{\hat{E}^{(b)}(Y_t|Y_t>q_{Y_t|\mathbf{X}_t=\mathbf{x}}(a),\mathbf{X}_t=\mathbf{x})}{E(Y_t|Y_t>q_{Y_t|\mathbf{X}_t=\mathbf{x}}(a),\mathbf{X}_t=\mathbf{x})}=1+o_p(1)
$
as $n(1-a) \rightarrow \infty$, therefore establishing consistency of the estimators.


\section{Monte Carlo study}\label{montecarlo}

We perform a Monte Carlo study to investigate the finite sample properties of the parameter estimator $\tilde{\gamma} = (\tilde{\sigma}_{\tilde{q}(a_N)},\tilde{k})^T$, the $a$-CVaR($\mathbf{x}$) estimator $\hat{q}_{Y_t|\mathbf{X}_t=\mathbf{x}}(a)$, the $a$-CES($\mathbf{x}$) estimator $\widehat{E}(Y_t|Y_t>q_{Y_t|\mathbf{X}_t=\mathbf{x}}(a),\mathbf{X}_t=\mathbf{x})$, as well as their bias-corrected versions given by $\tilde{\gamma}^{(b)} = (\tilde{\sigma}_{\tilde{q}(a_N)}^{(b)},\tilde{k}^{(b)})^T$, $\hat{q}^{(b)}_{Y_t|\mathbf{X}_t=\mathbf{x}}(a)$ and $\widehat{E}^{(b)}(Y_t|Y_t>q_{Y_t|\mathbf{X}_t=\mathbf{x}}(a),\mathbf{X}_t=\mathbf{x})$.  To simplify the notation, we put $\hat{q}_{Y_t|\mathbf{X}_t=\mathbf{x}}(a) \equiv \hat{q}$, $\widehat{E}(Y_t|Y_t>q_{Y_t|\mathbf{X}_t=\mathbf{x}}(a),\mathbf{X}_t=\mathbf{x}) \equiv \hat{E}$, $\hat{q}^{(b)}_{Y_t|\mathbf{X}_t=\mathbf{x}}(a) \equiv \hat{q}^{(b)}$, and $\widehat{E}^{(b)}(Y_t|Y_t>q_{Y_t|\mathbf{X}_t=\mathbf{x}}(a),\mathbf{X}_t=\mathbf{x}) \equiv \hat{E}^{(b)}$ with corresponding true values given by $q$ and $E$.  The underlying values of $a$ and $\mathbf{x}$ will be clear in context.  

We generate data from the following location-scale model 
\begin{equation}\label{eq:dgp} 
Y_t = m(Y_{t-1})+ h(t)^{1/2} \varepsilon_t, t= 1,\cdots,n.  
\end{equation}
We choose $m(Y_{t-1})$ to be $sin(0.5Y_{t-1})$ and consider $h(t) = h_i(Y_{t-1})+\theta h(t-1)$ for $i=1,2$, where $h_1(Y_{t-1}) = 1+ 0.01Y_{t-1}^2+0.5sin(Y_{t-1})$ and $h_2(Y_{t-1}) = 1 - 0.9exp(-2Y_{t-1}^2)$. The quadratic type heteroskedasticity function $h_1(\cdot)$ was considered in \cite{Cai2008}, where we add the $sin(\cdot)$ function to make the nonlinearity more prominent, and $h_2(\cdot)$ has been considered in \cite{Martinsfilho2006}.  $\theta$ is set to be $0$ or $0.5$.\footnote{We only report results for $\theta=0$.  All results for $\theta=0.5$ are available from the first author upon request.  However, in the text we discuss the results for $\theta=0.5$ and highlight the differences when needed.}  Our estimators are based on a model where $\theta=0$, but the model with $\theta=0.5$ and $h_1(\cdot)$ without the $sin(\cdot)$ function corresponds to the popular GARCH model, and it would be interesting to investigate the performance of our estimators under this structure.  Note also that $h_1$ is unbounded, therefore violating assumption A3.  Initial values of $Y_t$ and $h(t)$ are set to be zero and $Y_t$ is generated recursively according to equation (\ref{eq:dgp}).  We discard the first $1000$ observations so that the samples are not heavily influenced by the choice of initial values.

We generate $\varepsilon_t$ independently from a Student-t distribution with $v$ degrees of freedom.  It can be easily shown that $k_0= -\frac{1}{v}$, so we have $k_0 = -1/3$ for $v=3$ and $k_0 = -1/6$ for $v=6$.  We note that only the case where $v=6$ conforms to the assumptions needed to establish asymptotic normality of our estimators, but we consider the other case to investigate the behavior of the estimators when our asymptotic results may not hold.  Here, the variance of $\varepsilon_t$ is larger with $v=3$ and we expect that in this case estimation will be relatively more difficult.  In contrast, when $v=6$ the Student-t distribution resembles the normal distribution.  For identification purposes, we standardize $\varepsilon_t$ so that it has unit variance.\footnote{We have also performed our study using the log-gamma distribution, a density that is also in the domain of attraction of the Fr\'echet distribution.  Since its support is bounded from below, it is much less commonly used to model financial returns. Though the relative rankings regarding estimators' performances change somewhat in specific experiment designs, we do not report these results to save space and focus on the more popular Student-t distribution and a more detailed exposition.}

Implementation of our estimator requires the choice of bandwidths $h_{1n}$, $h_{2n}$ and $h_{3n}$. Since $h_{1n}$ and $h_{2n}$ are utilized to estimate the conditional mean and variance, we select them using the \it rule-of-thumb \rm data driven plug-in method of \cite{Ruppert1995} and denote them by $\hat{h}_{1n}$ and $\hat{h}_{2n}$. Specifically, $\hat{h}_{1n}$ and $\hat{h}_{2n}$ are obtained from the following regressand and regressor sequences $\{Y_t,Y_{t-1}\}_{t=1}^n$ and $\{(Y_t-\hat{m}(Y_{t-1}))^2,Y_{t-1}\}_{t=1}^n$, respectively.  We select $h_{3n}$ by using the  \it rule-of-thumb \rm bandwidth $\hat{h}_{3n} = 0.79R(Y_{t-1})n^{-1/5+\delta}$ as in (2.52) of \cite{Pagan1999}, where $R(Y_{t-1})$ is the sample interquartile range of $Y_{t-1}$, and we set $\delta=0.01$ so that it satisfies our assumption on the bandwidth. The second order Epanechnikov kernel is used for our estimators.

In estimating the parameters, we consider our estimators $\tilde{\gamma}=(\tilde{\sigma}_{\tilde{q}(a_N)},\tilde{k})^T$, our bias-corrected estimators $\tilde{\gamma}^{(b)} = (\tilde{\sigma}_{\tilde{q}(a_N)}^{(b)},\tilde{k}^{(b)})^T$, Smith type estimators $\hat{\gamma}=(\hat{\sigma}_{q_n(a_N)},\hat{k})^T$ and Smith type bias-corrected estimators $\hat{\gamma}^{(b)}=(\hat{\sigma}^{(b)}_{q_n(a_N)},\hat{k}^{(b)})^T$, where the bias correction is conducted as in $\tilde{\gamma}^{(b)}$.  Both $\hat{\gamma}$ and $\hat{\gamma}^{(b)}$ utilize the true conditional mean $m(\cdot)$, variance $h(\cdot)$ and $\varepsilon_t$ available in the simulation. Without having to estimate $m(\cdot)$ and $h(\cdot)$, we expect that Smith's estimators ($\hat{\gamma}$ and $\hat{\gamma}^{(b)}$) will perform best and serve as a benchmark to evaluate our estimators.  In estimating the conditional value-at-risk ($q$) and expected shortfall (E), we include our estimators ($\hat{q}$, $\hat{E}$), our bias-corrected estimators ($\hat{q}^{(b)}$, $\hat{E}^{(b)}$), the Smith type estimators ($q^s,E^s$), the Smith type bias-corrected estimators ($q^{s(b)}, E^{s(b)}$), where the bias correction is performed as in ($\hat{q}^{(b)}$, $\hat{E}^{(b)}$), and the estimators ($\dot{q},\dot{E}$) proposed by \cite{Cai2008}. We follow their instructions for implementation and utilize the theoretical optimal bandwidths available in the simulation for ($\dot{q},\dot{E}$) to minimize the noise. 

Figure 1 plots the true and estimated conditional value-at-risk and expected shortfall evaluated at the sample mean of $Y_{t-1}$ for values of $a$ between $0.95$ and $0.999$, since we are interested in higher order quantiles. The estimation utilizes $1000$ sample data points generated from equation (\ref{eq:dgp}) with $h_1(Y_{t-1}) = 1+ 0.01Y_{t-1}^2+0.5 sin(Y_{t-1})$, $\theta=0$ and Student-t distributed $\varepsilon_t$ with $v=3$ degrees of freedom. We use $N=round(c*1000^{0.8-0.01})= 164$ in constructing our estimates, where $c=0.7$ and $round(\cdot)$ gives the nearest integer. We note that all estimators are smooth functions of $a$, and they seem to capture the shape of the true value-at-risk and expected shortfall well. It seems more difficult to estimate expected shortfall than value-at-risk as the gap between the estimates and the true value is noticeably larger for expected shortfall. 

The performance of our estimators is fairly robust to our choice of $N$ in the simulations for $n=1000, 2000$ and $4000$. In the expression for $N$, we set $c=0.7$ so that we use less than $20\%$ of the total number of observations as tail observations in the second stage estimation, giving $N = 164, 284, $ and $491$, respectively.  Thus, with $n$ being doubled, the effective sample size $N$ in the second stage of our estimation is less than doubled, as required by the assumption on $N$. Each experiment is repeated $2000$ times. We summarize the performance of all parameter estimators in terms of their bias (B), standard deviation (S) and root mean squared error (R) in Table 1 for $\theta=0$.  

We consider the performance of the $a-$conditional value-at-risk and expected shortfall estimators for $a = 0.95, 0.99,$ and $0.999$ evaluated at $Y_n$, the most recent observation in the sample. Specifically, the performances in terms of the bias (B), standard deviation (S) and relative root mean squared error (R) for $\theta=0$ are detailed in Tables 2 and 3 for $v=3$ and $6$, respectively.  To facilitate comparison, we report the relative root mean squared error as the ratio of the root mean squared error of each estimator over that of the estimator with the smallest root mean squared error in each experiment design. To reduce the impact of extreme experiment runs, we truncate the smallest and largest $2.5\%$ estimates from the repetitions for all estimators. We give the $95\%$ empirical coverage probability for the bias-corrected estimators $q^{(b)}$, $q^{s(b)}$, $E^{(b)}$ and $E^{s(b)}$ in Table 4 for $\theta=0$.  As the results for $n=2000$ are qualitatively similar, we only report detailed results for $n=1000$ and $n=4000$. 

To implement the bias-corrected estimators, we need the second order parameter $\rho$, which is estimated by $\hat{\rho} = -\frac{1}{\hat{k}(N(c))\text{log}(2)}\text{log}(\frac{M_n(N((c/2))) - 2(\hat{k}(N((c/2))))^2}{M_n(N(c)) - 2(\hat{k}(N(c)))^2})$, where $\hat{k}(N(c))$, $M_n(N(c))$ are as defined in Theorem \ref{thm2}. Here $a_N(c) = 1 - \frac{N(c)}{n}$ and we choose $N(c) = cN\text{log}(n)$ for some positive constant $c$.  We let $c=0.25$ so that for the sample sizes and $N$ considered, $N(c)$ and $N((c/2))$ are  less than $n$.  The moments based estimator $\hat{k}(N_s)$ is utilized to construct the bias-corrected parameter estimate $\tilde{\gamma}^{(b)}$, which is then used to construct $\hat{q}^{(b)}$ and $\hat{E}^{(b)}$. We use the asymptotic distributions of $\hat{q}^{(b)}$ and $\hat{E}^{(b)}$ to construct confidence intervals, since they are asymptotically unbiased. Specifically, we estimate the $95\%$ confidence interval for the $a$-CVaR($\mathbf{x}$) as $\left(\hat{q}^{(b)}\left(1+Z_{0.975}\sqrt{\frac{\Sigma_1^{(b)}(\hat{k}(N_s),\hat{\rho})}{N}}\right)^{-1}, \hat{q}^{(b)}\left(1-Z_{0.975}\sqrt{\frac{\Sigma_1^{(b)}(\hat{k}(N_s),\hat{\rho})}{N}}\right)^{-1} \right)$, where $Z_{0.975}$ is the $97.5\%$ quantile for the standard normal distribution. For the $a$-CES($\mathbf{x}$), its $95\%$ confidence interval estimates are $\left(\hat{E}^{(b)}\left(1+ Z_{0.975} \sqrt{\frac{\Sigma_3^{(b)}(\hat{k}(N_s),\hat{\rho}) }{N}}\right)^{-1},\hat{E}^{(b)}\left(1- Z_{0.975} \sqrt{\frac{\Sigma_3^{(b)}(\hat{k}(N_s),\hat{\rho}) }{N}}\right)^{-1} \right)$, where $\mathcal{Z}$ is estimated by $\hat{\mathcal{Z}}=\hat{q}(a)/\tilde{q}(a_N)$.

In the case of estimating parameters, we notice that both $\hat{\gamma}$ and $\tilde{\gamma}$ overestimate $(\sigma_N,k_0)$, while the bias-corrected estimators $\hat{\gamma}^{(b)}$ and $\tilde{\gamma}^{(b)}$ often underestimate $(\sigma_N,k_0)$. As the sample size increases, all estimators' performance improves, in the sense that B, S and R decrease, with a few exceptions for B of $\hat{\gamma}^{(b)}$ and $\tilde{\gamma}^{(b)}$. This confirms the asymptotic results in the previous section.  When $k_0$ is decreased (smaller $v$ in Table 1), we generally find that all estimators exhibit smaller B, larger S and smaller R, since the drop in B dominates the increase in S, with a few exceptions in the case of estimating $\sigma_N$.  We think that this is related to the bias and variance trade-off for the parameter estimation.  As mentioned above, the variance of $\varepsilon_t$ without standardization is larger with smaller $k_0$, and the distribution of $\varepsilon_t$ exhibits heavier tail behavior, thus the more representative extreme observations have a larger probability to show up in a sample, which explains the lower bias.  It is generally harder to estimate $\sigma_N$ than $k_0$ when using $\hat{\gamma}$ and $\tilde{\gamma}$, as estimates of $\sigma_N$ exhibit larger R.  However, under this criterion, it is harder to estimate $k_0$ than $\sigma_N$,  when using $\hat{\gamma}^{(b)}$ and $\tilde{\gamma}^{(b)}$.  In terms of relative performance, we notice that $\hat{\gamma}^{(b)}$ and $\tilde{\gamma}^{(b)}$ are much better than  $\hat{\gamma}$ and $\tilde{\gamma}$, as they exhibit much lower B and R.  Thus, the bias-corrected parameter estimators significantly reduce B without inflating S. When $v=3$, $\hat{\gamma}$, $\hat{\gamma}^{(b)}$ generally outperform $\tilde{\gamma}$, $\tilde{\gamma}^{(b)}$ respectively, in terms of smaller B, S and R, though the difference diminishes with larger sample sizes. When $v=6$, $\tilde{\gamma}$, $\tilde{\gamma}^{(b)}$ frequently perform better than $\hat{\gamma}$, $\hat{\gamma}^{(b)}$ respectively, especially in estimating $\sigma_N$. Again the difference is fairly small and diminishes with larger sample sizes. The results suggest that our proposed estimators $\tilde{\gamma}$ and $\tilde{\gamma}^{(b)}$ are well supported by the nonparametric kernel estimators for the functions $m(Y_{t-1})$ and $h(Y_{t-1})$.

In the case of estimating conditional value-at-risk and expected shortfall, we observe that performances of all estimators generally improve with the sample sizes in terms of smaller B, S and R, with some exceptions for B.  This confirms the consistency of our estimators for conditional value-at-risk and expected shortfall.  In the case of estimating conditional value-at-risk, $\hat{q}$ and $q^s$ carry positive bias for $a=0.95$ and $0.99$, but exhibit negative bias for $a=0.999$.  $\dot{q}$ is frequently positively biased, while $\hat{q}^{(b)}$ and $q^{s(b)}$ are negatively biased.  In the case of estimating expected shortfall, all estimators are generally negatively biased.  There is mixed evidence on the impact an increase in $k_0$ has on the S of $(q^s, E^s)$,  $(q^{s(b)}, E^{s(b)})$, $(\hat{q}, \hat{E})$ and $(\hat{q}^{(b)}, \hat{E}^{(b)})$, especially for larger $a$ values.  This is expected since the distribution of $\varepsilon_t$ exhibits less heavy tails with larger $k_0$. The performance of $(\dot{q},\dot{E})$ does not seem to depend on $k_0$ in a clear fashion. With a few exceptions on B, we notice that it is more difficult to estimate the conditional expected shortfall relative to the value-at-risk, judged by the larger B, S and R for all estimators across different experiment designs. It is also harder to estimate higher order conditional value-at-risk and expected shortfall, as demonstrated by the larger B, S and R for all estimators, with some exceptions for B. 

Across all experiment designs, the best estimator for $q$ in terms of R is $q^s$ and the best estimator for $E$ is either $E^s$ or $E^{s(b)}$, where the latter is often the best when sample sizes are large and $v=3$. Thus, the root mean squared errors are constructed for the other estimators relative to $q^s,E^s$ or $ E^{s(b)}$. For the estimation of the $a$-CVaR($\mathbf{x}$), we notice that the biases of ($q^{s(b)}$, $\hat{q}^{(b)}$) do not seem to be smaller than those of ($q^s$, $\hat{q}$). As expected, S for $q^s$ and $q^{s(b)}$ are smallest, followed by that of $\hat{q}$ and $\hat{q}^{(b)}$, with exceptions when $v = 6$ and $h_1(Y_{t-1})$, where the S for $\hat{q}$ is smaller than that for $q^{s(b)}$. $\dot{q}$ always carries the largest S. In terms of R, $q^s$ is the best estimator, followed in order by $\hat{q}$, $q^{s(b)}$, $\hat{q}^{(b)}$ and $\dot{q}$, with exceptions for $v=3$ and $h_2(Y_{t-1})$.  In these cases the performance of $q^{s(b)}$ is better than that of $\hat{q}$.  For the estimation of the $a$-CES($\mathbf{x}$), we notice that the bias-corrected estimators ($E^{s(b)}, \hat{E}^{(b)}$) exhibit smaller B than their counterparts ($E^s, \hat{E}$), though with a cost of larger $S$. When $a>0.95$, the best estimators in terms of R are either $E^{s(b)}$ or $E^s$, followed in order by $\hat{E}$, $\hat{E}^{(b)}$ and $\dot{E}$.  When $a=0.95$, $E^{s(b)}$ is the best estimator, followed by $\hat{E}^{(b)}$ or $E^s$, and then by $\hat{E}$.  $\dot{E}$ does not always carry the largest R, on occasions it performs better than $\hat{E}$.  Thus, in terms of estimation performance, our proposed estimators $(\hat{q},\hat{E})$ and $(\hat{q}^{(b)},\hat{E}^{(b)})$ can offer finite sample improvement over $(\dot{q},\dot{E})$.  We notice that the improvement could be sizable when $a>0.95$. To illustrate, we plot in Figure 2 the relative root mean squared error of $\dot{q}$ and $\hat{q}$ $\left(\frac{\dot{q}}{\hat{q}}\right)$, and $\dot{E}$ and $\hat{E}$ $\left(\frac{\dot{E}}{\hat{E}}\right)$ across sample sizes $1000$ and $4000$ for $\theta=0$ and $a=0.999$. We observe that the relative root mean squared errors are all greater than one. Furthermore, as the sample size increases, the relative root mean squared error generally becomes larger, illustrating that the finite sample improvement of $(\hat{q},\hat{E})$ over $(\dot{q},\dot{E})$ gets magnified with sample sizes. As $v$ is increased, the advantage of $(\hat{q},\hat{E})$ over $(\dot{q},\dot{E})$ is more prominent. For example, in the case of estimating $q$, the relative root mean squared error of $\frac{\dot{q}}{\hat{q}}$ is over $1.9$ for $v=6$, so the reduction in the root mean squared error of $\hat{q}$ over $\dot{q}$ is more than $47\%$. In the case of estimating $E$, the relative root mean squared error $\frac{\dot{E}}{\hat{E}}$ is over $2.4$ for $v=6$, so the reduction in the root mean squared error of $\hat{E}$ over $\dot{E}$ is more than $58\%$. 

We conclude that our estimators $(\hat{q}, \hat{E})$ have good finite sample performance and can be especially useful when estimating higher order conditional value-at-risk and expected shortfall. $(\hat{q}^{(b)}, \hat{E}^{(b)})$ provide reasonable alternatives and their asymptotic distributions are bias free, which enable us to construct confidence intervals.  The $95\%$-empirical coverage probability (ECP) in Table 4 gives an indication of the performance of the confidence interval estimates. The ECP seems to improve when $v$ is decreased for all estimators at least when $h_1(Y_{t-1})$ is considered. The ECP for the bias-corrected estimators $(\hat{q}^{(b)},\hat{E}^{(b)})$ is similar to that for $(q^{s(b)},E^{s(b)})$, indicating that the estimation of $m$ and $h$ does not pose a significant challenge in constructing confidence intervals. The ECP for the $a$-CES($\mathbf{x}$) seems to be closer to the target of $95\%$. The ECP for the $a$-CVaR($\mathbf{x}$) can be relatively far from the target for lower values of $a$, but it gets much closer to the target when $a$ is larger. 

The simulation results for the estimators do not change qualitatively across different values of $\theta$, which suggests that accounting for the nonlinearity in the conditional mean and variance functions is important for estimating high order $q$ and $E$. Overall, the study suggests that utilizing Extreme Value Theory and properly accounting for nonlinearities seems to pay off in finite samples.

The choice of $N$ could be an important issue because the number of residuals exceeding the threshold is based on $\tilde{q}(a_N)$.  We need to choose a large $\tilde{q}(a_N)$ to reduce the bias from approximating the tail distribution with a GPD, but we need to keep $N$ large (or $\tilde{q}(a_N)$ small) to control the variance of the estimates.\footnote{Note that the number of exceedances $N_s$ over $\tilde{q}(a_N)$ is asymptotically of the same order as $N$, since $\sqrt{N} \left( \frac{N_s-N}{N} \right) =O_p(1)$ (Lemma 5).}  We suggested earlier that our estimators are relatively robust to the choice of $N$, and here we specifically illustrate the impact from different $N$'s on the performance of our estimators for the $99\%$ conditional value-at-risk and expected shortfall with a simulation. We set $n=1000$, $\sigma^2_1(Y_{t-1})=1+ 0.01Y_{t-1}^2+0.5sin(Y_{t-1})$, $\theta=0$ and use a Student-t distributed $\varepsilon_t$ with $v=3$. We graph the bias and root mean squared error of $\hat{q}$ and $\hat{E}$ against $N=20,25, \cdots, 300$ in the left panel of Figure 3, and plot those of $\hat{q}^{(b)}$ and $\hat{E}^{(b)}$ in the right panel. The other experiment designs give graphs of similar general pattern. We observe that $\hat{q}$ carries a small positive bias and $\hat{E}$ is generally negatively biased. As we have mentioned above, it is harder to estimate the conditional expected shortfall than the value-at-risk, judged by the larger bias and root mean squared error of $\hat{E}$. The performance of $\hat{q}$ is fairly robust in the range of $N$ considered, with slight improvement when $N$ is greater than $20.$  The bias of $\hat{E}$ seems to be smallest when $N$ is between $30$ and $50$, but its magnitude increases with smaller $N$, and grows steadily with larger $N$. The root mean squared error of $\hat{E}$ decreases sharply from $N=20$ to $60$ and drops gradually until $N=160$. It remains fairly stable for a wide range of $N$ and eventually increases slowly for $N$ greater than $220$. 

The performance of the bias-corrected estimators $\hat{q}^{(b)}$ and $\hat{E}^{(b)}$ could depend on the choice of $N$ in a delicate fashion, since it requires the estimation of the second order parameter $\rho$. From our discussion above, the estimation of $\rho$ depends on a fine-tune parameter $c$, which assures that for $N$ and $n$, $N(c)$ and $N((c/2))$ are less than $n$. We adopt the same $\rho$ estimate described above for all $N$ considered, so that we can focus attention on the impact of $N$ from the bias correction made other than the estimation of $\rho$. We observe that $\hat{q}^{(b)}$ and $\hat{E}^{(b)}$ are generally negatively biased. Again, it is harder to estimate conditional expected shortfall than value-at-risk, judged by the larger B and R of $\hat{E}^{(b)}$. The B and R of $\hat{q}^{(b)}$ are smallest for $N$ between $20$ and $60$, beyond which they start to increase but remain fairly small until $N=180$. Their performances start to deteriorate quickly for $N$ larger than $180$. On the other hand, the bias of $\hat{E}^{(b)}$ remains small for $N=30$ to $100$, beyond which it starts to increase. The root mean squared error drops sharply from $N=20$ to $30$, then it remains low for a wide range ($30 \leq N \leq 220$), beyond which it starts to increase. Thus, we conclude that the performance of $\hat{q}$, $\hat{E}$, $\hat{q}^{(b)}$ and $\hat{E}^{(b)}$ is fairly robust in a wide range of values for $N$. Relative to $\hat{q}$ and $\hat{E}$, the choice of $N$ is more crucial for the bias-corrected estimators, especially for $\hat{q}^{(b)}$.


\section{Empirical illustration with backtesting}

We illustrate the empirical applicability of our estimators using five historical daily series $\{Y_t\}$ on the following log returns of future prices (contracts expiring between 1 and 3 months): (1) Maize from August 10, 1998 to July 28, 2004. (2) Rice from August 1, 2002 to July 18, 2008. (3) Soybean from July 25, 2006 to July 6, 2012. (4) Soft wheat from August 15, 1996 to July 31, 2002.  The data are obtained from the Chicago Board of Trade.  We also obtain (5) Hard wheat from August 1, 1996 to July 18, 2002 from Kansas City Board of Trade. 

To backtest on a data set $\{Y_1,Y_2,\cdots,Y_m\}$, we utilize the previous $n$ observations $\{Y_{t-n+1},Y_{t-n+2},$ $\cdots,Y_t\}$ to estimate the $a$-CVaR by $\hat{q}_{Y|X=Y_t}(a)$ and the $a$-CES by $\hat{E}(Y|Y>q_{Y|X=Y_t}(a),X=Y_t)$ for $a=0.95, 0.99,$ and $0.995$, where $0<n<m$, $t \in T = \{n,n+1,\cdots, m-1\}$. We fix $m=1500$, $n=1000$, let $N = round(n^{0.8-0.01})=234$ and implement our estimators as in the simulation study.  We provide in Figure 4 the plot of log returns of Maize futures prices against time together with the $95\%$ conditional value-at-risk and expected shortfall estimates. Clearly our estimates respond quickly to changing prices.

To backtest the $a$-CVaR estimator, we define a violation as the event $\{Y_{t+1}>q_{Y|X=Y_t}(a)\}$. Under the null hypothesis that the dynamics of $Y_t$ are correctly specified, $I_t \equiv \chi_{\{Y_{t+1}>q_{Y|X=Y_t}(a)\}} \sim Bernoulli(1-a)$ where $\chi_A$ is the indicator function.  Consequently, $W = \sum_{t\in T} I_t \sim Binomial (m-n,1-a)$. We perform a two sided test with the alternative hypothesis that the quantile is not correctly estimated with too many or too few violations. Since $q_{Y|X=Y_t}(a)$ is not observed, we estimate it with $\hat{q}_{Y|X=Y_t}(a)$ and construct the empirical version of the test statistic as $\hat{W} = \sum_{t\in T}\chi_{\left\{Y_{t+1}>\hat{q}_{Y|X=Y_t}(a)\right\}}$. Under the null hypothesis, the standardized test statistic $\frac{\hat{W} - (m-n)(1-a)}{\sqrt{(m-n)(1-a)a}}$ is distributed asymptotically as a standard normal. We report the violation numbers together with the p-values based on the normal distribution for our estimator on the left part of Table 5. For all five daily series and across all values of $a$ considered, the actual number of violations is fairly close to the expected number, with large p-values indicating no rejection of the null hypothesis. The only relatively large deviation of the violation numbers from expected is for $a=0.95$ on Maize, but its p-value is still larger than $0.1$. 

If the dynamics of $Y_t$ are correctly specified, the violation sequences are expected to be independent and have a correct conditional coverage.  The above coverage test and the first-order Markov test proposed by \cite{Christoffersen1998} on the independence of the violation sequences are shown to have relatively small power in \cite{Christoffersen2004} and \cite{Christoffersen2009}. The duration based likelihood ratio tests ($T_{ind}$ and $T_{cc}$) proposed by \cite{Christoffersen2004}  have considerably better power in many cases. The tests are based on the duration of days between the violations of the value-at-risk. The first test statistic $T_{ind}$ corresponds to the independence assumption under the null, and the durations have an exponential distribution (memory-free distribution) but with a rate parameter that can be different from $1/(1-a)$, where $1-a$ denotes the value-at-risk coverage rate.  The second test statistic $T_{cc}$ corresponds to the conditional coverage assumption under the null, and the durations have an exponential distribution with a rate parameter equal to $1/(1-a)$. We implement the two duration based tests\footnote{We use the Matlab code by C. Hurlin available at  \href{http://www.runshare.org/CompanionSite/site.do?siteId=68}{http://www.runshare.org/CompanionSite/site.do?siteId=68}. Note the p-value entries for $T_{ind}$ and $T_{cc}$ in Table 5 for Soybean are not available at $a=0.995$ since with only two violations, and one being at the end of the evaluation sample, the log-likelihood can not be evaluated numerically.} constructed with our $\hat{q}_{Y|X=Y_t}(a)$ estimate and provide the corresponding p-values in the middle part of Table 5.  With exceptions on Rice and Soybean for $a=0.99$ and $0.995$, the p-values are reasonably large, indicating that the violations constructed with our $\hat{q}_{Y|X=Y_t}(a)$ estimate exhibit reasonable conditional coverage and independence.

To backtest the $a$-CES we consider the normalized difference between $Y_{t+1}$ and $E(Y|Y>q_{Y|X=Y_t}(a),X=Y_t)$ as $r_{t+1} = \frac{Y_{t+1} - E(Y|Y>q_{Y|X=Y_t}(a),X=Y_t)}{h^{1/2}(Y_t)} = \varepsilon_{t+1} - E(\varepsilon|\varepsilon>q(a))$. If the return dynamics are correctly specified, given that $Y_{t+1}>q_{Y|X=Y_t}(a)$, $r_{t+1}$ is independent and identically distributed with mean zero. Since $E(Y|Y>q_{Y|X=Y_t}(a),X=Y_t)$ is not observed, we use the estimated residuals $\{\hat{r}_{t+1}:t \in T \mbox{ and } Y_{t+1}> \hat{q}_{Y|X=Y_t}(a) \}$, where $\hat{r}_{t+1} = \frac{Y_{t+1} - \hat{E}(Y|Y>q_{Y|X=Y_t}(a),X=Y_t)}{\hat{h}^{1/2}(Y_t)}$. Without making specific distributional assumptions on the residuals, we perform a one-sided bootstrap test as described in Efron and Tibshirani (1993, pp.224-227) to test the null hypothesis that the mean of the residuals is zero against the alternative that the mean is greater than zero, since underestimating $a-$conditional expected shortfall is likely to be the direction of interest. The p-values of the test for the five series across all values of $a$ are provided on the right part of Table 5. Given a $5\%$ significance level for the test, the null hypothesis for our $a$-conditional expected shortfall estimator is not rejected for $a=0.99$ and $0.995$ for all series, but it is rejected for $a=0.95$. The empirical results seem to confirm our Monte Carlo study, in that our estimators can be especially useful in estimating higher order conditional value-at-risk and expected shortfall.


\section{Summary and conclusion}

The estimation of conditional value-at-risk and conditional expected shortfall has been the subject of much interest in both empirical finance and theoretical econometrics.  Perhaps the interest is driven by the usefulness of these measures for regulators, portfolio managers and other professionals interested in an effective and synthetic tool for measuring risk.  Most stochastic models and estimators proposed for conditional value-at-risk and expected shortfall are hampered in their use by tight parametric specifications that most certainly impact performance and usability.  In this paper we have proposed fully nonparametric estimators for value-at-risk and expected shortfall, showed their consistency and obtained their asymptotic distributions.  Our Monte Carlo study has revealed that our estimators outperform those proposed by \cite{Cai2008} indicating that the use of the approximations provided by Extreme Value Theory may indeed prove beneficial.

We see an important direction for future research related to the contribution in this paper.  The fact that we require $s \geq 2d$ presents a strong requirement on the smoothness of the location and scale functions.  This perverse manifestation of the curse of dimensionality requires a solution.  Perhaps restricting $m$ and $h$ to belong to a class of additive functions, such that $m(\mathbf{x})=\sum_{u=1}^d m_{u}(x_u)$ and $h(\mathbf{x})=\sum_{u=1}^d h_{u}(x_u)$ may be sufficient to substantially relax the restriction that $s \geq 2d$.

\section*{Appendix 1 - Figures and Tables}
\begin{figure}[H]
\centering
\includegraphics[width=6in,height=3.25in]{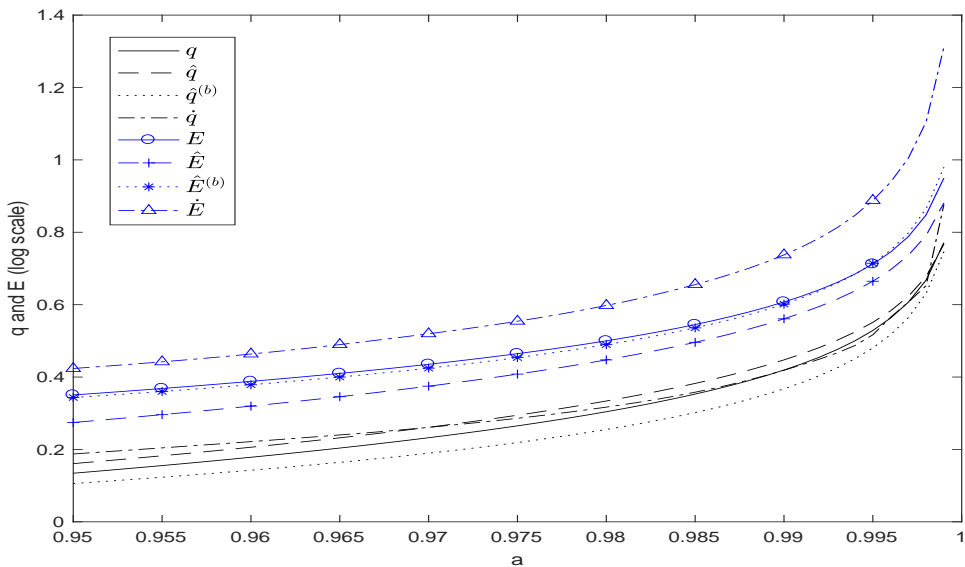}
\caption{Plots of true and estimated conditional value-at-risk ($q$) and expected shortfall ($E$) evaluated at the sample mean across different $a$, with $n=1000$, $h_1(Y_{t-1})=1+ 0.01Y_{t-1}^2+0.5sin(Y_{t-1})$, $\theta=0$ and Student-t distributed $\varepsilon_t$ with $v=3$.}
\end{figure}

\clearpage

\begin{figure}[H]
\centering
\includegraphics[width=6.5in,height=3.25in]{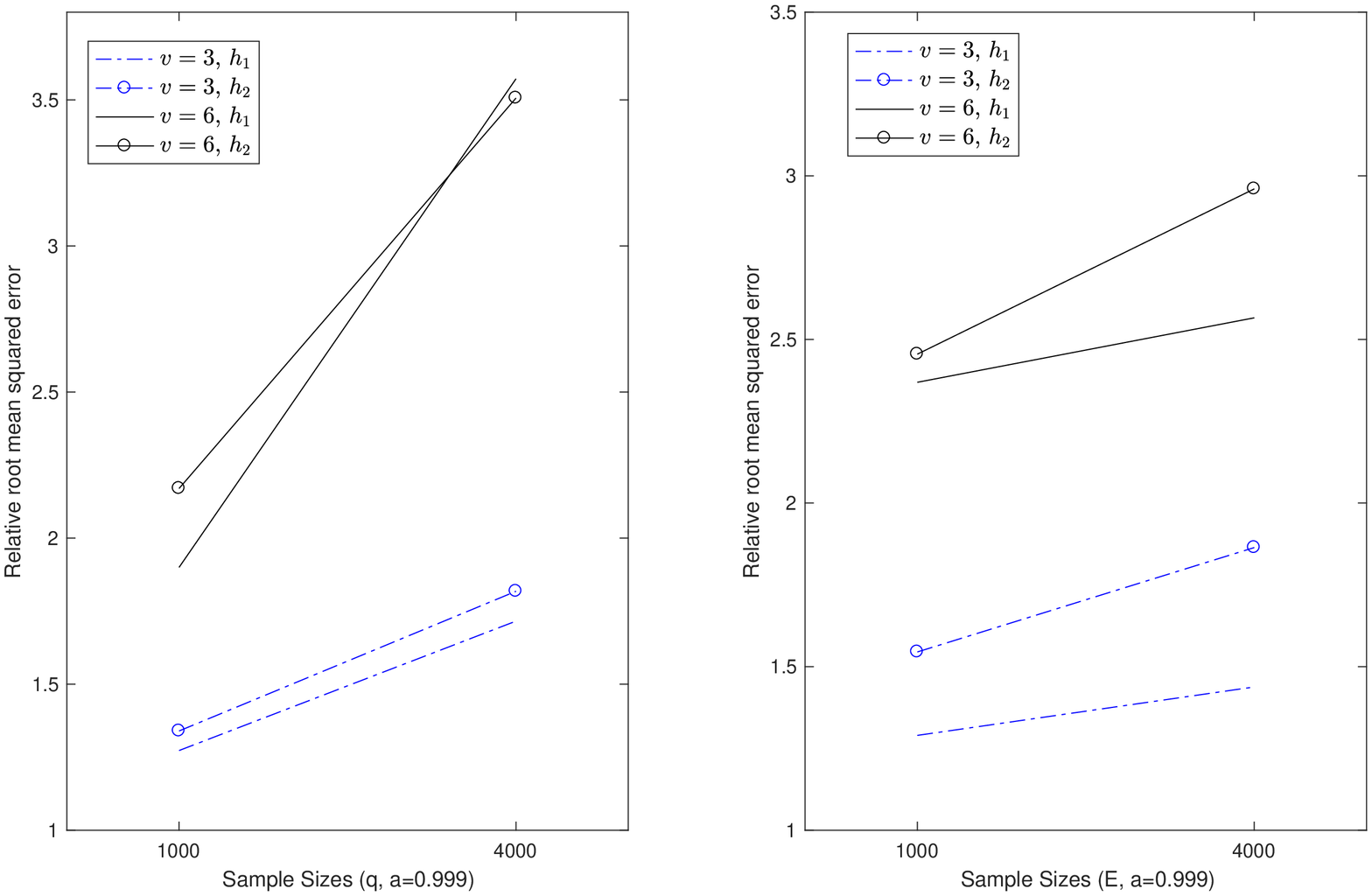}
\caption{Relative root mean squared error of $\frac{\dot{q}}{\hat{q}}$ (left) and $\frac{\dot{E}}{\hat{E}}$ (right) across sample sizes $1000$ and $4000$ for $\theta = 0$, Student-t distributed $\varepsilon_t$ with $v$ degrees of freedom, $h_i(Y_{t-1})$ for $i=1,2$ and $a=0.999$.}
\end{figure}

\begin{figure}[H]
\centering
\includegraphics[width=6.5in,height=3.25in]{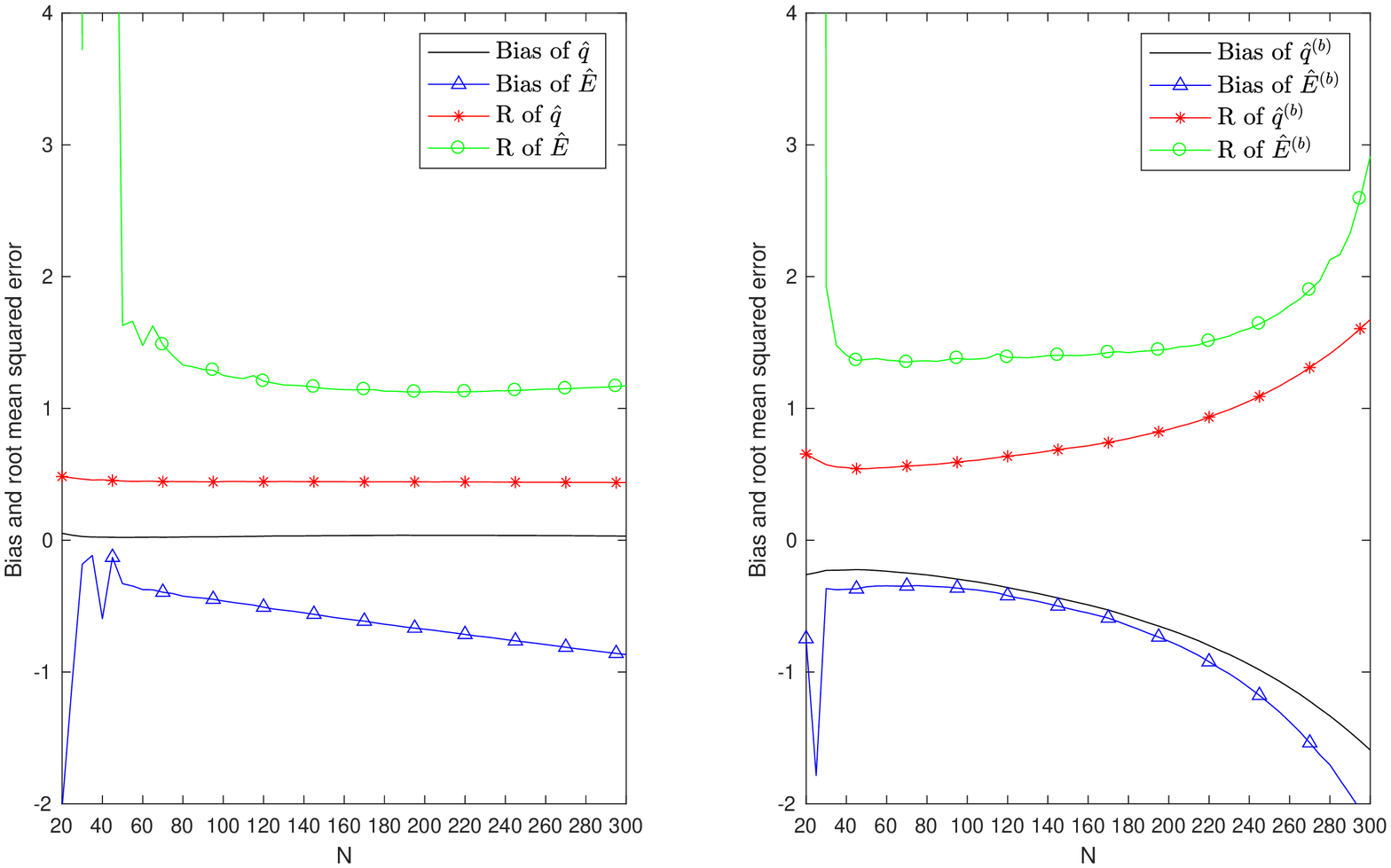}
\caption{Bias (B) and root mean squared error (R) of $99\%$ conditional value-at-risk and expected shortfall estimators against $N$, for $n=1000$, $h_1(y_{t-1})$, $\theta=0$ and Student-t distributed $\varepsilon_t$ with $v=3$.  Left panel: $(\hat{q},\hat{E})$.  Right panel: $(\hat{q}^{(b)},\hat{E}^{(b)})$.}
\end{figure}

\begin{figure}[H]
\centering
\includegraphics[width=7in,height=3.3in]{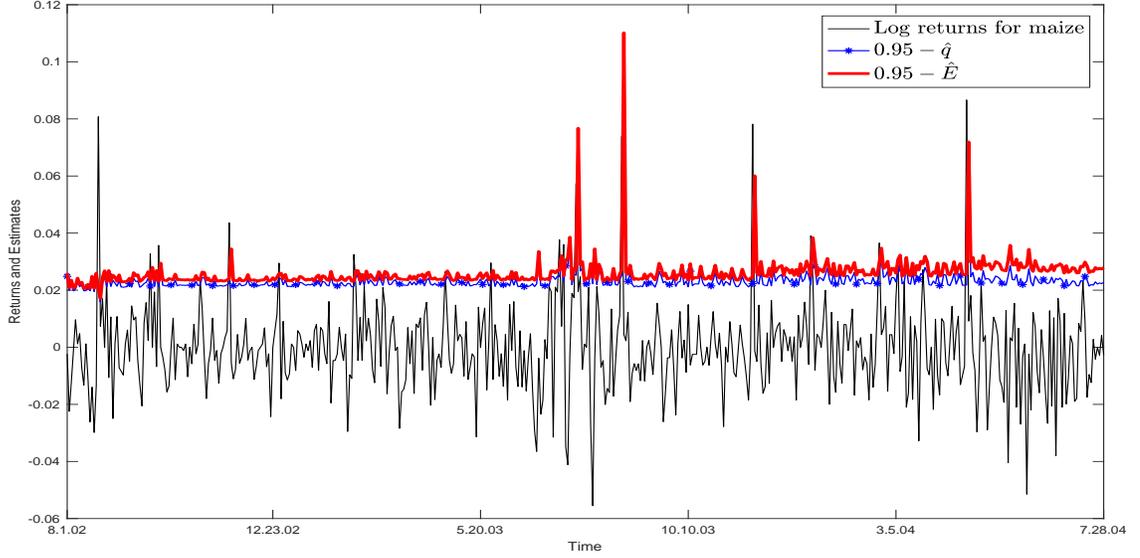}
\caption{Log returns for Maize future prices from Aug. 1, 2002 to July 28, 2004, together with the $95\%$ conditional value-at-risk ($\hat{q}$) and expected shortfall ($\hat{E}$) estimates.}
\end{figure}

\begin{table}[H]
\caption{Bias (B), Standard deviation (S) and Root mean squared error (R) for parameter estimators with sample size $n(\times 1000)$ and $\theta=0$, where $k_0=-1/v$.}
\begin{center}
\begin{tabular}{ccccccccccccccc}\hline \hline
\multicolumn{3}{c}{}&\multicolumn{6}{c}{$h_1(Y_{t-1})=1+0.01Y_{t-1}^2+0.5sin(Y_{t-1})$}& \multicolumn{6}{c}{$h_2(Y_{t-1}) = 1 - 0.9exp(-2Y_{t-1}^2)$} \\ \hline
\multicolumn{3}{c}{}&\multicolumn{3}{c}{$\sigma_N$}& \multicolumn{3}{c}{$k_0$}&\multicolumn{3}{c}{$\sigma_N$}& \multicolumn{3}{c}{$k_0$} \\ \hline
\multicolumn{1}{c}{}& \multicolumn{1}{c}{$v$}&\multicolumn{1}{c}{$n$}& \multicolumn{1}{c}{B}& \multicolumn{1}{c}{S}& \multicolumn{1}{c}{R}& \multicolumn{1}{c}{B}& \multicolumn{1}{c}{S}& \multicolumn{1}{c}{R}& \multicolumn{1}{c}{B} & \multicolumn{1}{c}{S}& \multicolumn{1}{c}{R}& \multicolumn{1}{c}{B}& \multicolumn{1}{c}{S}& \multicolumn{1}{c}{R}\\ \hline
$\hat{\gamma}$& 3&1&.294&.062&.300&.126&.102&.162&.294&.061&.300&.126&.101&.162  \\ \cline{4-15}
$\tilde{\gamma}$& 3&1&.320&.084&.331&.127&.107&.166&.274&.073&.283&.133&.110&.173  \\ \cline{4-15}
$\hat{\gamma}^{(b)}$& 3&1&-.019&.059&.062&-.030&.081&.087&-.019&.060&.063&-.030&.083&.088  \\ \cline{4-15}
$\tilde{\gamma}^{(b)}$& 3&1&-.011&.055&.057&-.028&.088&.092&-.043&.049&.065&-.024&.095&.098  \\ \cline{4-15}
$\hat{\gamma}$& 3&4&.251&.037&.253&.091&.057&.107&.250&.036&.252&.088&.057&.105  \\ \cline{4-15}
$\tilde{\gamma}$& 3&4&.266&.052&.271&.087&.061&.106&.211&.048&.217&.085&.069&.109  \\ \cline{4-15}
$\hat{\gamma}^{(b)}$& 3&4&.002&.038&.038&-.035&.041&.054&.004&.038&.039&-.037&.041&.056  \\ \cline{4-15}
$\tilde{\gamma}^{(b)}$& 3&4&.010&.034&.035&-.038&.045&.059&-.030&.032&.044&-.041&.054&.068  \\ \cline{1-15}
$\hat{\gamma}$& 6&1&.466&.073&.472&.161&.097&.188&.464&.068&.469&.161&.093&.186  \\ \cline{4-15}
$\tilde{\gamma}$& 6&1&.464&.072&.470&.169&.097&.195&.415&.068&.420&.163&.096&.189  \\ \cline{4-15}
$\hat{\gamma}^{(b)}$& 6&1&.025&.058&.063&-.050&.080&.094&.025&.055&.060&-.051&.076&.092  \\ \cline{4-15}
$\tilde{\gamma}^{(b)}$& 6&1&.026&.054&.060&-.043&.081&.091&.009&.049&.049&-.047&.080&.093  \\ \cline{4-15}
$\hat{\gamma}$& 6&4&.407&.038&.409&.124&.052&.134&.408&.038&.409&.124&.051&.134  \\ \cline{4-15}
$\tilde{\gamma}$& 6&4&.404&.038&.405&.125&.052&.136&.363&.036&.364&.121&.052&.132  \\ \cline{4-15}
$\hat{\gamma}^{(b)}$& 6&4&.051&.034&.061&-.059&.040&.072&.051&.034&.061&-.059&.039&.070  \\ \cline{4-15}
$\tilde{\gamma}^{(b)}$& 6&4&.051&.031&.059&-.058&.040&.070&.033&.029&.043&-.060&.040&.072  \\ \cline{1-15}
\hline \hline \\
\end{tabular} 
\end{center}
\label{table1}
\end{table}
\begin{table}[H]
\caption{Bias (B), Standard deviation (S) and Relative Root Mean Squared Error (R) for conditional value-at-risk (\rm q\rm) and expected shortfall (E) estimators with $v=3$, sample size $n(\times 1000)$, and $\theta=0$.}
\begin{center}
\begin{tabular}{ccccccccccc}\hline \hline 
\multicolumn{2}{c}{$h_1(Y_{t-1})$}&\multicolumn{3}{c}{$a=0.95$}& \multicolumn{3}{c}{$a=0.99$} &\multicolumn{3}{c}{$a=0.999$}\\ \hline
\multicolumn{1}{c}{}& \multicolumn{1}{c}{$n$}& \multicolumn{1}{c}{B}& \multicolumn{1}{c}{S}& \multicolumn{1}{c}{R}& \multicolumn{1}{c}{B}& \multicolumn{1}{c}{S}& \multicolumn{1}{c}{R} &  \multicolumn{1}{c}{B}& \multicolumn{1}{c}{S}& \multicolumn{1}{c}{R}\\ \cline{3-11}
$q^s$& 1&.008&.067&1&.018&.219&1&-.400&1.052&1 \\ \cline{3-11}
$\hat{q}$& 1&.024&.132&1.991&.043&.321&1.472&-.354&1.259&1.162 \\ \cline{3-11}
$q^{s(b)}$& 1&-.183&.086&2.999&-.499&.336&2.732&-1.043&1.482&1.610 \\ \cline{3-11}
$\hat{q}^{(b)}$& 1&-.183&.143&3.446&-.505&.414&2.967&-1.056&1.659&1.748 \\ \cline{3-11}
$\dot{q}$& 1&.036&.296&4.428&.148&.578&2.710&-.144&1.659&1.480 \\ \cline{3-11}
$q^s$& 4&.002&.035&1&.030&.110&1&-.219&.565&1 \\ \cline{3-11}
$\hat{q}$& 4&.013&.089&2.581&.064&.195&1.800&-.116&.735&1.228 \\ \cline{3-11}
$q^{s(b)}$& 4&-.081&.036&2.544&-.277&.160&2.810&-.517&.790&1.558 \\ \cline{3-11}
$\hat{q}^{(b)}$& 4&-.075&.094&3.455&-.261&.244&3.136&-.451&.960&1.751 \\ \cline{3-11}
$\dot{q}$& 4&-.004&.195&5.605&.056&.444&3.929&.139&1.270&2.108 \\ \hline
$E^s$& 1&-.481&.221&1.293&-.636&.616&1&-1.705&2.188&1\\ \cline{3-11}
$\hat{E}$& 1&-.467&.290&1.342&-.609&.743&1.085&-1.648&2.489&1.076 \\ \cline{3-11}
$E^{s(b)}$& 1&-.288&.291&1&-.590&.870&1.187&-.920&3.247&1.217\\ \cline{3-11}
$\hat{E}^{(b)}$& 1&-.292&.369&1.147&-.601&1.003&1.320&-.926&3.604&1.341 \\ \cline{3-11}
$\dot{E}$& 1&.088&.484&1.200&-.150&1.125&1.281&-.902&3.744&1.388 \\ \cline{3-11}
$E^s$& 4&-.428&.134&2.348&-.514&.342&1.163&-1.307&1.246&1.028\\ \cline{3-11}
$\hat{E}$& 4&-.406&.202&2.375&-.455&.458&1.217&-1.137&1.510&1.076 \\ \cline{3-11}
$E^{s(b)}$& 4&-.135&.135&1&-.290&.444&1&-.276&1.735&1\\ \cline{3-11}
$\hat{E}^{(b)}$& 4&-.116&.209&1.249&-.246&.565&1.162&.127&2.039&1.163 \\ \cline{3-11}
$\dot{E}$& 4&.089&.368&1.982&-.088&.793&1.504&-.564&2.658&1.547 \\ \hline \hline 
\multicolumn{2}{c}{$h_2(Y_{t-1})$}&\multicolumn{3}{c}{$a=0.95$}& \multicolumn{3}{c}{$a=0.99$} &\multicolumn{3}{c}{$a=0.999$}\\ \hline
\multicolumn{1}{c}{}& \multicolumn{1}{c}{$n$}& \multicolumn{1}{c}{B}& \multicolumn{1}{c}{S}& \multicolumn{1}{c}{R}& \multicolumn{1}{c}{B}& \multicolumn{1}{c}{S}& \multicolumn{1}{c}{R} &  \multicolumn{1}{c}{B}& \multicolumn{1}{c}{S}& \multicolumn{1}{c}{R}\\ \cline{3-11}
$q^s$& 1&.004&.035&1&.009&.112&1&-.211&.553&1 \\ \cline{3-11}
$\hat{q}$& 1&.007&.160&4.500&.018&.338&3.020&-.203&.995&1.717 \\ \cline{3-11}
$q^{s(b)}$& 1&-.106&.060&3.448&-.282&.208&3.124&-.587&.817&1.701 \\ \cline{3-11}
$\hat{q}^{(b)}$& 1&-.111&.177&5.898&-.298&.398&4.431&-.649&1.169&2.260 \\ \cline{3-11}
$\dot{q}$& 1&.021&.285&8.062&.238&.586&5.639&.183&1.348&2.301 \\ \cline{3-11}
$q^s$& 4&.001&.018&1&.018&.058&1&-.092&.300&1 \\ \cline{3-11}
$\hat{q}$& 4&.003&.143&7.976&.042&.279&4.649&-.017&.725&2.308 \\ \cline{3-11}
$q^{s(b)}$& 4&-.050&.025&3.118&-.156&.100&3.056&-.275&.428&1.619 \\ \cline{3-11}
$\hat{q}^{(b)}$& 4&-.048&.150&8.783&-.144&.309&5.615&-.238&.808&2.680 \\ \cline{3-11}
$\dot{q}$& 4&-.003&.198&11.013&.216&.516&9.217&.501&1.220&4.197 \\ \hline
$E^s$& 1&-.252&.146&1.234&-.333&.343&1&-.896&1.190&1\\ \cline{3-11}
$\hat{E}$& 1&-.258&.320&1.744&-.337&.647&1.526&-.900&1.811&1.357 \\ \cline{3-11}
$E^{s(b)}$& 1&-.167&.166&1&-.340&.480&1.231&-.542&1.731&1.217\\ \cline{3-11}
$\hat{E}^{(b)}$& 1&-.177&.338&1.619&-.371&.744&1.739&-.635&2.268&1.580 \\ \cline{3-11}
$\dot{E}$& 1&-.084&.358&1.559&-.754&.657&2.091&-1.914&2.469&2.097 \\ \cline{3-11}
$E^s$& 4&-.216&.099&2.181&-.251&.196&1.096&-.625&.687&1\\ \cline{3-11}
$\hat{E}$& 4&-.215&.274&3.203&-.221&.501&1.884&-.516&1.274&1.479 \\ \cline{3-11}
$E^{s(b)}$& 4&-.078&.076&1&-.162&.241&1&-.139&.919&1.001\\ \cline{3-11}
$\hat{E}^{(b)}$& 4&-.064&.257&2.435&-.129&.520&1.842&-.037&1.439&1.549 \\ \cline{3-11}
$\dot{E}$& 4&-.062&.315&2.946&-.654&.559&2.959&-1.825&1.799&2.758 \\ \hline \hline \\
\end{tabular} 
\end{center}
\label{table3}
\end{table}
\begin{table}[H]
\caption{Bias (B), Standard deviation (S) and Relative Root Mean Squared Error (R) for conditional value-at-risk (\rm q\rm) and expected shortfall (E) estimators with $v=6$, sample size $n(\times 1000)$, and $\theta=0$.}
\begin{center}
\begin{tabular}{ccccccccccc}\hline \hline
\multicolumn{2}{c}{$h_1(Y_{t-1})$}&\multicolumn{3}{c}{$a=0.95$}& \multicolumn{3}{c}{$a=0.99$} &\multicolumn{3}{c}{$a=0.999$}\\ \hline
\multicolumn{1}{c}{}& \multicolumn{1}{c}{$n$}& \multicolumn{1}{c}{B}& \multicolumn{1}{c}{S}& \multicolumn{1}{c}{R}& \multicolumn{1}{c}{B}& \multicolumn{1}{c}{S}& \multicolumn{1}{c}{R} &  \multicolumn{1}{c}{B}& \multicolumn{1}{c}{S}& \multicolumn{1}{c}{R}\\ \cline{3-11}
$q^s$& 1&.007&.063&1&.024&.144&1&-.167&.497&1 \\ \cline{3-11}
$\hat{q}$& 1&.007&.111&1.738&.004&.199&1.364&-.241&.543&1.134 \\ \cline{3-11}
$q^{s(b)}$& 1&-.322&.115&5.364&-.731&.304&5.426&-1.209&.842&2.811 \\ \cline{3-11}
$\hat{q}^{(b)}$& 1&-.324&.154&5.629&-.743&.337&5.597&-1.254&.855&2.895 \\ \cline{3-11}
$\dot{q}$& 1&.099&.417&6.716&.363&.695&5.376&.320&1.083&2.154 \\ \cline{3-11}
$q^s$& 4&-.003&.032&1&.026&.072&1&-.073&.257&1 \\ \cline{3-11}
$\hat{q}$& 4&-.004&.069&2.167&.018&.118&1.543&-.100&.306&1.208 \\ \cline{3-11}
$q^{s(b)}$& 4&-.167&.049&5.430&-.469&.152&6.401&-.745&.453&3.269 \\ \cline{3-11}
$\hat{q}^{(b)}$& 4&-.168&.089&5.913&-.473&.194&6.634&-.757&.489&3.380 \\ \cline{3-11}
$\dot{q}$& 4&.074&.364&11.580&.327&.658&9.535&.549&1.011&4.314 \\ \hline
$E^s$& 1&-.586&.158&1.227&-.643&.331&1&-1.035&.841&1\\ \cline{3-11}
$\hat{E}$& 1&-.601&.210&1.288&-.688&.381&1.087&-1.147&.876&1.082 \\ \cline{3-11}
$E^{s(b)}$& 1&-.527&.255&1.184&-.885&.558&1.446&-1.227&1.394&1.393\\ \cline{3-11}
$\hat{E}^{(b)}$& 1&-.546&.289&1.249&-.922&.582&1.507&-1.321&1.382&1.434 \\ \cline{3-11}
$\dot{E}$& 1&.002&.494&1&-.540&1.061&1.647&-1.218&3.194&2.563 \\ \cline{3-11}
$E^s$& 4&-.543&.114&1.643&-.561&.191&1&-.830&.477&1\\ \cline{3-11}
$\hat{E}$& 4&-.548&.160&1.692&-.577&.247&1.059&-.867&.535&1.065 \\ \cline{3-11}
$E^{s(b)}$& 4&-.313&.126&1&-.558&.294&1.064&-.673&.764&1.063\\ \cline{3-11}
$\hat{E}^{(b)}$& 4&-.320&.163&1.063&-.569&.330&1.110&-.699&.794&1.105 \\ \cline{3-11}
$\dot{E}$& 4&.030&.380&1.128&-.418&.937&1.729&-1.162&2.343&2.731\\ \hline \hline
\multicolumn{2}{c}{$h_2(Y_{t-1})$}&\multicolumn{3}{c}{$a=0.95$}& \multicolumn{3}{c}{$a=0.99$} &\multicolumn{3}{c}{$a=0.999$}\\ \hline
\multicolumn{1}{c}{}& \multicolumn{1}{c}{$n$}& \multicolumn{1}{c}{B}& \multicolumn{1}{c}{S}& \multicolumn{1}{c}{R}& \multicolumn{1}{c}{B}& \multicolumn{1}{c}{S}& \multicolumn{1}{c}{R} &  \multicolumn{1}{c}{B}& \multicolumn{1}{c}{S}& \multicolumn{1}{c}{R}\\ \cline{3-11}
$q^s$& 1&.002&.038&1&.010&.088&1&-.112&.296&1 \\ \cline{3-11}
$\hat{q}$& 1&-.014&.147&3.822&-.014&.244&2.768&-.146&.507&1.665 \\ \cline{3-11}
$q^{s(b)}$& 1&-.201&.093&5.741&-.450&.232&5.731&-.747&.551&2.929 \\ \cline{3-11}
$\hat{q}^{(b)}$& 1&-.213&.191&7.433&-.467&.359&6.665&-.772&.718&3.327 \\ \cline{3-11}
$\dot{q}$& 1&.066&.352&9.294&.398&.614&8.282&.574&.991&3.614 \\ \cline{3-11}
$q^s$& 4&-.002&.019&1&.015&.044&1&-.045&.154&1 \\ \cline{3-11}
$\hat{q}$& 4&-.017&.117&6.239&-.000&.183&3.957&-.055&.345&2.171 \\ \cline{3-11}
$q^{s(b)}$& 4&-.107&.044&6.103&-.290&.127&6.835&-.457&.294&3.381 \\ \cline{3-11}
$\hat{q}^{(b)}$& 4&-.115&.139&9.555&-.296&.257&8.480&-.459&.472&4.098 \\ \cline{3-11}
$\dot{q}$& 4&.065&.292&15.803&.418&.586&15.551&.784&.939&7.610 \\ \hline
$E^s$& 1&-.352&.152&1.022&-.389&.240&1&-.633&.548&1\\ \cline{3-11}
$\hat{E}$& 1&-.374&.280&1.243&-.419&.420&1.298&-.675&.785&1.236 \\ \cline{3-11}
$E^{s(b)}$& 1&-.327&.185&1&-.549&.374&1.454&-.774&.852&1.375\\ \cline{3-11}
$\hat{E}^{(b)}$& 1&-.346&.300&1.220&-.572&.516&1.685&-.806&1.032&1.564 \\ \cline{3-11}
$\dot{E}$& 1&-.291&.467&1.464&-.846&.756&2.481&-1.839&1.752&3.034 \\ \cline{3-11}
$E^s$& 4&-.323&.125&1.592&-.333&.156&1&-.492&.322&1\\ \cline{3-11}
$\hat{E}$& 4&-.337&.236&1.892&-.347&.323&1.290&-.498&.551&1.263 \\ \cline{3-11}
$E^{s(b)}$& 4&-.196&.095&1&-.347&.199&1.088&-.425&.449&1.051\\ \cline{3-11}
$\hat{E}^{(b)}$& 4&-.206&.210&1.354&-.352&.347&1.345&-.417&.639&1.297 \\ \cline{3-11}
$\dot{E}$& 4&-.232&.425&2.228&-.718&.669&2.672&-1.565&1.544&3.738\\ \hline \hline  \\
\end{tabular} 
\end{center}
\label{table4}
\end{table}

\begin{table}[H]
\caption{$95\%-$Empirical coverage probability for bias corrected $a$-conditional Value-at-Risk ($q$) and expected shortfall($E$) estimators with sample size $n(\times 1000)$, $\theta=0$.}
\begin{center}
\begin{tabular}{cccccccc}
\hline \hline 
\multicolumn{2}{c}{$h_1(Y_{t-1})$}&\multicolumn{3}{c}{$v=3$}&\multicolumn{3}{c}{$v=6$}\\
\cline{3-8}
\multicolumn{1}{c}{}&\multicolumn{1}{c}{$n$}&\multicolumn{1}{c}{$a=0.95$}&\multicolumn{1}{c}{$0.99$}&\multicolumn{1}{c}{$0.999$}&\multicolumn{1}{c}{$0.95$}&\multicolumn{1}{c}{$0.99$}&\multicolumn{1}{c}{$0.999$}\\
\cline{3-8}
\multicolumn{1}{c}{$q^{s(b)}$}&\multicolumn{1}{c}{$1$}&.439&.849&.986&.210&.850&.999\\ \cline{3-8}
\multicolumn{1}{c}{$\hat{q}^{(b)}$}&\multicolumn{1}{c}{$1$}&.438&.811&.963&.255&.860&.995\\ \cline{3-8}
\multicolumn{1}{c}{$q^{s(b)}$}&\multicolumn{1}{c}{$4$}&.413&.593&.979&.030&.138&.999\\ \cline{3-8}
\multicolumn{1}{c}{$\hat{q}^{(b)}$}&\multicolumn{1}{c}{$4$}&.391&.582&.935&.096&.159&.991\\ \cline{3-8}
\multicolumn{1}{c}{$E^{s(b)}$}&\multicolumn{1}{c}{$1$}&.995&.996&1&.999&1&1\\ \cline{3-8}
\multicolumn{1}{c}{$\hat{E}^{(b)}$}&\multicolumn{1}{c}{$1$}&.962&.981&.967&.987&.997&.998\\ \cline{3-8}
\multicolumn{1}{c}{$E^{s(b)}$}&\multicolumn{1}{c}{$4$}&.995&.999&1&.997&1&1\\ \cline{3-8}
\multicolumn{1}{c}{$\hat{E}^{(b)}$}&\multicolumn{1}{c}{$4$}&.955&.966&.964&.984&.997&.999\\
\hline\hline
\multicolumn{2}{c}{$h_2(Y_{t-1})$}&\multicolumn{3}{c}{$v=3$}&\multicolumn{3}{c}{$v=6$}\\ 
\cline{3-8}
\multicolumn{1}{c}{}&\multicolumn{1}{c}{$n$}&\multicolumn{1}{c}{$a=0.95$}&\multicolumn{1}{c}{$0.99$}&\multicolumn{1}{c}{$0.999$}&\multicolumn{1}{c}{$0.95$}&\multicolumn{1}{c}{$0.99$}&\multicolumn{1}{c}{$0.999$}\\
\cline{3-8}
\multicolumn{1}{c}{$q^{s(b)}$}&\multicolumn{1}{c}{$1$}&.395&.829&.989&.179&.847&.999\\ \cline{3-8}
\multicolumn{1}{c}{$\hat{q}^{(b)}$}&\multicolumn{1}{c}{$1$}&.383&.682&.933&.371&.801&.989\\ \cline{3-8}
\multicolumn{1}{c}{$q^{s(b)}$}&\multicolumn{1}{c}{$4$}&.304&.538&.973&.017&.102&1\\ \cline{3-8}
\multicolumn{1}{c}{$\hat{q}^{(b)}$}&\multicolumn{1}{c}{$4$}&.149&.397&.776&.155&.359&.966\\ \cline{3-8}
\multicolumn{1}{c}{$E^{s(b)}$}&\multicolumn{1}{c}{$1$}&.999&.996&.999&1&1&1\\ \cline{3-8}
\multicolumn{1}{c}{$\hat{E}^{(b)}$}&\multicolumn{1}{c}{$1$}&.920&.949&.942&.988&.995&.991\\ \cline{3-8}
\multicolumn{1}{c}{$E^{s(b)}$}&\multicolumn{1}{c}{$4$}&.999&.998&1&1&1&1\\ \cline{3-8}
\multicolumn{1}{c}{$\hat{E}^{(b)}$}&\multicolumn{1}{c}{$4$}&.761&.852&.905&.964&.988&.992\\ 
\hline \hline
\end{tabular} 
\end{center}
\label{table5}
\end{table}
\begin{table}[H]
\caption{Backtest results for $a$-conditional Value-at-Risk ($q$) and expected shortfall ($E$) on $m-n=500$ observations, expected violations = $(m-n)(1-a)$. $q$: Number of violations and p-value (in parentheses). Duration based tests for $q$: p-value for $T_{ind}$ and $T_{cc}$. $E$: p-value for exceedance residuals to have zero mean.}
\begin{center}
\noindent \begin{tabular}{lcccccccccccc}\hline \hline 
\multicolumn{1}{c}{\sc }&\multicolumn{3}{c}{$q$}&\multicolumn{6}{c}{$q$}& \multicolumn{3}{c}{$E$} \\ \cline{2-13}
\multicolumn{1}{c}{\sc }&\multicolumn{1}{c}{$a=0.95$}& \multicolumn{1}{c}{$0.99$} &\multicolumn{1}{c}{$0.995$}&\multicolumn{2}{c}{$0.95$}& \multicolumn{2}{c}{$0.99$} &\multicolumn{2}{c}{$0.995$}&\multicolumn{1}{c}{$0.95$}& \multicolumn{1}{c}{$0.99$} &\multicolumn{1}{c}{$0.995$}\\ \cline{2-13}
\multicolumn{1}{c}{}&\multicolumn{3}{c}{Expected violations}&\multicolumn{6}{c}{Duration based tests}& \multicolumn{3}{c}{\sc } \\
\multicolumn{1}{c}{\sc }& \multicolumn{1}{c}{$25$}& \multicolumn{1}{c}{$5$}& \multicolumn{1}{c}{$2.5$}& \multicolumn{1}{c}{$T_{ind}$}& \multicolumn{1}{c}{$T_{cc}$}& \multicolumn{1}{c}{$T_{ind}$}& \multicolumn{1}{c}{$T_{cc}$}& \multicolumn{1}{c}{$T_{ind}$}& \multicolumn{1}{c}{$T_{cc}$}& \multicolumn{3}{c}{}\\ \cline{2-13}
Maize & 18 (.151) & 5(1)& 2(.751)&.176&.096&.173&.356&.619&.494&0& .161& .735 \\ \cline{2-13}
Rice &29(.412)&4(.653)&2(.751)&.052&.126&.040&.076&.018&.033&0&.081&.248\\ \cline{2-13}
Soybean &21(.412)&3(.369)&2(.751)&.199&.259&.012&.014&-&-&0&.302&.244\\ \cline{2-13}
Soft Wheat &30(.305)&6(.653)&2(.751)&.839&.717&.283&.562&.712&.523&.001&.339&.273\\ \cline{2-13}
Hard Wheat &25(1)&5(1)&2(.751)&.294&.566&.055&.142&.739&.529&0&.082&.239\\ \hline \hline \\[.5in]
\end{tabular} 
\end{center}
\label{table6}
\end{table}


\section*{Appendix 2 - Proofs}
\sloppy We rely, throughout the proofs, on results from \cite{Smith1985} and \cite{Smith1987}.  For a nonstochastic positive sequence $q(a_N) \rightarrow \infty$ as $N \rightarrow \infty$ and $\sigma_N=\sigma(q(a_N))$, $k_0<0$ we have $E\left(\sigma_N \frac{\partial }{\partial \sigma} \logg (Z;\sigma_N,k_0) \right)=\frac{  \phi(q(a_N))}{(1-k_0^{-1}-\rho)} + o(\phi(q(a_N)))$, $E\left(\frac{\partial }{\partial k} \logg (Z;\sigma_N,k_0) \right)=\frac{k_0^{-1}  \phi(q(a_N))}{(-k_0^{-1}-\rho)(1-k_0^{-1}-\rho)} + o(\phi(q(a_N))$, $E\left(\sigma_N^2\frac{\partial ^2}{\partial \sigma^2} \logg (Z;\sigma_N,k_0) \right)=\frac{-k_0^{-1}}{2-k_0^{-1}} + O(\phi(q(a_N)))$, $E\left(\frac{\partial ^2}{\partial k^2} \logg (Z;\sigma_N, \right.$ $\left. k_0) \right)=-\frac{2k_0^{-2} }{(1-k_0^{-1})(2-k_0^{-1})} + O(\phi(q(a_N)))$ and $E\left(\sigma_N\frac{\partial ^2}{\partial \sigma \partial k} \logg (Z;\sigma_N,k_0) \right)= \frac{-k_0^{-2} }{(1-k_0^{-1})(2-k_0^{-1})} + O(\phi(q(a_N)))$,  where all expectations are taken with respect to the unknown distribution $F_{q(a_N)}$.  Evidently, these approximations are based on a sequence of thresholds $q(a_N)$ that approach the end point of the distribution $F$ as $N \rightarrow \infty$. 



\begin{proof}[\bf Theorem 2. \rm \it Proof.\rm]  a) Let $\tilde{r}_N=\frac{\tilde{\sigma}_{\tilde{q}(a_N)}}{\sigma_N} = 1+\delta_N \tau_1^*$, $\tilde{k}=k_0+\delta_N\tau_2^*$ and note that  
\begin{equation}\label{57}
\left(\begin{array}{cc}
\frac{1}{\delta_N^2}\frac{\partial}{ \partial \tau_1} L_{TN}(\tau_1^*,\tau_2^*)  \\
\frac{1}{\delta_N^2}\frac{\partial}{ \partial \tau_2} L_{TN}(\tau_1^*,\tau_2^*)\\ 
\end{array}\right)= \frac{1}{\delta_NN} \left(\begin{array}{cc}
\sum_{i=1}^N \frac{\partial}{ \partial r_N} \logg(\tilde{Z}_i;\tilde{r}_N\sigma_N,\tilde{k})\\
\sum_{i=1}^N \frac{\partial}{ \partial k} \logg(\tilde{Z}_i;\tilde{r}_N\sigma_N,\tilde{k})\\ 
\end{array}\right)=\left(\begin{array}{cc} 0  \\ 0\\ \end{array}\right).
\end{equation}
For some $\lambda_1,\, \lambda_2 \in (0,1)$ let $k^*=\lambda_2k_0+(1-\lambda_2)\tilde{k}$, $r^*_N=\lambda_1+(1-\lambda_1)\tilde{r}_N$, 
\begin{equation*}
H_N(r_N^*,k^*) = -\frac{1}{N}\sum_{i=1}^N\left(\begin{array}{cc}
\frac{\partial^2}{ \partial r_N^2}  \logg(\tilde{Z}_j;r_N^*\sigma_N,k^*)&  \frac{\partial^2}{ \partial k \partial r_N} \logg(\tilde{Z}_i;r_N^*\sigma_N,k^*)\\
\frac{\partial^2}{ \partial k \partial r_N} \logg(\tilde{Z}_i;r_N^*\sigma_N,k^*)& \frac{\partial^2}{ \partial k^2 } \logg(\tilde{Z}_i;r_N^*\sigma_N,k^*)\\ 
\end{array}\right) \mbox{ and }
\end{equation*}
\begin{equation*}
v_N(1,k_0) = \sqrt{N}\left(\begin{array}{cc}
 \frac{1}{N}\sum_{i=1}^N\frac{\partial}{ \partial r_N} \logg(\tilde{Z}_i;\sigma_N,k_0)\\
 \frac{1}{N}\sum_{i=1}^N\frac{\partial}{ \partial k} \logg(\tilde{Z}_i;\sigma_N,k_0)\\ 
\end{array}\right)=\sqrt{N}\left(\begin{array}{cc}
\delta_N (\tilde{I}_{1N}-I_{1N})+\delta_N I_{1N}\\
\delta_N ( \tilde{I}_{4N}-I_{4N})+\delta_N I_{4N}\\ 
\end{array}\right),
\end{equation*}
where $\tilde{I}_{1N}$, $I_{1N}$, $\tilde{I}_{4N}$, $I_{4N}$ are as defined in Theorem \ref{thm1}.  By a Taylor's expansion of the first order condition in (\ref{57}) around $(1,k_0)$ we have
\begin{equation}\label{60}
H_N(r^*_N,k^*) \sqrt{N}\left(\begin{array}{cc}
\tilde{r}_N-1  \\
\tilde{k}-k_0\\ 
\end{array}\right)= v_N(1,k_0).
\end{equation}
We start by investigating the asymptotic properties of $v_N(1,k_0)$.  Let $b_1=\frac{1-k_0}{k_0(1-2k_0)}$, $b_2=-\frac{1}{(1-k_0)(1-2k_0)}$ and observe that from Theorem \ref{thm1}, Lemma 5 and the fact that $\frac{q_n(a_N)}{q(a_N)}-1=o_p(1)$ we have 
\begin{eqnarray*}
v_N(1,k_0)&=&\left(\begin{array}{cc}
b_1\sqrt{N}\left( \frac{\tilde{q}(a_N)-q(a_N)}{q(a_N)} -\frac{q_n(a_N)-q(a_N)}{q(a_N)}\right)+\delta_N \sqrt{N}I_{1N}+o_p(1)  \\
b_2\sqrt{N}\left( \frac{\tilde{q}(a_N)-q(a_N)}{q(a_N)} - \frac{q_n(a_N)-q(a_N)}{q(a_N)}\right)+\delta_N \sqrt{N}I_{4N}+o_p(1)\\ 
\end{array}\right)
\end{eqnarray*}
By Lemma 6 and the fact that $N_1-N=O_p(N^{1/2})$
\begin{equation*}
\left(\begin{array}{cc}
\sqrt{N}\delta_N I_{1N}  \\
\sqrt{N}\delta_N I_{4N}\\ 
\end{array}\right)= 
\left(\begin{array}{cc}
b_1\sqrt{N}\frac{q_n(a_N)-q(a_N)}{q(a_N)}+\frac{1}{\sqrt{N}} \sum_{i=1}^N\frac{\partial}{\partial \sigma} \logg (Z_i';\sigma_N,k_0) \sigma_N+o_p(1)\\
b_2\sqrt{N}\frac{q_n(a_N)-q(a_N)}{q(a_N)}+\frac{1}{\sqrt{N}} \sum_{i=1}^N\frac{\partial}{\partial k} \logg (Z_i';\sigma_N,k_0)+o_p(1)\\ 
\end{array}\right)
\end{equation*}
where $Z_i'=\varepsilon_i-q(a_N)$ for $\varepsilon_i > q(a_N)$.  Hence, by letting $b_\sigma=E\left(\frac{\partial}{\partial \sigma} \logg (Z_i';\sigma_N,k_0) \sigma_N\right)$ and $b_k=E\left(\frac{\partial}{\partial k} \logg (Z_i';\sigma_N,k_0) \right)$ we have 
\begin{equation}\label{new29}
v_N(1,k_0)-\sqrt{N}\left(\begin{array}{cc}
b_\sigma  \\
b_k\\ 
\end{array}\right)=\left(\begin{array}{cc}
b_1\sqrt{N} \frac{\tilde{q}(a_N)-q(a_N)}{q(a_N)} +\frac{1}{\sqrt{N}}\left( \sum_{i=1}^N\frac{\partial}{\partial \sigma} \logg (Z_i';\sigma_N,k_0) \sigma_N-b_\sigma \right)+o_p(1)  \\
b_2\sqrt{N} \frac{\tilde{q}(a_N)-q(a_N)}{q(a_N)} +\frac{1}{\sqrt{N}} \left(\sum_{i=1}^N\frac{\partial}{\partial k} \logg (Z_i';\sigma_N,k_0) - b_k\right)+o_p(1)\\ 
\end{array}\right).
\end{equation}
Note that we can write 
\begin{align*}
\frac{1}{\sqrt{N}}\left( \sum_{i=1}^N\frac{\partial}{\partial \sigma} \logg (Z_i';\sigma_N,k_0) \sigma_N-b_\sigma \right)&=\frac{1}{\sqrt{N}}\sumtn \left( \frac{\partial}{\partial \sigma} \logg (Z_t';\sigma_N,k_0) \sigma_N-b_\sigma \right)\chi_{\{\varepsilon_t>q(a_N)\}}=\sumtn Z_{t1},\\
\frac{1}{\sqrt{N}}\left( \sum_{i=1}^N\frac{\partial}{\partial k} \logg (Z_i';\sigma_N,k_0)-b_k \right)&=\frac{1}{\sqrt{N}}\sumtn \left( \frac{\partial}{\partial k} \logg (Z_t';\sigma_N,k_0)-b_k \right)\chi_{\{\varepsilon_t>q(a_N)\}}=\sumtn Z_{t2}.
\end{align*}
Also, from Lemma 5, $\sqrt{N} \frac{\tilde{q}(a_N)-q(a_N)}{q(a_N)}$ is distributed asymptotically as $k_0\sumtn   (n(1-F(y_n)))^{-1/2}(q_{1n}-E(q_{1n}))+o_p(1) = \sumtn Z_{t3} + o_p(1)$ where $q_{1n}=\frac{1}{h_{3n}}\int_{-\infty}^{y_n}K_3\left( \frac{y-\varepsilon_t}{h_{3n}} \right)dy$ and $y_n=q(a_N)(1+N^{-1/2}z)$ for arbitrary $z$.  It can be easily verified that $E(Z_{t1})=E(Z_{t2})=E(Z_{t3})=0$.  In addition,
$V(Z_{t1})
=\frac{1}{n}E\left(\frac{\partial}{\partial \sigma_N} \logg (Z_i';\sigma_N,k_0) \sigma_N-b_\sigma\right)^2
=\frac{1}{n}\left(\frac{1}{1-2k_0}+o(1)\right)$,
where the last equality follows from the results listed in section 3.1.  Using similar arguments we obtain 
$
V(Z_{t2})=\frac{1}{n}\left(\frac{2}{(1-k_0)(1-2k_0)}+o(1)\right)
$ 
and from Lemma 5 we have that $V(Z_{t3})=\frac{1}{n}k_0^3F(y_n)+o(h_{3n})$.  We now define the vector $\psi_n=\sumtn ( Z_{t1}, Z_{t2},Z_{t3})^T$ and for arbitrary $0 \neq \lambda \in \mathds{R}^3$ we consider $\lambda^T \psi_n = \sumtn ( \lambda_1Z_{t1}+\lambda_2Z_{t2}+\lambda_3Z_{t3})=\sumtn Z_{tn}$.  From above, we have that $E(Z_{tn})=0$ and $V(Z_{tn})=\sum_{l=1}^3 \lambda_d^2E(Z_{td}^2)+2 \sum_{1 \leq d<d'\leq 3}\lambda_d\lambda_{d'}E(Z_{td}Z_{td'})$.  First, we consider $E(Z_{t1}Z_{t2})$ which can be written as 
$
E(Z_{t1}Z_{t2})=\frac{1}{n}T_{1n}-\frac{N}{n^2}b_\sigma b_k
$
where $T_{1n}=E\left(\frac{\partial}{\partial \sigma_N} \logg (Z_i';\sigma_N,k_0) \sigma_N\frac{\partial}{\partial k} \logg (Z_i';\sigma_N,k_0) \right)$.  Since $b_\sigma=\frac{\phi(\varepsilon_{(n-N)})}{1-\frac{1}{k_0}-\rho}+o(\phi(\varepsilon_{(n-N)}))$ and $b_k=\frac{ \phi(\varepsilon_{(n-N)})}{k_0(-\frac{1}{k_0}-\rho)(1-\frac{1}{k_0}-\rho)}+o(\phi(\varepsilon_{(n-N)}))$ we have that 
$
E(Z_{t1}Z_{t2})=\frac{1}{n}T_{1n}-O\left(\frac{(N^{1/2}\phi(\varepsilon_{(n-N)}))^2}{n^2}\right)=\frac{1}{n}T_{1n}-\frac{1}{n^2}O(1)
$
since $N^{1/2}\phi(\varepsilon_{(n-N)})=O(1)$.  Now, 
from \cite{Smith1987} we have that  $E\left( \left(1-\frac{k_0Z_t'}{\sigma_N}\right)^{-2}\left(\frac{k_0Z_t'}{\sigma_N}\right)^2\right)=\frac{2k_0^2}{(1-k_0)(1-2k_0)}+O(\phi(\varepsilon_{(n-N)}))$,  
and from Lemma 7 we have that 
$E\left( log \left(1-\frac{k_0Z_t'}{\sigma_N}\right)\left(1-\frac{k_0Z_t'}{\sigma_N}\right)^{-1}\left(\frac{k_0Z_t'}{\sigma_N}\right)\right)=\frac{k_0^3-2k_0^2}{(1-k_0)^2}+O(\phi(\varepsilon_{(n-N)})).
$
Combined with the orders obtained for the other components of the expectation give
$
E(Z_{t1}Z_{t2})=-\frac{1}{n(k_0-1)(2k_0-1)}+\frac{1}{n}\phi(\varepsilon_{(n-N)})O(1)-\frac{1}{n^2}O(1)$.  We now turn to $E(Z_{t1}Z_{t3})$ which can be written as
$
E(Z_{t1}Z_{t3})=T_{2n}-k_0E\left(N^{-1/2} \left( \frac{\partial}{\partial \sigma_N} \logg (Z_i';\sigma_N,k_0) \sigma_N\right)\chi_{\varepsilon_t>q(a_N)}\right)E(q_{1n})(n(1-F(y_n)))^{-1/2},
$
where $T_{2n}=E\left( N^{-1/2} \left( \frac{\partial}{\partial \sigma_N} \logg (Z_i';\sigma_N,k_0) \sigma_N\right)\chi_{\varepsilon_t>q(a_N)}(n(1-F(y_n)))^{-1/2}k_0q_{1n}\right)$.  We note that 
\begin{align*}
E\left( N^{-1/2} \left( \frac{\partial}{\partial \sigma_N} \logg (Z_i';\sigma_N,k_0) \sigma_N\right)\chi_{\{\varepsilon_t>q(a_N)\}}\right)&= \frac{\sqrt{N}}{n}b_\sigma=\frac{\sqrt{N}}{n}O(\phi(\varepsilon_{(n-N)})),
\end{align*}
from Lemma 5, $E(q_{1n})=F(y_n)+O(h_{3n}^{m+1})=O(1)$ and since $(n(1-F(y_n)))^{-1/2}$ is asymptotically equivalent to $N^{-1/2}$, the second term in the covariance expression is of order $\frac{\sqrt{N}}{n}O(\phi(\varepsilon_{(n-N)}))O(1)N^{-1/2}=n^{-1}O(\phi(\varepsilon_{(n-N)}))$.  We now turn to $T_{2n}$, the first term in the covariance expression.  Since $(n(1-F(y_n)))^{-1/2}$ is asymptotically equivalent to $N^{-1/2}$, we have by the Cauchy-Schwartz inequality
\begin{align*}
T_{2n}&=\frac{1}{n}E\left(\frac{\partial}{\partial \sigma_N} \logg (Z_i';\sigma_N,k_0) \sigma_N q_{1n} \right) \leq \frac{1}{n}\left|E\left(\frac{\partial}{\partial \sigma_N} \logg (Z_i';\sigma_N,k_0) \sigma_N q_{1n} \right)\right|\\
&\leq  \frac{1}{n}\left(E\left(\left(\frac{\partial}{\partial \sigma_N} \logg (Z_i';\sigma_N,k_0) \sigma_N\right)^2 \right)E(q^2_{1n}) \right)^{1/2}=\frac{1}{n}o(1).
\end{align*}
Hence, $E(Z_{t1}Z_{t3})=\frac{1}{n}o(1)$.  In a similar manner we obtain $E(Z_{t2}Z_{t3})=\frac{1}{n}o(1)$.  Hence,  $nV(Z_{tn})=\lambda^TV_1\lambda+o(1)$, where
$
V_1(k_0)=\left(\begin{array}{ccc}
\frac{1}{1-2k_0}&  -\frac{1}{(k_0-1)(2k_0-1)}&0\\
-\frac{1}{(k_0-1)(2k_0-1)}& \frac{2}{(k_0-1)(2k_0-1)}&0\\ 
0& 0&k_0^2\\
\end{array}\right)
$.
By Liapounov's CLT $\sumtn Z_{tn} \convd \mathcal{N}(0,\lambda^TV_1(k_0)\lambda)$ provided that $\sumtn E(|Z_{tn}|^3) \rightarrow 0$.  To verify this condition, it suffices to show that:\\ 
$
(i) \, \sumtn E(|Z_{t1}|^3) \rightarrow 0;\, (ii) \, \sumtn E(|Z_{t2}|^3) \rightarrow 0; (iii)  \, \sumtn E(|Z_{t3}|^3) \rightarrow 0
$. 
$(iii)$ was verified in Lemma 5, so we focus on $(i)$ and $(ii)$.  For $(i)$, note that $\sumtn E(|Z_{t1}|^3)\leq \frac{1}{\sqrt{N}}E\left(\left|(1/k_0-1)(1-k_0Z_t'/\sigma_N)^{-1}k_0Z_t'/\sigma_N-1\right|^3\right)\rightarrow 0$ provided $E(-(1-k_0Z_t'/\sigma_N)^{-3}(k_0Z_t'/\sigma_N)^3)<C$, which is easily verified by noting that $-(1-k_0Z_t'/\sigma_N)^{-3}\\(k_0Z_i'/\sigma_N)^3<-(1-k_0Z_t'/\sigma_N)^{-3}(1-k_0Z_t'/\sigma_N)^3=1.
$  
Lastly, 
$$\sumin E(|Z_{t2}|^3)\leq\frac{1}{\sqrt{N}}E\left( \left| -(1/k_0^2)log(1-k_0Z_t'/\sigma_N)+(1/k_0)(1-1/k_0)(1-k_0Z_t'/\sigma_N)^{-1}k_0Z_t'/\sigma_N\right|^3\right)\rightarrow 0
$$
provided $E\left(log(1-k_0Z_t'/\sigma_N)^3\right)<C$ given the bound we obtained in case $(i)$.  By FR1 a) and integrating by parts we have 
\begin{align*}
E\left(\text{log}\left(1-k_0\frac{Z_t'}{\sigma_N}\right)^3\right)&=-\int_0^\infty \text{log}\left(1-k_0\frac{z}{\sigma_N}\right)^3 dF_{\varepsilon_{(n-N)}}(z)\\
&=-\frac{1-F(\varepsilon_{(n-N)}(1+\frac{z}{\varepsilon_{(n-N)}}))}{1-F(\varepsilon_{(n-N)})}\left(\text{log}(1+\frac{z}{\varepsilon_{(n-N)}})\right)^3|_0^\infty \\
&+ \int_0^\infty \frac{L(\varepsilon_{(n-N)}(1+z/\varepsilon_{(n-N)}))}{L(\varepsilon_{(n-N)})}(1+z/\varepsilon_{(n-N)})^{1/k_0}3(\text{log}(1+z/\varepsilon_{(n-N)}))^2\\
&\times (1+z/\varepsilon_{(n-N)})^{-1}(1/\varepsilon_{(n-N)})dz=\tau_{1n}+\tau_{2n}.
\end{align*}
Three repeated applications of L'H\^opital's rule and Proposition 1.15 in \cite{Resnick1987} give $\tau_{1n}=0$.  For $\tau_{2n}$ we have that given FR 1 a) and again integrating by parts and letting $t=1+z/\varepsilon_{(n-N)}$
\begin{eqnarray*}
\tau_{2n}&=&\int_1^\infty3(log(t))^2t^{\frac{1}{k_0}-1}dt+\phi(\varepsilon_{(n-N)})\int_1^\infty3(log(t))^2t^{\frac{1}{k_0}-1}\frac{1}{\rho}(t^\rho-1)dt+o(\phi(\varepsilon_{(n-N)})).
\end{eqnarray*}
It is easy to verify that $\int_1^\infty3(log(t))^2t^{\frac{1}{k_0}-1}dt=-6k_0^3$ and consequently $\tau_{2n}=-6k_0^3+O(\phi(\varepsilon_{(n-N)}))$ which verifies $(ii)$.
By the Cramer-Wold theorem we have that $\psi_n \convd \mathcal{N}(0,V_1(k_0))$.  Consequently, for any vector $\gamma \in \mathds{R}^2$ we have $\gamma^T\left(v_N(1,k_0) - \sqrt{N} \left(\begin{array}{c}
b_\sigma  \\
b_k\\ 
\end{array}\right)\right) \convd \mathcal{N}(0,\gamma^TV_2 (k_0)\gamma)$ where 
$
V_2(k_0)=\left(\begin{array}{cc}
\frac{k_0^2-4k_0+2}{(2k_0-1)^2}&-\frac{1}{k_0(k_0-1)}  \\
-\frac{1}{k_0(k_0-1)}&\frac{2k_0^3-2k_0^2+2k_0-1}{k_0^2(k_0-1)^2(2k_0-1)}\\ 
\end{array}\right)
$.
Again, by the Cramer-Wold theorem $\left(v_N(1,k_0) - \sqrt{N} \left(\begin{array}{c}
b_\sigma  \\
b_k\\ 
\end{array}\right)\right) \convd \mathcal{N}(0,V_2(k_0) )$.  Hence, given equation (\ref{60}), provided that $H_N(r_N^*,k^*) \convp H(k_0)$ we have 
\begin{align*}
\sqrt{N} \left(\begin{array}{c}
\tilde{r}_N-1  \\
\tilde{k}-k_0\\ 
\end{array}\right)-H^{-1}(k_0)\sqrt{N}\left(\begin{array}{c}
b_\sigma  \\
b_k\\ 
\end{array}\right)&=H^{-1}(k_0)\left(v_N(1,k_0) - \sqrt{N} \left(\begin{array}{c}
b_\sigma  \\
b_k\\ 
\end{array}\right)\right)\\ &\convd N\left(0, H^{-1}(k_0)V_2(k_0)H^{-1}(k_0)\right).
\end{align*}
To see that $H_N(r_N^*,k^*) \convp H(k_0)$, first observe that whenever $(\tau_1,\tau_2)\in S_T$ we have $(\tilde{r}_N,\tilde{k})\in S_R=\{(\sigma, \,k): \| (\frac{\sigma}{\sigma_N}-1, \,k-k_0)\|_E<\delta_N\}$ and consequently  $(r_N^*,k^*)\in S_R$.  In addition, from Theorem \ref{thm1} and the results from \cite{Smith1987} we have $H_N(\tilde{r}_N,\tilde{k}) \convp H(k_0)$ uniformly on $S_R$.  By Theorem 21.6 in \cite{Davidson1994} we conclude that $H_N(r_N^*,k^*) \convp H(k_0)$.


\noindent b) Let $\tilde{r}_N^{(b)}=\frac{\tilde{\sigma}^{(b)}_{\tilde{q}(a_N)}}{\sigma_N}$ and note that using (\ref{60}) we can write
\begin{equation*}
\left(\begin{array}{cc}
\tilde{r}_N^{(b)}-1  \\
\tilde{k}^{(b)}-k_0\\ 
\end{array}\right)= H_N^{-1}(r^*_N,k^*)\left(\frac{1}{\sqrt{N}}v_N(1,k_0) - (M_n(N_s)-2 (\hat{k}(N_s))^2 ) \left( \begin{array}{c} \frac{1}{(1-\tilde{k}^{-1}-\hat{\rho})\hat{d}}\\ \frac{1}{\tilde{k}(-\tilde{k}^{-1}-\hat{\rho})(1-\tilde{k}^{-1}-\hat{\rho})\hat{d}}\end{array}\right) \right).
\end{equation*}
Since $\hat{\rho}-\rho=o_p(1)$, $\tilde{k}-k_0=o_p(1)$ and $\tilde{r}_N-1=o_p(1)$, we have
$$
\left( \begin{array}{c} \frac{1}{(1-\tilde{k}^{-1}-\hat{\rho})\hat{d}}\\ \frac{1}{\tilde{k}(-\tilde{k}^{-1}-\hat{\rho})(1-\tilde{k}^{-1}-\hat{\rho})\hat{d}}\end{array}\right) \convp  \left( \begin{array}{c} \frac{1}{(1-k_0^{-1}-\rho)d}\\ \frac{1}{k_0(-k_0^{-1}-\rho)(1-k_0^{-1}-\rho)d}\end{array}\right)
$$
where $d=\frac{2k_0^4\rho}{(1+ k_0 \rho )^2}$.  In addition, using the arguments in the proof of part a) we have
$$
\sqrt{N}(\hat{k}(N_s)-k_0)=k_0Q_n-\sqrt{N}P_{1n}+k_0\sqrt{N}\frac{\phi(q(a_N))}{(-k_0^{-1}-\rho)}+o_p(1) \mbox{ and }
$$
$$
\sqrt{N}(M_n(N_s) - 2k_0^2)=-2k_0Q_n+\sqrt{N}P_{2n}+2 \left( \frac{1}{(k_0^{-1}+\rho)^2}-k_0^2\right)\sqrt{N}\frac{\phi(q(a_N))}{\rho}+o_p(1),
$$
where $Q_n=\sumtn (n(1-F(y_n)))^{-1/2}(q_{1n}-E(q_{1n}))$, $P_{1n}=\frac{1}{N_1}\sumtn \left( \text{log} \left( \frac{\varepsilon_t}{q(a_N)} +k_0+k_0\frac{\phi(q(a_N))}{(-k_0^{-1}-\rho)}\right)  \right) \chi_{\{\varepsilon_t >q(a_N)\}}$ and $P_{2n}=\frac{1}{N_1}\sumtn \left( \text{log} ^2\left( \frac{\varepsilon_t}{q(a_N)} \right)-2k_0^2-2\frac{\phi(q(a_N))}{\rho}( \frac{1}{(k_0^{-1}+\rho)^2}-k_0^2) \right) \chi_{\{\varepsilon_t >q(a_N)\}}$.  As a consequence, we obtain
\begin{align}\label{32nova}
\sqrt{N}(M_n(N_s) - 2(\hat{k}(N_s))^2)&=-2(1+2k_0)k_0Q_n +4k_0\sqrt{N}P_{1n}+\sqrt{N}P_{2n}+ \frac{2k_0^4\rho}{(1+k_0\rho)^2}\sqrt{N} \phi(q(a_N))\notag\\
& +\sqrt{N} \phi(q(a_N))o_p(1)+o_p(1).
\end{align}
We rewrite (\ref{new29}) as 
$
\frac{1}{\sqrt{N}}v_N(1,k_0) =\left( \begin{array}{c} b_\sigma +b_1 \frac{k_0}{\sqrt{N}}Q_n + P_{3n}\\ b_k +b_2 \frac{k_0}{\sqrt{N}}Q_n + P_{4n} \end{array}\right) +o_p(N^{-1/2})
$
where $P_{3n}=\frac{1}{N}\left( \sum_{i=1}^N\frac{\partial}{\partial \sigma} \logg (Z_i';\sigma_N,k_0) \sigma_N-b_\sigma \right)$ and $P_{4n}=\frac{1}{N}\left( \sum_{i=1}^N\frac{\partial}{\partial k} \logg (Z_i';\sigma_N,k_0) -b_k \right)$.  Since, $b_\sigma=\frac{\phi(\varepsilon_{(n-N)})}{1-\frac{1}{k_0}-\rho}+o(\phi(\varepsilon_{(n-N)}))$ and $b_k=\frac{ \phi(\varepsilon_{(n-N)})}{k_0(-\frac{1}{k_0}-\rho)(1-\frac{1}{k_0}-\rho)}+o(\phi(\varepsilon_{(n-N)}))$ we have,
\begin{align}
&\sqrt{N}\left(\begin{array}{cc}
\tilde{r}_N^{(b)}-1  \\
\tilde{k}^{(b)}-k_0\\ 
\end{array}\right)= H_N^{-1}(r^*_N,k^*)\notag\\
&\times\sqrt{N} \left( \begin{array}{c} -\frac{2(1+\rho k_0)^2}{k_0^2\rho((1-\rho)k_0-1)} P_{1n} -\frac{(1+\rho k_0)^2}{2k_0^3\rho((1-\rho)k_0-1)}P_{2n}+P_{3n}+\left(  \frac{1-k_0}{k_0(1-2k_0)}+\frac{(1+2k_0)(1+\rho k_0)^2}{k_0^3\rho((1-\rho)k_0-1)} \right) \frac{k_0}{\sqrt{N}}Q_n \\ \frac{2(1+\rho k_0)}{k_0^2\rho((1-\rho)k_0-1)}P_{1n} +\frac{(1+\rho k_0)}{2k_0^3\rho((1-\rho)k_0-1)}P_{2n}+ P_{4n}-\left(\frac{1}{(1-k_0)(1-2k_0)}+\frac{(1+2k_0)(1+\rho k_0)}{k_0^3\rho((1-\rho)k_0-1)} \right)\frac{k_0}{\sqrt{N}}Q_n\end{array}\right).\label{trintae3}
\end{align}
Note that $\sqrt{N}P_{3n}=\sumtn Z_{t1}$, $\sqrt{N}P_{4n}=\sumtn Z_{t2}$, $k_0Q_n=\sumtn Z_{t3}$ from part a).  We put $\sqrt{N}P_{2n}=\sumtn Z_{t4}$, $\sqrt{N}P_{1n}=\sumtn Z_{t5}$ and observe that,
\begin{align*}
\sqrt{N}P_{1n}&=(1+o_p(1))\frac{1}{\sqrt{N}}\sumtn \left( \text{log}    \left(\frac{\varepsilon_t}{q(a_N)}\right)+k_0+k_0\frac{\phi(q(a_N))}{(-k_0^{-1}-\rho)}\right)\chi_{\{\varepsilon_t>q(a_N)\}}\\
&=\frac{1}{\sqrt{N}}\sumtn \left( \text{log}    \left(\frac{\varepsilon_t}{q(a_N)}\right)  - E \left( \text{log}    \left(\frac{\varepsilon_t}{q(a_N)}\right)\right)   \right)\chi_{\{\varepsilon_t>q(a_N)\}}+o_p(1)\\
\sqrt{N}P_{2n}&=(1+o_p(1))\frac{1}{\sqrt{N}}\sumtn \left( \text{log}^2 \left(\frac{\varepsilon_t}{q(a_N)}\right)-2k_0^2-2\frac{\phi(q(a_N))}{\rho} \left( \frac{1}{(k_0^{-1}+\rho)^2} -k_0^2 \right) \right)\chi_{\{\varepsilon_t>q(a_N)\}}\\
&= \frac{1}{\sqrt{N}}\sumtn \left( \text{log}^2 \left(\frac{\varepsilon_t}{q(a_N)}\right)-E\left( \text{log}^2 \left(\frac{\varepsilon_t}{q(a_N)}\right) \right) \right)\chi_{\{\varepsilon_t>q(a_N)\}}+o_p(1)
\end{align*}
where the second equalities in both expressions follow from the fact that $\sqrt{N}o(\phi(q(a_N)))=o(1)$, and thus $E(Z_{t4})=E(Z_{t5})=0$.  Using arguments similar to those in part a) of the proof we obtain, $V(Z_{t4})=\frac{1}{n}(20k_0^4+o(1))$, $V(Z_{t5})=\frac{1}{n}(k_0^2+o(1))$, $E(Z_{t4}Z_{t5})=\frac{1}{n}(-4k_0^3+o(1))$, $E(Z_{t1}Z_{t4})=\frac{1}{n}(\frac{4k_0^2-2k_0^3}{(1-k_0)^2}+o(1))$, $E(Z_{t1}Z_{t5})=\frac{1}{n}(-\frac{k_0}{1-k_0}+o(1))$, $E(Z_{t2}Z_{t4})=\frac{1}{n}(\frac{4k_0^3-6k_0^2}{(1-k_0)^2}+o(1))$, $E(Z_{t2}Z_{t5})=\frac{1}{n}(\frac{k_0}{1-k_0}+o(1))$, $E(Z_{t3}Z_{t4})=E(Z_{t3}Z_{t5})=\frac{1}{n}o(1)$ and $E(Z_{t4}Z_{t5})=\frac{1}{n}(-4k_0^3+o(1))$.  

Now, integrating by parts $E\left( \text{log}^6\left(\frac{\varepsilon_t}{q(a_N)} \right)\right)=720k_0^6+O(\phi(q(a_N)))< \infty$.  Consequently, $\sumtn E|Z_{t4}|^3 \leq \frac{C}{\sqrt{N}}E\left( \left| \text{log}^2\left( \frac{\varepsilon_t}{q(a_N)}\right) \right|^3\right)=o(1)$ and $\sumtn E|Z_{t5}|^3 \leq \frac{C}{\sqrt{N}}E\left( \left| \text{log}\left( \frac{\varepsilon_t}{q(a_N)}\right) \right|^3\right)=o(1)$.  Consequently, by Liapounov's CLT and the Cramer-Wold device we have
$$
\sqrt{N}
\left( \begin{array}{c}
	\frac{\tilde{\sigma}^{(b)}_{\tilde{q}(a_N)}}{\sigma_{N}}-1\\
	\tilde{k}^{(b)}-k_0
\end{array}\right) \stackrel{d}{\rightarrow}\mathcal{N}\left(\left( \begin{array}{c}
	0\\
	0
\end{array}\right) ,H^{-1}(k_0)V_2^{(b)}(k_0,\rho)H^{-1}(k_0)\right)
$$
since $H_N^{-1}(r^*_N,k^*) \convp H^{-1}(k_0)$.
\end{proof}


\begin{proof}[\bf Theorem 3. \rm \it Proof.\rm]  a) Let $a \in (0,1)$, $a_N<a$ and write $q(a)=q(a_N)Z_{N,a}$.  By assumption $\frac{1-a}{1-a_N}=C$, where $C$ is an arbitrary constant satisfying $0<C<1$, which we set at $C=\mathcal{Z}^{1/k_0}$ for $\mathcal{Z}>0$.  Then, if $u(x)=q(1-x^{-1})$ for $x>1$, by FR2
$
\underset{n \rightarrow \infty}{\text{lim}}\frac{q(a)}{q(a_N)}=\underset{n \rightarrow \infty}{\text{lim}}\frac{u(1/(1-a))}{u(1/(1-a_N))}=\underset{n \rightarrow \infty}{\text{lim}}\frac{u\left(\mathcal{Z}^{-1/k_0}\frac{1}{1-a_N}\right)}{u\left(\frac{1}{1-a_N}\right)}=\mathcal{Z}
$.  Consequently, $Z_{N,a} \rightarrow \mathcal{Z}$ as $n \rightarrow \infty$.  Now, we write
\begin{align*}
\frac{\hat{q}(a)}{q(a)}-1
&=\frac{1}{Z_{N,a}}\left(1+\frac{\tilde{q}(a_N)}{q(a_N)}-1 \right) \left( 1+  \left( \frac{\tilde{\sigma}_{\tilde{q}(a_N)}}{\tilde{k}\tilde{q}(a_N)} - \frac{\sigma_N}{k_0 q(a_N)} +\frac{\sigma_N}{k_0 q(a_N)}  \right)\left(1-\left(\frac{n(1-a)}{N}\right)^{\tilde{k}} \right)   \right)-1.
\end{align*}
By Theorem \ref{thm2} and the fact that $\frac{n(1-a)}{N}=\mathcal{Z}^{1/k_0}$ we have
\begin{equation}\label{vinte9n}
\left(\frac{n(1-a)}{N}\right)^{\tilde{k}}=\left(\frac{n(1-a)}{N}\right)^{k_0}\left(1+(\tilde{k}-k_0)\text{log} \frac{n(1-a)}{N}+o_p(N^{-1/2})\right).
\end{equation}
Let $h(\sigma,k,q)=\text{log}(1-\frac{\sigma}{kq})$, and since $\sigma_N=-k_0 q(a_N)$, we have 
$
\frac{\tilde{\sigma}_{\tilde{q}(a_N)}}{\tilde{k}\tilde{q}(a_N)} - \frac{\sigma_N}{k_0 q(a_N)} =-2 \left( exp \left( h(\tilde{\sigma}_{\tilde{q}(a_N)},\tilde{k},\tilde{q}(a_N)) -h(\sigma_N,k_0,q(a_N)) \right) -1\right)$.  By the MVT, there exists $(\sigma^*,k^*,q^*)$ such that 
\begin{align*}
h(\tilde{\sigma}_{\tilde{q}(a_N)},\tilde{k},\tilde{q}(a_N))-h(\sigma_N,k_0,q(a_N))&=
\left(
\begin{array}{ccc}
\sigma_N D_1 h(\sigma^*,k^*,q^*)  & D_2 h(\sigma^*,k^*,q^*)   & q(a_N) D_3 h(\sigma^*,k^*,q^*)   \\ 
\end{array}
\right)\\
&\times \left(
\begin{array}{c}
\tilde{r}_N-1  \\
\tilde{k}-k_0   \\
 \frac{\tilde{q}(a_N)}{q(a_N)}-1   
\end{array}
\right).
\end{align*}
Since $\sigma_N D_1 h(\sigma^*,k^*,q^*) =\left( 1+\frac{\sigma^*/\sigma_N}{k^*q^*/k_0q(a_N)}\right)^{-1}\frac{q(a_N)k_0}{k^*q^*} \convp \frac{1}{2}$, $D_2 h(\sigma^*,k^*,q^*) \convp -\frac{1}{2k_0}$ and $q(a_N) D_3 h(\sigma^*,k^*,q^*) \convp -1/2$ by the MVT, Theorem \ref{thm2} and Lemma 5, we have  
\begin{align}\label{trintan}
\frac{\tilde{\sigma}_{\tilde{q}(a_N)}}{\tilde{k}\tilde{q}(a_N)} - \frac{\sigma_N}{k_0 q(a_N)} &=\left(
\begin{array}{ccc}
-1  &1/k_0   &1   \\ 
\end{array}
\right)\left(
\begin{array}{c}
\tilde{r}_N-1  \\
\tilde{k}-k_0   \\
\frac{\tilde{q}(a_N)}{q(a_N)}-1   
\end{array}
\right)+o_p(N^{-1/2}).
\end{align}
Letting $h_1(k,c)=k\,\text{log}(1+c)$ for $c \geq 0$, we have by FR2, Theorem \ref{thm2}, the MVT and the fact that $Z_{N,a} \rightarrow \mathcal{Z}$,
\begin{align}\label{trinta1n}
\frac{1}{Z_{N,a}} \left( \frac{n(1-a)}{N} \right)^{k_0}
&=1+k_0k(\mathcal{Z})\phi(q(a_N))+o(\phi(q(a_N)))+o_p(N^{-1/2}).
\end{align}
Using equations (\ref{vinte9n}), (\ref{trintan}) and (\ref{trinta1n}) we write,
\begin{align*}
\frac{\hat{q}(a)}{q(a)}-1
&=\mathcal{Z}^{-1} \left( \frac{\tilde{q}(a_N)}{q(a_N)}-1 \right) + k_0c_b^T \left(
\begin{array}{c}
\tilde{r}_N-1  \\
\tilde{k}-k_0     
\end{array}
\right)+k_0k(\mathcal{Z})\phi(q(a_N))+o(\phi(q(a_N))) +o_p(N^{-1/2}),
\end{align*}
where $c_b^T=\left(\begin{array}{cc}
-k_0^{-1}(\mathcal{Z}^{-1}-1)&  k_0^{-2}log(\mathcal{Z})+k_0^{-2}(\mathcal{Z}^{-1}-1)\\
\end{array}\right)$.
From the proof of Lemma 5 we have that
$$
\mathcal{Z}^{-1} \sqrt{N}\left( \frac{\tilde{q}(a_N)}{q(a_N)}-1 \right)=\mathcal{Z}^{-1}k_0\sum_{t=1}^n \frac{1}{\sqrt{n(1-F(y_n))}}(q_{1n}-E(q_{1n}))+o_p(1)=\mathcal{Z}^{-1}\sum_{t=1}^nZ_{t3}+o_p(1).
$$
In addition, from the proof of Theorem \ref{thm2} (adopting its notation) we have that
\begin{align*}
\sqrt{N}\left(
\begin{array}{c}
\tilde{r}_N-1  \\
\tilde{k}-k_0     
\end{array}
\right)-H^{-1}(k_0)\sqrt{N}\left(
\begin{array}{c}
b_\sigma  \\
b_k    
\end{array}
\right)
&=H^{-1}(k_0)\left( \begin{array}{ccc} 1& 0& b_1\\0&1&b_2 \end{array}\right)\psi_n+o_p(1)
\end{align*}
where $\psi_n=\sumtn ( Z_{t1}, Z_{t2},Z_{t3})^T$. 
Hence, letting 
$\eta^T=\left(
\begin{array}{cc}
c_b^T H^{-1}(k_0)&  c_b^T H^{-1}(k_0) \left( \begin{array}{c}b_1\\b_2\end{array}\right)+(\mathcal{Z}k_0)^{-1}
\end{array}
\right)
$
we can write 
$
\sqrt{N}\left(\frac{\hat{q}(a)}{q(a)}-1 - k_0k(\mathcal{Z})\phi(q(a_N))-k_0c_b^TH^{-1}(k_0)\left(
\begin{array}{c}
b_\sigma  \\
b_k    
\end{array}
\right) \right)
=k_0\eta^T  \psi_n +o_p(1),
$
where $k_0\eta^T\psi_n  \convd \mathcal{N}(0,k_0^2\eta^TV_1(k_0)\eta)$.  Since $k_0k(\mathcal{Z})\sqrt{N}\phi(q(a_N)) \rightarrow  k_0 k(\mathcal{Z})\mu (-k_0^{-1}-\rho)$ and $k(\mathcal{Z})=\frac{\mathcal{Z}^\rho-1}{\rho}$ we have that  $k_0k(\mathcal{Z})\sqrt{N}\phi(q(a_N)) \rightarrow  k_0\mu (-k_0^{-1}-\rho)\frac{\mathcal{Z}^\rho-1}{\rho}$.  Hence, given the structure of $V_1(k_0)$, we conclude 
$
\sqrt{N}\left(\frac{\hat{q}(a)}{q(a)}-1 \right)-\mu_1 \convd\mathcal{N}(0,\Sigma_1(k_0))
$, where $\Sigma_1(k_0)=k_0^2\left( c_b^TH^{-1}(k_0)c_b+k_0^2\left(c_b^TH^{-1}(k_0) \left( \begin{array}{c}b_1\\b_2\end{array}\right)  \right)^2+2k_0\mathcal{Z}^{-1}c_b^TH^{-1}(k_0) \left( \begin{array}{c}b_1\\b_2\end{array}\right)+ \mathcal{Z}^{-2}\right)$ and $
\mu_1=k_0 \left(\mu (-k_0^{-1}-\rho)\frac{\mathcal{Z}^\rho-1}{\rho} +c_b^TH^{-1}(k_0) \underset{\ntoi}{\text{lim}}\sqrt{N} \left(
\begin{array}{c}
b_\sigma  \\
b_k    
\end{array}
\right) \right)
$. 

\noindent b) Since $\hat{\mathcal{Z}}=\frac{\hat{q}(a)}{\tilde{q}(a_N)} \convp \mathcal{Z}$, $\hat{\rho}-\rho=o_p(1)$ and $\hat{d}-d=o_p(1)$, we have $\frac{\hat{\mathcal{Z}}^{\hat{\rho}}-1}{\hat{\rho}\hat{d}}\convp \frac{\mathcal{Z}^{\rho}-1}{\rho d}$.  From equation (\ref{32nova}) we can write
$
\hat{B}_q=k(\mathcal{Z})\phi(q(a_N))+\frac{\mathcal{Z}^\rho-1}{\rho d}\left( (-2-4k_0)k_0\frac{Q_n}{\sqrt{N}}+P_{2n}+4k_0P_{1n}\right)+o_p(N^{-1/2}).
$
Using the MVT as in part a) we have
\begin{equation}\label{trintaesete}
\frac{(1+\hat{B}_q)^{-k_0}}{(1+k(\mathcal{Z})\phi(q(a_N))+o(\phi(q(a_N))))^{-k_0}}=1-k_0\left(\frac{\mathcal{Z}^\rho-1}{\rho d}\right)\left( (-2-4k_0)k_0\frac{Q_n}{\sqrt{N}}+P_{2n}+4k_0P_{1n}\right)+o_p(N^{-1/2}).
\end{equation}
Also, from part a), substituting $\tilde{k}$ by $\tilde{k}^{(b)}$ in equation (\ref{vinte9n}) we have 
\begin{equation}\label{37meio}
\left(\frac{n(1-a)}{N}\right)^{\tilde{k}^{(b)}}=\left(\frac{n(1-a)}{N}\right)^{k_0}\left(1+(\tilde{k}^{(b)}-k_0)\text{log} \frac{n(1-a)}{N}+o_p(N^{-1/2})\right).
\end{equation}
Now, since $\hat{B}_q=O_p(N^{-1/2})$, by the MVT
\begin{equation}\label{trintaeoito}
\frac{(1+\hat{B}_q)^{-\tilde{k}^{(b)}}}{(1+\hat{B}_q)^{-k_0}}=1-(\tilde{k}^{(b)}-k_0)\text{log}(1+\hat{B}_q)+o_p(N^{-1/2}).
\end{equation}
By equations (\ref{trintaesete}) and (\ref{trintaeoito}), and since $\text{log}(1+\hat{B}_q)=O_p(N^{-1/2})$ we have 
\begin{equation}\label{quarenta}
\frac{(1+\hat{B}_q)^{-\tilde{k}^{(b)}}}{(1+k(\mathcal{Z})\phi(q(a_N))+o(\phi(q(a_N))))^{-k_0}}=1-k_0\left(\frac{\mathcal{Z}^\rho-1}{\rho d}\right)\left( (-2-4k_0)k_0\frac{Q_n}{\sqrt{N}}+P_{2n}+4k_0P_{1n}\right)+o_p(N^{-1/2})
\end{equation}
and consequently we have $(1+\hat{B}_q)^{-\tilde{k}^{(b)}}\convp 1$.  Now, we write
\begin{align*}
\frac{\hat{q}^{(b)}(a)}{q(a)}-1
&=\frac{1}{Z_{N,a}}\left(1+\frac{\tilde{q}(a_N)}{q(a_N)}-1 \right) \left( 1+  \left( \frac{\tilde{\sigma}_{\tilde{q}(a_N)}}{\tilde{k}\tilde{q}(a_N)} - \frac{\sigma_N}{k_0 q(a_N)} +\frac{\sigma_N}{k_0 q(a_N)}  \right)\left(1-\left(\frac{n(1-a)}{N}\right)^{\tilde{k}^{(b)}} \right.\right.\\
&\times \left.\left.(1+\hat{B}_q)^{-\tilde{k}^{(b)}} \right)   \right)-1.
\end{align*}
Using equations \eqref{trintan}, \eqref{37meio} and \eqref{quarenta} we have 
\begin{align*}
\sqrt{N} \left(\frac{\hat{q}^{(b)}(a)}{q(a)}-1\right)&=\left( \frac{N}{n(1-a)}\right)^{k_0} \sqrt{N}\left( \frac{\tilde{q}(a_N)}{q(a_N)}-1 \right)+k_0c_b^T \sqrt{N}\left(
\begin{array}{c}
\tilde{r}^{(b)}_N-1  \\
\tilde{k}^{(b)}-k_0     
\end{array}
\right)\\
&-k_0\left(\frac{\mathcal{Z}^\rho-1}{\rho d}\right)\left( (-2-4k_0)k_0Q_n+\sqrt{N}P_{2n}+4k_0\sqrt{N}P_{1n}\right)+o_p(1).
\end{align*}
Given equation \eqref{trintae3}  and the fact that $\sqrt{N}\left( \frac{\tilde{q}(a_N)}{q(a_N)}-1 \right)=k_0Q_n+o_p(1)$ we have, using the notation in Theorem \ref{thm2}, that for $\mathcal{q}_n^T=\left( \begin{array}{ccccc} \sumtn Z_{t1}& \sumtn Z_{t2}&\sumtn Z_{t3}&\sumtn Z_{t4}&\sumtn Z_{t5}\end{array}\right)$, 
$
\sqrt{N} \left(\frac{\hat{q}^{(b)}(a)}{q(a)}-1\right)=c_q^T\mathcal{q}_n+o_p(1)
$
where $c_q^T=k_0c_b^TH^{-1}(k_0)A(k_0,\rho)+v(k_0,\rho)$ and 
$$
v(k_0,\rho)=\left( \begin{array}{ccccc} 0 & 0 & \mathcal{Z}^{-1}+(\mathcal{Z}^{\rho}-1)\frac{(1+2k_0)(1+\rho k_0)^2}{k_0^3 \rho^2}& -(\mathcal{Z}^{\rho}-1)\frac{(1+\rho k_0)^2}{2k_0^3 \rho^2}& -2(\mathcal{Z}^{\rho}-1)\frac{(1+\rho k_0)^2}{k_0^2 \rho^2} \end{array}\right).
$$
Since from part b) of Theorem \ref{thm2}, $\mathcal{q}_n \convd \mathcal{N}(0,V^{(b)}(k_0))$ the proof is complete.
\end{proof}

\begin{proof}[\bf Theorem 4. \rm \it Proof.\rm] a) We write
$
\frac{\hat{E}(\varepsilon_t|\varepsilon_t>q(a))}{q(a)/(1+k_0)}-1
=\left(\frac{\hat{q}(a)}{q(a)}-1\right)\left(\frac{k_0-\tilde{k}}{1+\tilde{k}} \right) + \frac{\hat{q}(a)}{q(a)}-1 +\frac{k_0-\tilde{k}}{1+\tilde{k}}.
$
From part a) of Theorems \ref{thm2} and \ref{thm3} we have $\frac{\tilde{k}-k_0}{1+\tilde{k}}=O_p(N^{-1/2})$ and $\frac{\hat{q}(a)}{q(a)}-1=O_p(N^{-1/2})$.  Hence,
$
\sqrt{N}\left( \frac{\hat{q}(a)/(1+\tilde{k})}{q(a)/(1+k_0)}-1\right)=\left(\begin{array}{cc}
1  &-(1+k_0)^{-1}\\ 
\end{array}\right)
\left(\begin{array}{c}
\sqrt{N}\left( \frac{\hat{q}(a)}{q(a)}-1\right)  \\
\sqrt{N}\left( \tilde{k}-k_0\right) 
\end{array}\right)+o_p(1).
$
From part a) of Theorem \ref{thm3} we have
$
\sqrt{N}\left(\frac{\hat{q}(a)}{q(a)}-1 - k_0k(\mathcal{Z})\phi(q(a_N))-k_0c_b^TH^{-1}(k_0)\left(
\begin{array}{c}
b_\sigma  \\
b_k    
\end{array}
\right) \right)
=k_0\eta^T  \psi_n +o_p(1),
$
and from part a) of Theorem \ref{thm2},
$
\sqrt{N}(\tilde{k}-k_0)-\sqrt{N}\left(\begin{array}{cc}
0 & 1 \\
\end{array}\right)H^{-1}(k_0)\left(\begin{array}{c}
b_\sigma  \\
b_k\\ 
\end{array}\right)=\left(\begin{array}{cc}
\left(\begin{array}{cc}
0& 1 \\
\end{array}\right)H^{-1}(k_0) & \left(\begin{array}{cc}
0& 1 \\
\end{array}\right)H^{-1}(k_0) \left(\begin{array}{c}
b_1\\
b_2\\ 
\end{array}\right)\\
\end{array}\right)\psi_n
+o_p(1),
$
where $\psi_n=\left(\begin{array}{ccc}
\sumtn Z_{t1}&\sumtn Z_{t2}&\sumtn Z_{t3} \end{array}\right)^T$.  Hence, 
\begin{align*}
&\left(\begin{array}{c}
\sqrt{N}\left( \frac{\hat{q}(a)}{q(a)}-1- k_0k(\mathcal{Z})\phi(q(a_N))-k_0c_b^TH^{-1}(k_0)\left(
\begin{array}{c}
b_\sigma  \\
b_k    
\end{array}
\right)\right)  \\
\sqrt{N}\left( \tilde{k}-k_0-\left(\begin{array}{cc}
0 & 1 \\
\end{array}\right)H^{-1}(k_0)\left(\begin{array}{c}
b_\sigma  \\
b_k\\ 
\end{array}\right)\right) 
\end{array}\right)= \left( \begin{array}{c}  k_0\eta^T\\\theta_T\end{array}\right) \psi_n+o_p(1),
\end{align*}
for $\theta^T = \left(\begin{array}{cc}
 \left(\begin{array}{cc}
0 & 1  
\end{array}\right) H^{-1}(k_0) &  \left(\begin{array}{cc}
0 & 1  
\end{array}\right) H^{-1}(k_0) \left(\begin{array}{c}
b_1  \\
b_2\\ 
\end{array}\right) 
\end{array}\right)$.  From Theorem \ref{thm3} part a) $\psi_n \convd \mathcal{N}(0,V_1(k_0))$, hence we conclude that  
\begin{align*}
\sqrt{N}\left( \frac{\hat{q}(a)/(1+\tilde{k})}{q(a)/(1+k_0)}-1\right)&-
\left(
\begin{array}{cc}
1 & -(1+k_0)^{-1}
\end{array}
\right)
\left(
\begin{array}{c}
\sqrt{N} \left(k_0k(\mathcal{Z})\phi(q(a_N))+k_0c_b^TH^{-1}(k_0)\left(
\begin{array}{c}
b_\sigma  \\
b_k    
\end{array}
\right) \right) \\
\sqrt{N}\left(\begin{array}{cc}
0 & 1 \\
\end{array}\right)H^{-1}(k_0)\left(\begin{array}{c}
b_\sigma  \\
b_k\\ 
\end{array}\right)
\end{array}\right)\\
&\convd N\left(0, \left(
\begin{array}{cc}
1 & -(1+k_0)^{-1}
\end{array}
\right) \left( \begin{array}{c}  k_0\eta^T\\\theta^T\end{array}\right)V_1(k_0)\left( \begin{array}{c}  k_0\eta^T\\\theta^T\end{array}\right)^T\left(
\begin{array}{cc}
1 & -(1+k_0)^{-1}
\end{array}
\right)^T\right).
\end{align*}
Additional algebra, gives
\begin{eqnarray*}
\sqrt{N}\left(\frac{\hat{q}(a)/(1+\tilde{k})}{q(a)/(1+k_0)}-1\right)&\convd& N\left( k_0 \frac{(\mathcal{Z}^\rho-1)\mu(-k_0^{-1}-\rho)}{\rho} + k_0c_b^TH^{-1}(k_0) \underset{n \rightarrow \infty}{\mbox{lim}}\sqrt{N} \left(\begin{array}{c}
b_\sigma  \\
b_k\\ 
\end{array}\right) \right.\\
&-& \left. \frac{1}{1+k_0}\left(\begin{array}{cc}
0 & 1 \\
\end{array}\right)H^{-1}(k_0)\underset{n \rightarrow \infty}{\mbox{lim}}\sqrt{N}\left(\begin{array}{c}
b_\sigma  \\
b_k\\ 
\end{array}\right), \Sigma_2 (k_0)\right),
\end{eqnarray*}
where 
$
\Sigma_2(k_0)=\left(\begin{array}{c}
k_0 \eta^T-\frac{1}{1+k_0}\theta^T 
\end{array}\right)V_1(k_0)\left(\begin{array}{c}
k_0 \eta -\frac{1}{1+k_0}\theta 
\end{array}\right)$.


\noindent b) As in part a) we write
$
\frac{\hat{E}^{(b)}(\varepsilon_t|\varepsilon_t>q(a))}{q(a)/(1+k_0)}-1=\left(\frac{\hat{q}^{(b)}(a)}{q(a)}-1\right)\left(\frac{k_0-\tilde{k}^{(b)}}{1+\tilde{k}^{(b)}} \right) + \frac{\hat{q}^{(b)}(a)}{q(a)}-1 +\frac{k_0-\tilde{k}^{(b)}}{1+\tilde{k}}.
$
From part b) of Theorems \ref{thm2} and \ref{thm3} we have $\frac{\tilde{k}^{(b)}-k_0}{1+\tilde{k}^{(b)}}=O_p(N^{-1/2})$ and $\frac{\hat{q}^{(b)}(a)}{q(a)}-1=O_p(N^{-1/2})$.  Hence,
\begin{eqnarray*}
\sqrt{N}\left( \frac{\hat{q}^{(b)}(a)/(1+\tilde{k}^{(b)})}{q(a)/(1+k_0)}-1\right)
&=&\left(\begin{array}{cc}
1  &-(1+k_0)^{-1}\\ 
\end{array}\right)
\left(\begin{array}{c}
\sqrt{N}\left( \frac{\hat{q}^{(b)}(a)}{q(a)}-1\right)  \\
\sqrt{N}\left( \tilde{k}^{(b)}-k_0\right) 
\end{array}\right)+o_p(1).
\end{eqnarray*} 
From parts b) of Theorems \ref{thm2} and \ref{thm3} we have 
$
\sqrt{N}\left( \frac{\hat{q}^{(b)}(a)}{q(a)}-1\right)=c_q^T\mathcal{q}_n+o_p(1)
$
and
$
\sqrt{N}\left( \tilde{k}^{(b)}-k_0\right)=\left(\begin{array}{cc}0&1\end{array}\right)H^{-1}(k_0)A(k_0,\rho)\mathcal{q}_n+o_p(1)
$.  Hence, since $\mathcal{q}_n \convd \mathcal{N}(0,V^{(b)}(k_0))$, we conclude that,
$
\sqrt{N}\left(\frac{\hat{E}^{(b)}(\varepsilon_t|\varepsilon_t>q(a))}{q(a)/(1+k_0)}-1\right)\convd\mathcal{N}\left(0,\Sigma_2^{(b)}(k_0,\rho)\right)$, where
$
\Sigma_2^{(b)}(k_0,\rho)= \left( c_q^T-\frac{1}{1+k_0} \left(\begin{array}{cc}0&1\end{array}\right)H^{-1}(k_0)A(k_0,\rho)   \right)V^{(b)}(k_0)\left(c_q^T-\frac{1}{1+k_0} \left(\begin{array}{cc}0&1\end{array}\right)H^{-1}(k_0)A(k_0,\rho) \right)^T.
$
\end{proof}

\begin{proof}[\bf Theorem 5. \rm \it Proof.\rm] a) Since $\hat{q}^{(b)}_{Y_t|\mathbf{X}_t=\mathbf{x}}(a)=\hat{m}(\mathbf{x})+h^{1/2}(\mathbf{x})\hat{q}^{(b)}(a)$, we write
\begin{align*}
\frac{\hat{q}^{(b)}_{Y_t|\mathbf{X}_t=\mathbf{x}}(a)}{q_{Y_t|\mathbf{X}_t=\mathbf{x}}(a)}-1&=\frac{\hat{m}(\mathbf{x})-m(\mathbf{x})}{m(\mathbf{x})+h^{1/2}(\mathbf{x})q(a)}+\frac{\left( \hat{h}^{1/2}(\mathbf{x})-h^{1/2}(\mathbf{x}) \right)}{\left(\frac{m(\mathbf{x})}{q(a)}+h^{1/2}(\mathbf{x})\right)}\frac{\hat{q}^{(b)}(a)}{q(a)}+\frac{h^{1/2}(\mathbf{x})}{\left(\frac{m(\mathbf{x})}{q(a)}+h^{1/2}(\mathbf{x})\right)}\left(\frac{\hat{q}^{(b)}(a)-q(a)}{q(a)}\right).
\end{align*}
From Lemma 3, the fact that $q(a) \rightarrow \infty$ as $n \rightarrow \infty$ and assumption A3 2), we have $\frac{\hat{m}(\mathbf{x})-m(\mathbf{x})}{m(\mathbf{x})+h^{1/2}(\mathbf{x})q(a)}=o_p(L_{1n})$.  Given A5 and $n(1-a) \propto N$ we have $\sqrt{n(1-a)}\frac{\hat{m}(\mathbf{x})-m(\mathbf{x})}{m(\mathbf{x})+h^{1/2}(\mathbf{x})q(a)}=o_p(1)$.  By Corollary 1 (in the online supplement), A5 and the fact that $m(\mathbf{x})$ is bounded for fixed $\mathbf{x}$ we have $\sqrt{n(1-a)}\frac{\left( \hat{h}^{1/2}(\mathbf{x})-h^{1/2}(\mathbf{x}) \right)}{\left(\frac{m(\mathbf{x})}{q(a)}+h^{1/2}(\mathbf{x})\right)}=o_p(1)$.  From part b) of Theorem \ref{thm3} we have $\frac{\hat{q}^{(b)}(a)}{q(a)}=1+o_p(1)$, which gives $\sqrt{n(1-a)}\frac{\left( \hat{h}^{1/2}(\mathbf{x})-h^{1/2}(\mathbf{x}) \right)}{\left(\frac{m(\mathbf{x})}{q(a)}+h^{1/2}(\mathbf{x})\right)}\frac{\hat{q}^{(b)}(a)}{q(a)}=o_p(1)$.  Lastly, since $q(a) \rightarrow \infty$ as $n\rightarrow \infty$, for fixed $\mathbf{x}$ we have $\frac{h^{1/2}(\mathbf{x})}{\left(\frac{m(\mathbf{x})}{q(a)}+h^{1/2}(\mathbf{x})\right)}\rightarrow 1$ and by part a) of Theorem \ref{thm3}  $\sqrt{n(1-a)}\left(\frac{\hat{q}^{(b)}(a)-q(a)}{q(a)}\right) \convd \mathcal{N}(0,\Sigma_1^{(b)}(k_0,\rho))$.

\noindent b) We write
\begin{align*}
&\frac{\hat{E}^{(b)}\left(Y_t|Y_t>q_{Y_t|\mathbf{X}_t=\mathbf{x}}(a),\mathbf{X}_t=\mathbf{x})\right)}{E\left(Y_t|Y_t>q_{Y_t|\mathbf{X}_t=\mathbf{x}}(a),\mathbf{X}_t=\mathbf{x})\right)}-1=\frac{\hat{m}(\mathbf{x})-m(\mathbf{x})}{m(\mathbf{x})+h^{1/2}(\mathbf{x})E(\varepsilon_t|\varepsilon_t>q(a))}+\frac{\hat{h}^{1/2}(\mathbf{x})-h^{1/2}(\mathbf{x})}{\left(\frac{m(\mathbf{x})}{E(\varepsilon_t|\varepsilon_t>q(a))}+h^{1/2}(\mathbf{x}) \right)}
\end{align*}
\begin{align*}
&\times \frac{(\hat{E}^{(b)}(\varepsilon_t|\varepsilon_t>q(a))+\hat{B}_E)-E(\varepsilon_t|\varepsilon_t>q(a))}{E(\varepsilon_t|\varepsilon_t>q(a))}+ \frac{h^{1/2}(\mathbf{x})}{\left(\frac{m(\mathbf{x})}{E(\varepsilon_t|\varepsilon_t>q(a))}+h^{1/2}(\mathbf{x})\right)}\\
&\times \left(\frac{\hat{E}^{(b)}(\varepsilon_t|\varepsilon_t>q(a))+\hat{B}_E-E(\varepsilon_t|\varepsilon_t>q(a))}{E(\varepsilon_t|\varepsilon_t>q(a))}\right)
+\frac{E(\varepsilon_t|\varepsilon_t>q(a))(\hat{h}^{1/2}(\mathbf{x})-h^{1/2}(\mathbf{x}))}{m(\mathbf{x})+h^{1/2}(\mathbf{x})E(\varepsilon_t|\varepsilon_t>q(a))}
\end{align*}
As in part a), since $m(\mathbf{x})+h^{1/2}(\mathbf{x})E(\varepsilon_t|\varepsilon_t>q(a)) \rightarrow \infty$ as $n \rightarrow \infty$, given Lemma 3 and A5 and $n(1-a) \propto N$, $\frac{\hat{m}(\mathbf{x})-m(\mathbf{x})}{m(\mathbf{x})+h^{1/2}(\mathbf{x})E(\varepsilon_t|\varepsilon_t>q(a))}=o_p(N^{-1/2})$. Similarly, $\frac{E(\varepsilon_t|\varepsilon_t>q(a))(\hat{h}^{1/2}(\mathbf{x})-h^{1/2}(\mathbf{x}))}{m(\mathbf{x})+h^{1/2}(\mathbf{x})E(\varepsilon_t|\varepsilon_t>q(a))}=o_p(N^{-1/2})$ and $\frac{\hat{h}^{1/2}(\mathbf{x})-h^{1/2}(\mathbf{x})}{\left(\frac{m(\mathbf{x})}{E(\varepsilon_t|\varepsilon_t>q(a))}+h^{1/2}(\mathbf{x}) \right)} \frac{(\hat{E}^{(b)}(\varepsilon_t|\varepsilon_t>q(a))+\hat{B}_E)-E(\varepsilon_t|\varepsilon_t>q(a))}{E(\varepsilon_t|\varepsilon_t>q(a))}=o_p(N^{-1/2})$.  Now, consider
\begin{align*}
&\frac{\hat{E}^{(b)}(\varepsilon_t|\varepsilon_t>q(a))+\hat{B}_E-E(\varepsilon_t|\varepsilon_t>q(a))}{E(\varepsilon_t|\varepsilon_t>q(a))}=\left(\frac{\hat{E}^{(b)}(\varepsilon_t|\varepsilon_t>q(a))}{\frac{q(a)}{1+k_0}}-1\right)\left(\frac{E(\varepsilon_t|\varepsilon_t>q(a))}{\frac{q(a)}{1+k_0}}\right)^{-1}\\
&-\frac{\left(  \left( \frac{\phi(q(a))}{(\rho+k_0^{-1}+1)(1+k_0^{-1})} +o\left(\phi(q(a)) \right) \right)q(a)-\hat{B}_E \right)}{E(\varepsilon_t|\varepsilon_t>q(a))}=\hat{I}_1^b-\hat{I}_2^b.
\end{align*}
 By Lemma 8, $\frac{E(\varepsilon_t|\varepsilon_t>q(a))}{\frac{q(a)}{1+k_0}}=1+o(1)$ and by part b) of Theorem \ref{thm4}, 
 $\sqrt{N}\left(\frac{\hat{E}^{(b)}(\varepsilon_t|\varepsilon_t>q(a))}{\frac{q(a)}{1+k_0}}-1\right)= \left( c_q^T-\frac{1}{1+k_0} \left(\begin{array}{cc}0&1\end{array}\right)H^{-1}(k_0)A(k_0,\rho) \right)\mathcal{q}_n+o_p(1)
 $, which is asymptotically $\mathcal{N}(0, \Sigma_2^{(b)}(k_0,\rho))$.  Hence $\sqrt{N}\hat{I}_1^b \convd \mathcal{N}(0, \Sigma_2^{(b)}(k_0,\rho))$.  We note that if $\hat{\rho}$, $\hat{\mathcal{Z}}$, $\hat{d}$ and $\tilde{k}$ are as defined in Theorem \ref{thm2}, then $\frac{\hat{\mathcal{Z}}^{\hat{\rho}}}{\hat{d}(\hat{\rho}+\tilde{k}^{-1}+1)(1+\tilde{k}^{-1})}=\frac{\mathcal{Z}^{\rho}}{d(\rho+k^{-1}+1)(1+k^{-1})}+o_p(1)$.  Since, $E(\varepsilon_t|\varepsilon_t>q(a))=\frac{q(a)}{1+k_0}+q(a)\left( \frac{\phi(q(a))}{(\rho+k_0^{-1}+1)(1+k_0^{-1})} +o\left(\phi(q(a)) \right) \right)$ we write 
 $
 \hat{I}_2^b=\frac{q(a)\left( \frac{\phi(q(a))}{(\rho+k_0^{-1}+1)(1+k_0^{-1})} +o\left(\phi(q(a)) \right)\right)-\hat{B}_E}{\frac{q(a)}{1+k_0}+q(a)\left( \frac{\phi(q(a))}{(\rho+k_0^{-1}+1)(1+k_0^{-1})} +o\left(\phi(q(a)) \right) \right)}
 $.  Given that $\frac{\hat{q}^{(b)}(a)}{q(a)}=1+O_p(N^{-1/2})$ we have that 
 \begin{align*}
 \frac{\hat{B}_E}{q(a)}
 &=\frac{\mathcal{Z}^{\rho}}{d(\rho+k_0^{-1}+1)(1+k_0^{-1})}\left( (-2-4k_0)k_0\frac{Q_n}{\sqrt{N}}+P_{2n}+ 4k_0P_{1n}+d \phi(q(a_N))+o_p(N^{-1/2})\right).
 \end{align*}
 Furthermore, as in Theorem \ref{thm3}, $q(a)=q(a_N)Z_{N,a}$ with $Z_{N,a} \rightarrow \mathcal{Z}$ and $\frac{\phi(q(a))}{\phi(q(a_N))}=\frac{\phi(q(a_N)Z_{N,a})}{\phi(q(a_N))}$.  Since, $\phi$ is regularly varying with index $\rho< 0$, $\frac{\phi(q(a_N)Z_{N,a})}{\phi(q(a_N))} \rightarrow \mathcal{Z}^\rho$ as $n\rightarrow \infty$, hence $\frac{\phi(q(a_N))\mathcal{Z}^\rho}{(\rho+k^{-1}+1)(1+k_0^{-1})}=\frac{\phi(q(a))}{(\rho+k_0^{-1}+1)(1+k_0^{-1})}(1+o_p(1))$.  Consequently,
  \begin{align*}
 \frac{\hat{B}_E}{q(a)} &=\frac{\phi(q(a))}{(\rho+k_0^{-1}+1)(1+k^{-1})}+ \frac{\mathcal{Z}^{\rho}(-2-4k_0)k_0}{d(\rho+k_0^{-1}+1)(1+k_0^{-1})}\frac{Q_n}{\sqrt{N}}
  +\frac{\mathcal{Z}^{\rho}}{d(\rho+k_0^{-1}+1)(1+k_0^{-1})}P_{2n}\\
 &+\frac{4k_0\mathcal{Z}^{\rho}}{d(\rho+k_0^{-1}+1)(1+k_0^{-1})}P_{1n} +o_p(N^{-1/2}), \mbox{ and }\\
\sqrt{N}\hat{I}_2^b 
 &=-(1+k_0)\left(\frac{\mathcal{Z}^{\rho}(-2-4k_0)k_0}{d(\rho+k_0^{-1}+1)(1+k_0^{-1})}Q_n+\frac{\mathcal{Z}^{\rho}}{d(\rho+k_0^{-1}+1)(1+k_0^{-1})}\sqrt{N}P_{2n} \right.\\
 &+\left. \frac{4k_0\mathcal{Z}^{\rho}}{d(\rho+k_0^{-1}+1)(1+k_0^{-1})}\sqrt{N}P_{1n}+o_p(1) \right).
 \end{align*}
Thus, letting $\upsilon_1(k_0,\rho) = \left( \begin{array}{ccccc} 0&0&\frac{\mathcal{Z}^{\rho}k_0(-2-4k_0)}{d(\rho+k_0^{-1}+1)}&\frac{k_0\mathcal{Z}^{\rho}}{d(\rho+k_0^{-1}+1)}&\frac{4k_0^2\mathcal{Z}^{\rho}}{d(\rho+k_0^{-1}+1)}
 \end{array}\right)$ we have
$$
\sqrt{N}\left( \frac{\hat{E}^{(b)}\left(Y_t|Y_t>q_{Y_t|\mathbf{X}_t=\mathbf{x}}(a),\mathbf{X}_t=\mathbf{x})\right)}{E\left(Y_t|Y_t>q_{Y_t|\mathbf{X}_t=\mathbf{x}}(a),\mathbf{X}_t=\mathbf{x})\right)}-1\right)\convd \mathcal{N}\left(0,\Sigma_3^{(b)}(k_0,\rho)\right), \mbox{ where } 
$$
\begin{align*}
\Sigma_3^{(b)}(k_0,\rho)&=\left( c_q^T-\frac{1}{1+k_0} \left(\begin{array}{cc}0&1\end{array}\right)H^{-1}(k_0)A(k_0,\rho)+\upsilon_1(k_0,\rho)\right)^T V^{(b)}(k_0)\\
& \times \left( c_q^T-\frac{1}{1+k_0} \left(\begin{array}{cc}0&1\end{array}\right)H^{-1}(k_0)A(k_0,\rho) +\upsilon_1(k_0,\rho) \right).
\end{align*}
\end{proof}
\setlength{\baselineskip}{12pt}
\bibliographystyle{elsart-harv}
\bibliography{martins_filho_yao_torero_final(2016).bbl}
\end{document}